\documentclass[aps,pra,twocolumn,showpacs,floatfix,
superscriptaddress, showpacs]{revtex4-2}
  \usepackage[T1]{fontenc}
  \usepackage{amsmath}
  \usepackage{amsbsy}
  \usepackage{amssymb}
\usepackage[utf8]{inputenc}
\usepackage[english]{babel}
\usepackage[pdftex,dvipsnames,usenames]{xcolor}
\usepackage[colorlinks=true,urlcolor=blue,citecolor=blue,linkcolor=blue]{hyperref}
\usepackage{lipsum}

\usepackage{graphicx, color}

\newcommand{\ket}[1]{\left|#1\right\rangle}

\begin{document}
\title{Time correlations in atmospheric quantum channels}

\author{M. Klen}
\affiliation{Bogolyubov Institute for Theoretical Physics, NAS of Ukraine, Vul. Metrologichna 14-b, 03680 Kyiv, Ukraine}

\author{D. Vasylyev}
\affiliation{German Aerospace Center, Institute for Solar-Terrestrial Physics, D-17235 Neustrelitz, Germany}

\author{W. Vogel}
\affiliation{Institut f\"ur Physik, Universit\"at Rostock, Albert-Einstein-Stra\ss{}e 23, D-18059 Rostock, Germany}

\author{A. A. Semenov}

\affiliation{Bogolyubov Institute for Theoretical Physics, NAS of Ukraine, Vul. Metrologichna 14-b, 03680 Kyiv, Ukraine}
\affiliation{Institute of Physics, NAS of Ukraine, Prospect Nauky 46, 03028 Kyiv, Ukraine}
\affiliation{Department of Theoretical and Mathematical Physics, Kyiv Academic University, Boulevard Vernadskogo  36, 03142  Kyiv, Ukraine}

\begin{abstract}
Efficient transfer of quantum information between remote parties is a crucial challenge for quantum communication over atmospheric channels. 
Random fluctuations of the channel transmittance are a major disturbing factor for its practical implementation.
We study correlations between channel transmittances at different moments of time and focus on two transmission protocols.
The first is related to the robustness of both discrete- and continuous-variable entanglement between time-separated light pulses, showing a possibility to enlarge the effective dimension of the Hilbert space.
The second addresses a selection of high-transmittance events by testing them with bright classical pulses followed by quantum light.
Our results show a high capacity of the time-coherence resource for encoding and transferring quantum states of light in atmospheric channels.  
\end{abstract}

\maketitle

\section{Introduction} 


Quantum technologies are opening up fascinating perspectives, but they also face a serious problem: the fragility of quantum information against environmental noise.
For instance, in various tasks of quantum communication \cite{gisin02,Xu2020,Pirandola2020,Renner2023,Gottesman01,bennett93,Braunstein1998,Zukowski1993}, as well as in fundamental studies \cite{ursin07,fedrizzi09}, we use quantum light as a natural carrier for the transmission of quantum states over large distances.
The nature of this noise is well studied and theoretically described, but importantly, it is strictly dependent on the type of communication channel---an optical fiber or free space.
The latter has many practical advantages, such as the possibility of establishing satellite-mediated global communication, communication with moving objects and through hard-to-access regions, and so on; see, e.g., Refs.~\cite{ursin07,elser09,heim10,fedrizzi09,capraro12,yin12,ma12,peuntinger14,vasylyev17,Jin2019,nauerth13,wang13,bourgoin15,vallone15,dequal16,vallone16,carrasco-casado16,takenaka17,liao17,yin17,gunthner17,ren17,yin17b,Liao2018,Vasylyev2019,Ecker2021,Ecker2022} for various implementations.
However, in this scenario, quantum protocols are seriously affected by atmospheric turbulence and stray light. 


Theoretical description of quantum light distributed through the turbulent atmosphere relies on methods \cite{Tatarskii, Tatarskii2016, Fante1975,Fante1980,Andrews_book} of classical optics.
Nevertheless, the theoretical techniques are different depending on the light degrees of freedom and measurements involved in the transmission protocol.
For example, a group of protocols \cite{usenko12,guo2017,papanastasiou2018, derkach2020b,hosseinidehaj2019,ShiyuWang2018,chai2019, Derkach2021,pirandola2021,hosseinidehaj2021,pirandola2021b,pirandola2021c,peuntinger14,Wang2018,hofmann2019,zhang2017,villasenor2021,bohmann16,hosseinidehaj15a,elser09,heim10,usenko12,peuntinger14,heim14} deals with continuous-variable (CV) quantum states of a quasimonochromatic field.
In this case, the quantum state is modified by fluctuating losses caused by passing the randomly shaped beam through the transmitter aperture \cite{perina73,milonni04,semenov09,semenov12,vasylyev12,vasylyev16,vasylyev18,semenov2018,klen2023}.
A similar description \cite{semenov10,gumberidze16} is employed for transmission protocols involving the polarization degree of freedom, often used, e.g., for Bell inequality tests \cite{ursin07,fedrizzi09}. 
 
Spatial structure of light beams represents another resource for encoding quantum states of light sent through the turbulent atmosphere.
Firstly, spatial modes that are different from Gaussian beams, can show a better transmittance for particular realizations of free space channels \cite{Klug2023,Bachmann2023}.
Secondly, single photons and photon pairs can also encode more quantum information, by considering, for example, the optical angular momentum \cite{paterson05,pors11,Roux2013,Roux2015,Leonhard2015,Lavery2018,Sorelli2019}.
Thirdly, higher-order light modes, such as Hermite-Gaussian modes, have the potential to increase the effective dimensionality of the Hilbert space of transferred states. 


An alternative approach is to use the time instance of lightpulse generation as a degree of freedom for encoding quantum states.
For example, in Refs.~\cite{Bulla2023a,Bulla2023b} time-bin entanglement is considered for photons propagating in different directions.
In a sense, time encoding was also used in the experiment reported in Ref.~\cite{fedrizzi09}, where both entangled photons were copropagated with a small time delay through the 144~km quantum channel in the Canary Islands.
In this case, one can say that the effective dimension of the Hilbert space for the transmitted light is doubled by using two light pulses (time modes) \cite{Cozzolino2019}.    


As shown in Refs.~\cite{semenov10,gumberidze16} and Ref.~\cite{bohmann16}, correlations between the transmittances of two modes are essential for preserving discrete- and continuous-variable entanglement, respectively.
In the case of pulse copropagation, they correspond to time correlations between the transmittances of two (or more) pulses.
In this paper, we study the dependence of these correlations on the time between pulses and their impact on the entanglement robustness.
This protocol can also be considered as a way to enlarge the effective dimension of the Hilbert space \cite{Cozzolino2019} for the transmitted states.    

We also consider time correlations as a resource for adaptive real-time selection protocols, proposed and implemented in Refs.~\cite{Tang2013,vallone2015} in the context of quantum-key distribution.
In this scenario, a bright pulse of classical light is used to test the channel transmittance.
If it exceeds a predetermined value, the quantum light pulses are then sent and analyzed at the receiver.
Specifically, we consider preserving nonclassical properties of photocounting statistics.
As is discussed in Refs.~\cite{semenov09,semenov2018}, the sub-Poissonian character of the photon-number statistics is sensitive to atmospheric turbulence.
However, an application of recently proposed methods \cite{semenov2021,kovtoniuk2023} has shown that the photocounting statistics, or more specifically click statistics, retain their nonclassical properties even when estimated by detectors with realistic photon-number resolution.  

The rest of the paper is organized as follows.
In Sec.~\ref{Sec:TransProtocols}, we describe the transmission protocols considered in this paper and introduce the input-output relations between quantum states at the transmitter and receiver.
Numerical simulations of the two-time probability distribution of transmittance (PDT), which is the main channel characteristic for the discussed protocols, are reported in Sec.~\ref{Sec:TwoTimePDT}.
In Secs.~\ref{Sec:Gaussian} and \ref{Sec:Bell} we consider the distribution of Gaussian and discrete-variable entanglement of two consecutive pulses, respectively.
The adaptive real-time selection protocol is considered in Sec.~\ref{Sec:Preselection}.
Summary and concluding remarks are given in Sec.~\ref{Sec:Conclusions}.
Supplemental Material \cite{supplement} include the \textsf{PYTHON 3} code and numerically simulated data.

\section{Transmission protocols}
\label{Sec:TransProtocols}

The first group of transmission protocols considered in this paper assumes encoding quantum states in two optical pulses separated by the time interval $\tau$.
Each pulse can be considered as a quasimonochromatic mode, e.g., prepared in the form of a Gaussian beam; see Ref.~\cite{klen2023}. 
Moreover, one can additionally use the polarization degree of freedom for each pulse, such that the total number of modes is four.
We aim to find the time intervals $\tau$ for which quantum correlations between two pulses, such as entanglement, are still preserved.

Atmospheric turbulence randomly changes the shapes of propagating beams.
As a result, the light pulses pass through the receiver aperture with fluctuating transmittances $\eta_0$ and $\eta_\tau$ at the time instances $t=0$ and $t=\tau$, respectively.
It is worth noting that the transmittances for two polarization modes at the same time instance are almost perfectly correlated due to the negligible depolarization effect in the atmosphere \cite{Tatarskii}.
The methods of Refs.~\cite{semenov09,semenov12,vasylyev12,vasylyev16,vasylyev18,semenov2018,klen2023} yield the input-output relation between the quantum states of the pulses at the receiver and the transmitter,
    \begin{align}\label{Eq:PInOut}
    \hat{\rho}_\mathrm{out}=\int_\Xi d\boldsymbol{\eta}\mathcal{P}\left(\boldsymbol{\eta}\right) \mathcal{L}\left[\boldsymbol{\eta}\right]\!(\hat{\rho}_{\mathrm{in}}).
    \end{align}
Here $\boldsymbol{\eta}=\begin{pmatrix} \eta_0 & \eta_\tau \end{pmatrix}$ is the vector of two transmitances, $\Xi=[0,1]\times[0,1]$ is the domain of their integration, $\hat{\rho}_\mathrm{in}$ and $\hat{\rho}_\mathrm{out}$ are the density operators at the transmitter and receiver, respectively, $\mathcal{L}\left[\boldsymbol{\eta}\right]\!(\hat{\rho}_\mathrm{in})$ is the superoperator describing the effect of linear losses with transmittances $\boldsymbol{\eta}$ on the density operator $\hat{\rho}_\mathrm{in}$, and $\mathcal{P}\left(\boldsymbol{\eta}\right)$ is the two-time probability distribution of transmittances (PDT). 

The action of the superoperator $\mathcal{L}\left[\boldsymbol{\eta}\right]$ can be conveniently expressed in the Glauber-Sudarshan $P$ representation \cite{glauber63c,sudarshan63}.
If $P_\mathrm{in}(\boldsymbol{\alpha})$ corresponds to the density operator $\hat{\rho}_\mathrm{in}$, then it is given by
    \begin{align}\label{eq:Losses1}
        \mathcal{L}\left[\boldsymbol{\eta}\right]P_\mathrm{in}(\alpha_0,\alpha_\tau)=
        \frac{1}{\eta_0\eta_\tau}P_\mathrm{in}\left(\frac{\alpha_0}{\sqrt{\eta_0}},\frac{\alpha_\tau}{\sqrt{\eta_\tau}}\right).
    \end{align}
If two polarization modes are involved for each of time instances, then 
    \begin{align}
        \mathcal{L}\left[\boldsymbol{\eta}\right]P_\mathrm{in}&(\alpha_{\textrm{h}0},\alpha_{\textrm{v}0},\alpha_{\textrm{h}\tau},\alpha_{\textrm{v}\tau})=\\
        &\frac{1}{\eta_0^2\eta_\tau^2}P_\mathrm{in}\left(\frac{\alpha_{\textrm{h}0}}{\sqrt{\eta_0}},\frac{\alpha_{\textrm{v}0}}{\sqrt{\eta_0}},\frac{\alpha_{\textrm{h}\tau}}{\sqrt{\eta_\tau}},\frac{\alpha_{\textrm{v}\tau}}{\sqrt{\eta_\tau}}\right).\nonumber
    \end{align}
Here the indices h and v indicate the modes with horizontal and vertical polarization, respectively.

\begin{figure}[ht!]
        \centering
        \includegraphics[width=1\linewidth]{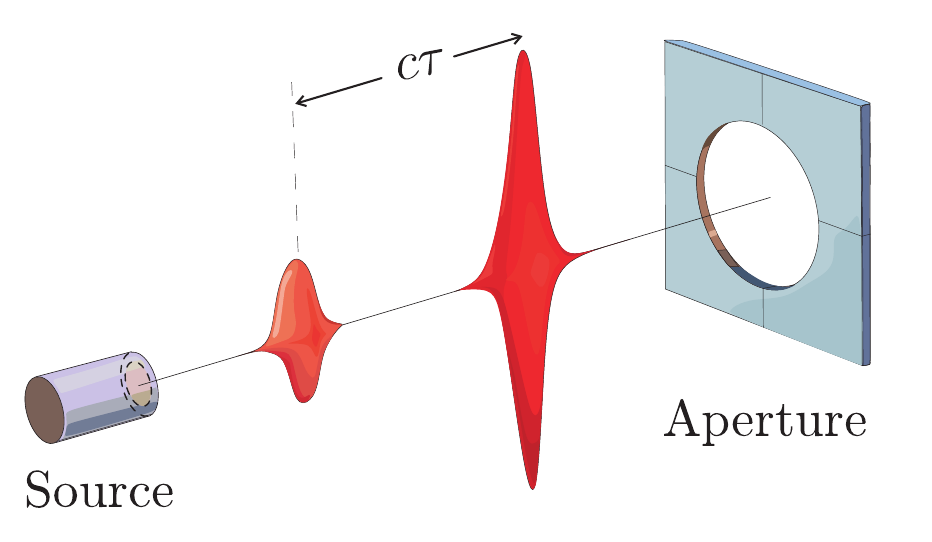}
        \caption{Sketch of the adaptive real-time selection protocol.
        A bright classical pulse is sent through the channel to test the transmittance at the time instance $t=0$. 
        If it exceeds the predetermined value $\eta_\textrm{min}$, then a quantum-light pulse is sent at $t=\tau$.
        The distance between pulses is then $c\tau$ with $c$ being the speed of light.}
        \label{fig:preselection}
    \end{figure}

The second group of transmission protocols (see Refs.~\cite{Tang2013,vallone2015}) assumes that the channel transmittance $\eta_0$ is measured with a classical pulse and, if it exceeds a predetermined threshold $\eta_\mathrm{min}$, it is followed by a pulse of quantum light whose transmittance is $\eta_\tau$; see Fig.~\ref{fig:preselection}.
A key question here is the time interval $\tau$ for which nonclassical properties of light are preserved.
Input-output relations for this case are given by the conditional version of Eq.~(\ref{Eq:PInOut}),
    \begin{align}
         \hat{\rho}_\mathrm{out}=\int\limits_0^1 d\eta_\tau\mathcal{P}\left(\eta_\tau|\eta_0 \geq  \eta_\mathrm{min}\right) \mathcal{L}\left[\eta_\tau\right]\!(\hat{\rho}_{\mathrm{in}}).
    \end{align}
Here   
    \begin{align}\label{eq:condPDT}
        \mathcal{P}(\eta_\tau|\eta_0 \geq  \eta_\mathrm{min}) = \frac{1}{\overline{\mathcal{F}}(\eta_\mathrm{min})} \int_{\eta_\mathrm{min}}^{1} \mathrm{d}\eta_0 \mathcal{P}(\eta_\tau,\eta_0).
    \end{align}
is the conditional PDT and    
     \begin{align}\label{eq:preseleff}
     \overline{\mathcal{F}}(\eta_\mathrm{min}) = \int_{\eta_\mathrm{min}}^{1} \mathrm{d}\eta_0 \mathcal{P}(\eta_0)
     \end{align}
is the single-time complementary cumulative probability distribution function (exceedance) at the point $\eta=\eta_\textrm{min}$, which characterizes the overall efficiency of the adaptive real-time selection procedure.
The function 
    \begin{align}\label{eq:Single-Time-PDT}
       \mathcal{P}(\eta_0)=\int\limits_0^{\infty}d\eta_\tau \mathcal{P}(\eta_\tau,\eta_0) 
    \end{align}
is the single-time PDT, considered in Refs.~\cite{semenov09,semenov12,vasylyev12,vasylyev16,vasylyev18,semenov2018,klen2023}.

\section{Two-time PDT}
\label{Sec:TwoTimePDT}
As discussed in Refs.~\cite{semenov09,semenov12,vasylyev12,vasylyev16,vasylyev18,semenov2018,klen2023}, the PDT has the same form in both quantum and classical optics, and, consequently, its derivation is based on purely classical methods. 
Atmospheric turbulence leads to random fluctuations in space and time of the refractive index $n(\mathbf{r},z;t)$.
The random value of the transmittance for the pulse at the time instance $t$ reads
    \begin{align}\label{Eq:Efficiency}
        \eta_t=\int_\mathcal{A}d^2\mathbf{r} \left|u_t(\mathbf{r},z_\mathrm{ap})\right|^2.
    \end{align}
Here $|u_t(\mathbf{r},z)|$ is the field amplitude amplitude, the coordinate $z$ is chosen in the propagation direction, $\mathbf{r}$ is the two-dimensional vector of the transverse coordinates, $z=z_{\mathrm{ap}}$ is the channel length,  and $\mathcal{A}$ is the receiver-aperture opening area.
As discussed in Appendix~A of Ref.~\cite{klen2023}, the transmittance (\ref{Eq:Efficiency}) does not depend on the pulse (quasimonochromatic mode) time shape (spectrum).
The field amplitude $u_t(\mathbf{r},z)$ is a solution to the paraxial equation with the inhomogeneous refractive index $n_t(\mathbf{r},z)=1+\delta n_t(\mathbf{r},z)$,
		\begin{align}\label{Eq:Paraxial}
			2ik\frac{\partial u_t(\mathbf{r},z)}{\partial z}+\Delta_\mathbf{r}
			u_t(\mathbf{r},z)+2k^2\delta n_t(\mathbf{r},z) u_t(\mathbf{r},z)=0,
		\end{align}
where $k$ is the wave number.
Since the evolution time of the atmosphere is several orders of magnitude larger than the pulse propagation time, the refractive index for a given pulse can be considered independent of time.
However, the time evolution of $n_t(\mathbf{r},z)$ between successive pulses should be taken into account.
Moreover, $\delta n_t(\mathbf{r},z)$ is a fluctuating part of the refractive index with zero mean.

According to Taylor's frozen turbulence hypothesis \cite{taylor1938}, the time evolution of the refractive index can be described by a wind-driven shift, compared to which the proper time evolution of the turbulence eddies is much slower.
We assume that the longitudinal and transverse components of the wind velocity are comparable.
As discussed in Ref.~\cite{Tatarskii2016}, the longitudinal component does not contribute significantly to the radiation-field statistics if its ratio to the transverse component is much smaller than $\sqrt{z_{\mathrm{ap}}/\lambda}$, where $\lambda=2\pi/k$ is the wavelength.
In the considered scenario, this conclusion is also supported by another argumentation.
During the time when the refractive-index field is shifted transversely by the size of the aperture opening, its longitudinal shift is insignificant compared to the channel length.
Therefore, we consider only the transverse component of the wind shift.

To solve the paraxial equation (\ref{Eq:Paraxial}) numerically, we use the phase-screen method \cite{Fleck1976,Frehlich2000,Lukin_book,Schmidt_book}; see also Refs.~\cite{Lavery2018,Sorelli2019,Klug2023,Bachmann2023} for its recent applications.
This method can be summarized as follows: (i) the propagation distance is divided into $M$ intervals; (ii) at the center of each interval, one samples a random phase screen that corresponds to a given realization of the refractive index; (iii) the field amplitude is simulated according to the vacuum propagation [$\delta n_t(\mathbf{r},z)=0$] between the phase screens; (iv) on each phase screen, the $\mathbf{r}$-dependent phase is incremented.  
Similar to Ref.~\cite{klen2023}, we apply the sparse-spectrum model \cite{Charnotskii2013a,Charnotskii2013b,Charnotskii2020} of the phase-screen method.
In the context of this paper this model has two advantages.
Firstly, it enables the generation of long phase screens, which are directly used to model the wind shift. 
Secondly, the phase-perturbation statistics obtained with this model are in good agreement with the analytical expression over the entire length of the phase screen.

The two-time PDT is a more general characteristic of atmospheric quantum channels than the single-time PDT (\ref{eq:Single-Time-PDT}) (cf. Refs.~\cite{semenov09,semenov12,vasylyev12,vasylyev16,vasylyev18,semenov2018,klen2023}) since it includes information about time correlations.
We sample the two-time PDT by using the sparse-spectrum model of the phase-screen method.
Let us choose the long sides of the phase screens to be directed along the $x$ axis, i.e., along the direction of the transverse component of the wind velocity.
We start simulations with the time instance $t=0$.
In this case, the beam crosses the phase screens near one of their short sides.
This gives us the field amplitude $u_0(\mathbf{r},z_\mathrm{ap})$ at the receiver plane.
Next, according to Taylor's hypotheses, we shift the phase screens along the $x$-axis on the distance $s=v\tau$, where $v$ is the transverse wind velocity and repeat simulations in order to obtain $u_\tau(\mathbf{r},z_\mathrm{ap})$.
Finally, we apply Eq.~(\ref{Eq:Efficiency}) to find a random vector $\boldsymbol{\eta}=\begin{pmatrix} \eta_0 & \eta_\tau \end{pmatrix}$.
This procedure is repeated multiple times with newly generated phase screens in order to accumulate a sampling set of $\boldsymbol{\eta}$, which we use for the estimation of the two-time PDT.

Our procedure involves a method to overcome the computational complexity associated with the generation of long phase screens. 
Within the sparse-spectrum model of the phase-screen method, we first generate and store the spectra for each of $M$ phase screens. 
These spectra are used to generate the square-shaped phase screens centered on the $z$-axis for an arbitrary time instance $t$. 
This procedure is equivalent to the generation and subsequent shifting of long phase screens.
The sampled data and \textsf{PYTHON 3} codes can be found in the Supplemental Material \cite{supplement}.

Our model assumes that the transverse wind velocity is constant for the whole propagation distance.
In real conditions, it may vary randomly for different parts of the quantum channel and during the data-collection time.
These conditions are related to a particular micro-meteorological situation.
We restrict our consideration to the simplest scenario with the aim to understand the general role of time correlations in quantum communication through atmospheric channels.
Moreover, our results can be easily extended to any particular wind-velocity distribution.

We simulate the two-time PDTs for channels with $z_\mathrm{ap}=50~\textrm{km}$.
Turbulence is described by the modified von K\'arm\'an--Tatarskii spectrum \cite{Fante1975} with three different values of the refractive-index structure constant  ($C_n^2=1\times10^{-16}~\textrm{m}^{-2/3}$, $C_n^2=2\times10^{-16}~\textrm{m}^{-2/3}$, and $C_n^2=3\times10^{-16}~\textrm{m}^{-2/3}$)  and the inner and outer turbulence scales $\ell_0=1~\textrm{mm}$ and $L_0=80~\textrm{m}$, respectively. 
The wavelength is $\lambda=808~\textrm{nm}$.
This implies that the Rytov parameter $\sigma_{\mathrm{R}}^2$, characterizing the strength of the scintillation, is $5.5$, $11$, $16.5$.
At the transmitter side, $z=0$, the beam is chosen in a Gaussian form,
	\begin{align}\label{Eq:BoundaryConditions}
			u_t(\mathbf{r},0)=\sqrt{\frac{2}{\pi
			W_0^2}}\exp\Bigl[-\frac{\mathbf{r}^2}{W_0^2}{-}\frac{ik
			}{2F_0}\mathbf{r}^2\Bigr].
	\end{align}
Here the beam-spot and wavefront radii are chosen to be $W_0=8~\textrm{cm}$ and $F_0=50~\textrm{km}$, respectively.
In the Supplemental Material \cite{supplement} we also present the results for other channels.

For the numerical simulations we use a spatial grid with 2048 points along each axis.
The spatial grid step is $1~\textrm{mm}$.
The number of spectral rings is 1024.
The inner and outer bounds of the spectrum are $K_\mathrm{min}=1/15 L_0$ and $K_\mathrm{max}=2/\ell_0$.
The number of phase screens is 15, as recommended in Refs.~\cite{Schmidt_book,Martin1988}.
The number of samples is $5\times10^4$.
For details see Ref.~\cite{klen2023}.

\begin{figure}[h!]
    \includegraphics[width=1\linewidth]{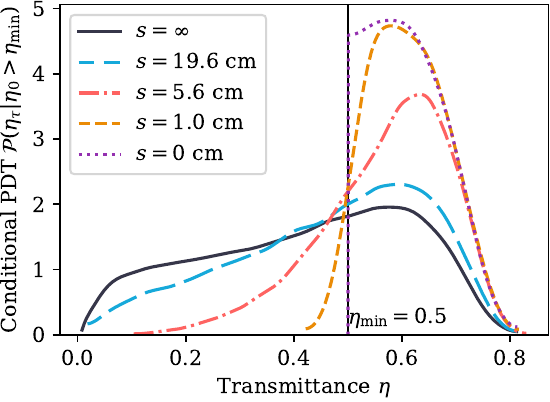}
    \caption{The conditional PDT (\ref{eq:condPDT}) is shown for transmittance threshold $\eta_\mathrm{min}=0.5$, refractive-index structure constant $C_n^2=2\times10^{-16}~\textrm{m}^{-2/3}$, aperture radius $R_\mathrm{ap}=30~\textrm{cm}$, and for different values of the wind-driven shift $s$.}
    \label{fig:condPDT}
\end{figure}

The results of the simulations for the conditional PDT [cf. Eq.~(\ref{eq:condPDT})] are given in Fig.~\ref{fig:condPDT}. 
This distribution preserves mostly its initial shape at $t=0$, if the magnitude of the wind-driven shift $s$ is within the range up to one centimeter.
This corresponds to one millisecond for the time $\tau$ under the typical transverse wind velocity $v=10~\textrm{m/s}$.
Unless explicitly stated otherwise, we will use this value of wind velocity in the following considerations, assuming that  the time $\tau$ for different values of $v$ can be obtained as $\tau=s/v$.
If the magnitude of $s$ is in the range of a few centimeters (milliseconds for $\tau$), then the contribution of the transmittance values $\eta<\eta_\mathrm{min}$ becomes significant.
For tens of centimeters for $s$ (tens of milliseconds for $\tau$), the correlations vanish and the conditional PDT takes the form of the single-time PDT (\ref{eq:Single-Time-PDT}).
 
\begin{figure}[h!]
    \includegraphics[width=1\linewidth]{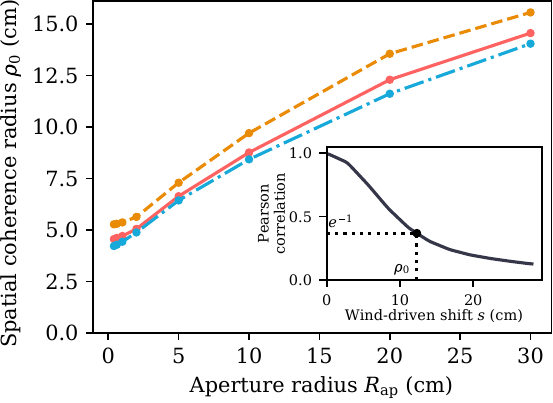}
    \caption{\label{fig:eta0_etat_corr}Aperture-averaged spatial coherence radius $\rho_0$ as a function of the aperture radius is shown.
    The inset demonstrates the Pearson correlation coefficient between transmittances as a function of $s$  with the indication of the coherence radius $\rho_0$.
	The dashed, solid, and dot-dashed lines correspond to $C_n^2=1\times10^{-16}~\textrm{m}^{-2/3}$, $C_n^2=2\times10^{-16}~\textrm{m}^{-2/3}$, and $C_n^2=3\times10^{-16}~\textrm{m}^{-2/3}$, respectively.}
\end{figure}

The simulations also reveal that the Pearson correlation coefficient between two transmittances decreases with the time $\tau$ (wind-driven shift $s$); see inset in Fig.~\ref{fig:eta0_etat_corr}.
The aperture-averaged spatial coherence radius $\rho_0$ can be defined as the value of $s$, for which this correlation is equal to $e^{-1}$; see, e.g., Ref.~\cite{Andrews_book}. 
This quantity can be considered as a maximum wind-driven shift (time) for which correlations are preserved.
It depends on the radius of the receiver aperture as a monotonically increasing function; see Fig.~\ref{fig:eta0_etat_corr}.
Hence, we can expect that the time $\tau$ for which nonclassical properties of the radiation are preserved increases with the receiver-aperture radius $R_\textrm{ap}$.
This result has direct impact on the scenarios of quantum-state transmission considered in the following sections.

\section{Gaussian entanglement between pulses}
\label{Sec:Gaussian}
Let us consider the following scenario.
The two-mode squeezed vacuum state (TMSVS)
    \begin{align}\label{eq:TWSVS}
       \ket{\xi}=\cosh^{-2}\xi\sum\limits_{n}^{+\infty}(-\tanh \xi)^n\ket{n,n} 
    \end{align}
is generated on the transmitter side, where $\ket{n,n}$ is the two-mode Fock state and $\xi$ is the squeezing parameter.
One mode is sent at the time instance $t=0$.
The other mode is stored in a quantum memory (see, e.g., Refs.~\cite{Julsgaard2004,Lvovsky2009,Hedges2010,Jensen2011,Yan2018,Ma2022}) and sent through the atmosphere at the time instance $t=\tau$. 
At the receiver station, pulses are analyzed.
If the corresponding measurement involves homodyne detection, then the local oscillator is also sent in the same spatial but orthogonally polarized mode \cite{elser09,heim10,semenov12}.

We aim to determine the Gaussian entanglement of the state at the receiver station.
For this purpose we will use the Simon inseparability criterion \cite{simon00}.
Applying it to the input state (\ref{eq:TWSVS}) and the input-output relation (\ref{Eq:PInOut}) [cf. also Eq.~(\ref{eq:Losses1})], we get that the state at the receiver preserves Gaussian entanglement iff the Simon certifier
    \begin{align}
        \mathcal{W}=&\sinh^2\xi\Big[-\left\langle\sqrt{\eta_0\eta_\tau}\right\rangle^2\cosh^2\xi+\left\langle\eta_0\right\rangle\left\langle\eta_\tau\right\rangle\sinh^2\xi\Big]\nonumber\\
        &\times\Big[1-\frac{\left\langle\sqrt{\eta_0\eta_\tau}\right\rangle^2}{4}\sinh^22\xi\nonumber\\
        &+\sinh^2\xi\left(\left\langle\eta_0\right\rangle+\left\langle\eta_\tau\right\rangle+\left\langle\eta_0\right\rangle\left\langle\eta_\tau\right\rangle\sinh^2\xi\right)\Big]\label{eq:Simon}
    \end{align}
is negative, i.e., $\mathcal{W}<0$.
The expectation values $\left\langle\sqrt{\eta_0\eta_\tau}\right\rangle$ and the first moments of transmitances, $\left\langle\eta_0\right\rangle$ and $\left\langle\eta_\tau\right\rangle$, are calculated from the numerically simulated data.

Deterministic losses in the channel ($0.1~\textrm{dB/km}$, see Ref.~\cite{scarani09}) and losses at the optical system should also be included in the transmittances $\eta_0$ and $\eta_\tau$.  
In addition, we incorporate the quantum-memory effect by primarily considering the writing and reading mapping losses while assuming zero excess noise.
Under the conditions considered, the multiplier in the second brackets of Eq.~(\ref{eq:Simon}) is always positive.
Therefore, the sign of the Simon certifier $\mathcal{W}$ is determined by the multiplier in the first brackets of Eq.~(\ref{eq:Simon}).
In this case, all deterministic losses, including those related to the quantum memory, do not change the sign of $\mathcal{W}$.  

\begin{figure}[h!]
        \includegraphics[width=1\linewidth]{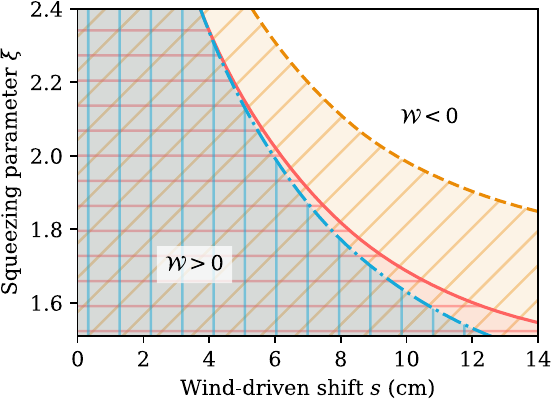}
        \caption{\label{fig:GE} Squeezing parameter $\xi$ vs. the wind-driven shift $s$ for the value of the Simon certifier $\mathcal{W}=0$ is shown.
        The dashed, solid, and dot-dashed lines correspond to $C_n^2=1\times10^{-16}~\textrm{m}^{-2/3}$, $C_n^2=2\times10^{-16}~\textrm{m}^{-2/3}$, and $C_n^2=3\times10^{-16}~\textrm{m}^{-2/3}$, respectively.
        The hatched (shaded) areas correspond to the domains, where Gaussian entanglement is preserved.
        The aperture radius is $R_\textrm{ap}=20~\textrm{cm}$.}    
    \end{figure}

Gaussian entanglement in the turbulent atmosphere has been analyzed in Ref.~\cite{bohmann16} in the context of fully correlated and anticorrelated transmittances.
In the considered case, we deal with the intermediate scenario in which the correlations depend on the time $\tau$ between pulses.
In Fig.~\ref{fig:GE} we show the domain in the space of the squeezing parameter $\xi$ and the wind-driven shift $s$, where Gaussian entanglement is preserved.
Counterintuitively, increasing the squeezing parameter decreases the maximum possible wind-driven shift for which entanglement is still retained.
Hence, the strong squeezing cannot be considered as a resource for preserving entanglement in atmospheric channels.
However, even for $\xi=2$ ($17.4~\textrm{dB}$ of squeezing) and for $C_n^2=2\times10^{-16}~\textrm{m}^{-2/3}$ Gaussian entanglement exists up to $s=6.4~\textrm{cm}$ ($\tau=6.4~\textrm{ms}$ for $v=10~\textrm{m/s}$).
Therefore, Gaussian entanglement between light pulses is highly stable even if the time interval between them significantly exceeds $1~\textrm{ms}$.
At the same time, the absolute value of the Simon certifier (\ref{eq:Simon}) can be small due to the reading mapping losses of the quantum memory.
 
\begin{figure}[h!]
        \includegraphics[width=1\linewidth]{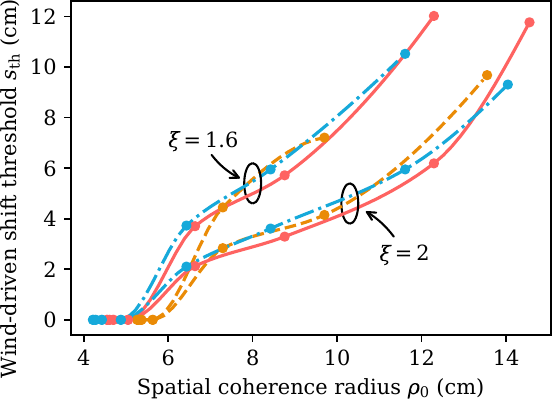}
        \caption{\label{fig:Q-Cl-CohL} 
        Threshold value of the wind-driven shift, $s_\mathrm{th}$ (for which $\mathcal{W}=0$), is shown as a function of the coherence radius $\rho_0$ for different values of the squeezing parameter $\xi$ and the refractive-index structure constant.
        The dashed, solid, and dot-dashed lines correspond to $C_n^2=1\times10^{-16}~\textrm{m}^{-2/3}$, $C_n^2=2\times10^{-16}~\textrm{m}^{-2/3}$, and $C_n^2=3\times10^{-16}~\textrm{m}^{-2/3}$, respectively.
        }    
    \end{figure}

Let us consider the threshold value of the wind-driven shift, $s_\mathrm{th}$, for which $\mathcal{W}=0$.
Obviously entanglement is witnessed, i.e., $\mathcal{W}<0$, only for $s<s_\mathrm{th}$.
This threshold value can be considered as a function of the aperture radius $R_\mathrm{ap}$.
On the other hand, the coherence radius $\rho_0$ is also a function of $R_\mathrm{ap}$; see Fig.~\ref{fig:eta0_etat_corr}.
Therefore, we can consider $s_\mathrm{th}$ as a parametric function of $\rho_0$, which also depends on the channel parameters and the squeezing parameter $\xi$. 
As shown in Fig.~\ref{fig:Q-Cl-CohL}, this is a monotonically increasing nonlinear function.
As expected, $s_\mathrm{th}$ is smaller for larger squeezing.

\section{Discrete-variable entanglement between pulses}
\label{Sec:Bell}

In this scenario, a four-mode state is generated at the transmitter side: two polarization modes (horizontal and vertical) for each time instances, $t=0$ and $t=\tau$.
In the ideal scenario, one can consider the polarization-encoded Bell state of two pulses,
    \begin{align}
        &\ket{\mathcal{B}}=\frac{1}{\sqrt{2}}\Big(\ket{\mathrm{h}}_0
        \ket{\mathrm{v}}_\tau-
        \ket{\mathrm{v}}_0\ket{\mathrm{h}}_\tau
        \Big)\label{eq:BellState2}\\\
        &=\frac{1}{\sqrt{2}}\Big(\ket{1}_\mathrm{h0}\ket{0}_\mathrm{v0}
        \ket{0}_\mathrm{h\tau}\ket{1}_\mathrm{v\tau}-\ket{0}_\mathrm{h0}\ket{1}_\mathrm{v0}
        \ket{1}_\mathrm{h\tau}\ket{0}_\mathrm{v\tau}\Big).\nonumber
    \end{align}
Here the indices in the second line indicate the polarization and time modes, and $\ket{\textrm{h}}_t$ and $\ket{\textrm{v}}_t$ are the single-photon states in the horizontally and vertically polarized modes, respectively, for the time instance $t$.
In the case of using the parametric down-conversion (PDC) source, the state at the transmitter (cf. Refs.~\cite{Kok2000,Ma2007}) is given by
    \begin{align}
        \ket{\mathrm{PDC}}=(\cosh\xi)^{-2}\sum\limits_{n=0}^{+\infty}
        \sqrt{n+1}\tanh^n\xi\left|\Phi_n\right\rangle,\label{eq:PDC1}
    \end{align}
where $\xi$ is the squeezing parameter and
    \begin{align}
        &\left|\Phi_n\right\rangle=\label{eq:PDC2}\\
        &\frac{1}{\sqrt{n+1}}\sum\limits_{m=0}^{n}(-1)^m
        \left|n{-}m\right\rangle_\mathrm{h0}\left|m\right\rangle_\mathrm{v0}\left|m\right\rangle_\mathrm{h\tau}\left|n{-}m\right\rangle_\mathrm{v\tau}\nonumber
    \end{align}
is a multipair state such that $\ket{\Phi_1}=\ket{\mathcal{B}}$.
Measurements with such a state can be treated as measurements with the Bell state \cite{Semenov2011} in the framework of the so-called squash model \cite{Beaudry2008,Moroder2010,Fung2011}.
This model claims that a consistent mapping of continuous-variable PDC states to discrete-variable Bell states is possible only if one assigns a random value to the measurement outcome at the side where two detectors are clicked simultaneously.

In fact, we describe a scenario, which is similar to the experiment in Ref.~\cite{fedrizzi09}.
In that case, the entangled photons copropagate with the time delay $\tau$, their polarization is analyzed, and the Bell parameter $\mathcal{B}$ in the Clauser-Horn-Shimony-Holt (CHSH) form \cite{CHSH} is estimated.
If the parameter $\mathcal{B}>2$, then Bell nonlocality and, consequently, entanglement is witnessed.
In particular, this means that the channel preserves the quantum correlations between two qubits and the transferred quantum state belong to the four-dimensional Hilbert space $\mathbb{C}^4$ or is mapped onto it by the squash model.
Thus, using two light pulses instead of one, whose quantum states belong to the two-dimensional Hilbert space $\mathbb{C}^2$, increases the dimensionality of the Hilbert space \cite{Cozzolino2019}.

The time between pulses, $\tau$, in the considered experiment is $50~\textrm{ns}$.
This is significantly less than the channel correlation time related to the aperture-averaged coherence radius $\rho_0$ (see Fig.~\ref{fig:eta0_etat_corr}) such that the transmittances remain perfectly correlated.
As discussed in Refs.~\cite{semenov10,gumberidze16}, the correlations of the channel transmittances are a key factor for Bell-type experiments in atmospheric channels.
Therefore, it is important to find dependencies of the measured Bell parameter on such time $\tau$ between pulses when the correlations become imperfect. 

We have performed the analysis by directly following calculations in Ref.~\cite{gumberidze16} but assuming partially correlated transmittances obeying the numerically simulated two-time PDT.
Similarly to the case of Gaussian entanglement, we have accounted for the quantum-memory effect as well as deterministic losses in the atmosphere and in the receiver optical system.
The latter also involves $3~\textrm{dB}$ of losses at the beam splitter, which sorts the received photons; see Ref.~\cite{fedrizzi09} for details.
The time-dependent reading mapping losses of the quantum memory are described by an exponential decay, which is a reasonable approximation to the model discussed in Refs.~\cite{Yan2018,Ma2022}.
For the considered scheme, it is also important to account for the noise counts related to dark counts and stray light; see Refs.~\cite{Pratt1969,Karp1970,Lee2005,Semenov2008}.

\begin{figure}[ht!]
        \includegraphics[width=1\linewidth]{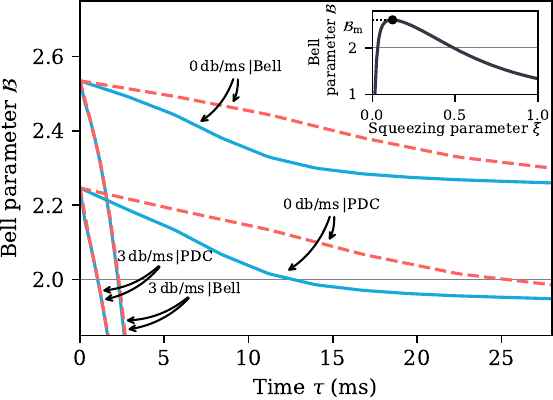}
        \caption{Bell parameter $\mathcal{B}$ vs the time $\tau$ between entangled pulses is shown for the decay rate of the quantum-memory reading mapping efficiency of $3~\textrm{dB/ms}$ and $0~\textrm{dB/ms}$.
        The solid and dashed lines corresponds to the transverse wind velocity $v=10~\textrm{m/s}$ and $v=5~\textrm{m/s}$, respectively.
        The lines marked by "PDC" and "Bell" correspond to the PDC [cf. Eq.~(\ref{eq:PDC1})] and Bell [cf. Eq.~(\ref{eq:BellState2})] states, respectively.
        For the PDC states, the maximum value $\mathcal{B}_\textrm{m}$ of the Bell parameter, as is sketched in the inset, is chosen.
        The refractive-index structure constant is $C_n^2=2\times10^{-16}~\textrm{m}^{-2/3}$.
        The mean number of noise counts is $5\times 10^{-4}$.
        The aperture radius is $R_\textrm{ap}=10~\textrm{cm}$. 
        The rest of the deterministic losses are $9.42~\textrm{dB}$.}
        \label{fig:bell}
    \end{figure}

The dependence of the Bell parameter on the time $\tau$ between the pulses is shown in Fig.~\ref{fig:bell}.
For the PDC source [cf. Eq.~(\ref{eq:PDC1})] we have found the maximum values of the Bell parameter, $\mathcal{B}_\textrm{m}$, over all values of the squeezing parameter $\xi$; see inset in Fig.~\ref{fig:bell}.
In contrast to the case of Gaussian entanglement, in the considered scenario the dependence on the wind velocity $v$ plays a more important role due to the trade-off with the time-dependent reading mapping losses of the quantum memory.
To separate the effect of turbulence from the effect of quantum memory, we consider two scenarios: with zero and $3~\textrm{dB/ms}$ losses of the latter.
Our results in Fig.~\ref{fig:bell} show that the turbulence itself preserves discrete-variable entanglement for large values of the time $\tau$ between pulses.
However, time-dependent reading mapping losses of the quantum memory play a key destructive role in this scenario.
Therefore, the application of this protocol requires the further development of efficient quantum memory.


\section{Adaptive real-time selection for nonclassical states}
\label{Sec:Preselection}

This protocol assumes that the channel transmittance $\eta_0$ is tested by a strong classical pulse at the time instance $t=0$.
If this transmittance exceeds a predetermined threshold value $\eta_\textrm{min}$, the pulse of nonclassical light at the time instance $t=\tau$ is sent and analyzed; see Fig.~\ref{fig:preselection}.
We aim to find such time $\tau$, for which nonclassical properties of the light are still preserved at the receiver.
This technique has been proposed and implemented in  Refs.~\cite{Tang2013,vallone2015} to increase the efficiency of quantum-key distribution protocols.

We consider amplitude-squeezed coherent states $\ket{\alpha_0,\xi}=\hat{D}(\alpha_0)\hat{S}(\xi)\ket{0}$. 
Here, $\hat{D}(\alpha_0)$ and $\hat{S}(\xi)$ are the displacement and squeeze operators, respectively, with $\alpha_0\in\mathbb{R}$ and $\xi<0$.
This state represents a well-known example characterized by nonclassical photocounting statistics with negative values of the Mandel $Q$ parameter \cite{mandel_book,Schnabel2017},
    \begin{align}
        Q=\frac{\left\langle \Delta n^2\right\rangle}{\left\langle n\right\rangle}-1.
    \end{align}
Here $\left\langle n\right\rangle$ and $\left\langle \Delta n^2\right\rangle$ are the expectation value and variance of the number of photons, respectively.    
As is shown in Refs.~\cite{semenov09,semenov2018}, the Mandel $Q$ parameter for the state at the receiver becomes positive for $\langle \hat{n} \rangle_\mathrm{in}/|Q_\mathrm{in}|>
\langle\eta^2\rangle/\langle\Delta\eta^2\rangle$.
Here $Q_\mathrm{in}$ is the Mandel $Q$ parameter and $\langle \hat{n} \rangle_\mathrm{in}$ is the mean number of photons at the transmitter.
For the considered experiment, the statistical moments of the transmittance are estimated with the conditional PDT given by Eq.~(\ref{eq:condPDT}).

The Mandel $Q$ parameter is applicable only to the detectors with the ideal photon-number resolution.
Let us consider a more realistic case of click detectors that involve spatial or temporal splitting of the light beam and then detecting each part of them by on-off detectors; see Refs.~\cite{paul1996,castelletto2007,schettini2007,blanchet08,achilles03,fitch03,rehacek03}. 
The number of triggered detectors is equal to the number of clicks.
The click-number distribution for the coherent state $\ket{\alpha}$, i.e., $Q$ symbols of the positive operator-valued measure (POVM) (cf. Ref.~\cite{sperling12a}) for such detectors reads
    \begin{align}
        \Pi(n|\alpha)=\binom{N}{n}\left(1-e^{-|\alpha|^2/N}\right)^ne^{-(N-n)|\alpha|^2/N},
    \end{align}
where $N$ is the total number of detectors.
A measure of nonclassicality of click statistics for such detectors, known as the Binomial $Q$ parameter, was introduced in Ref.~\cite{sperling12c},
        \begin{align}
            Q_{N}=N\frac{\left\langle\Delta c^2\right\rangle}{\left\langle c\right\rangle(N-\left\langle c\right\rangle)}-1.
        \end{align}
Here $\left\langle c\right\rangle$ and $\left\langle\Delta c^2\right\rangle$ are the expectation value and the variance for the number of clicks, respectively.
For $N\rightarrow+\infty$ this parameter becomes the Mandel $Q$ parameter.
The values $Q_{N}<0$ witness nonclassicality of photocounting (click) statistics obtained with click detectors.

For our purposes, we also witness nonclassicality of click statistics with the method proposed in Ref.~\cite{semenov2021}. 
This technique can be applied to the detectors with realistic photon-number resolution.
As discussed in Refs.~\cite{semenov2021}, the click statistic is nonclassical---i.e., it cannot be reproduced with the classical electromagnetic radiation \cite{titulaer65,mandel86,mandel_book,vogel_book,agarwal_book,Schnabel2017,sperling2018a,sperling2018b,sperling2020}---iff there exists such $\lambda(n)$ that the inequality
    \begin{align} \label{eq:ineq}
    	\sum_{n=0}^{N-1} \lambda(n) P(n) \leq \sup_{\alpha \in \mathbb{C}} \sum_{n=0}^{N-1} \lambda(n) \Pi(n|\alpha),
    \end{align}
is violated.
Here $P(n)$ is the click distribution for the state under study.
We use the optimal set of $\lambda(n)$, representing the tight subset of inequalities (\ref{eq:ineq}) (cf. Ref.~\cite{kovtoniuk2023}).

\begin{figure}[ht!]
        \includegraphics[width=1\linewidth]{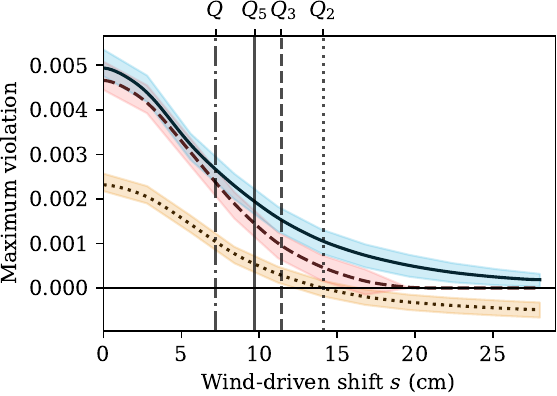}
        \caption{Maximum violation of inequality (\ref{eq:ineq}), i.e. the difference of its left- and right-hand sides, for click detectors vs the wind-driven shift $s$ in the adaptive real-time selection technique is shown for the amplitude-squeezed state with $\alpha_0=1.15$ and $\xi=0.59$.
        The symbols $Q$ and $Q_N$ indicate the thresholds of $s$ for which $Q=0$ (dash-dotted line) and $Q_N=0$, respectively.
        The dotted, dashed, and solid vertical lines and curves correspond to $N=2$, $N=3$, and $N=5$, respectively.
        The refractive-index structure constant is $C_n^2=2\times10^{-16}~\textrm{m}^{-2/3}$.
        The aperture radius is $R_\mathrm{ap}=30~\textrm{cm}$, the overall deterministic losses are $6~\textrm{dB}$.
        The threshold transmittance is $\eta_\textrm{min}=0.1$. 
        The confidence intervals correspond to $10^6$ selected samples.
        The overall selection efficiency (\ref{eq:preseleff}) is $\overline{\mathcal{F}}(\eta_\mathrm{min}) =0.58$.}
        \label{fig:nc}
    \end{figure}

The maximum violation of inequality (\ref{eq:ineq}), i.e., the difference of left- and right-hand sides, as a function of the time $\tau$ (the wind-driven shift $s$) between the classical test-pulse and nonclassical pulse  and the values of $s$ related to $Q=0$ and $Q_\textrm{N}=0$ are shown in Fig.~\ref{fig:nc}.
For the given conditions, the Mandel $Q$ parameter becomes positive for $s=7.2~\textrm{cm}$ ($\tau=7.2~\textrm{ms}$).
However, the click statistics are still nonclassical even far beyond this threshold for the considered values of parameters.
This can be observed even with click detectors having small $N$ employing both methods: one based on the parameter $Q_N$ and the other on inequalities (\ref{eq:ineq}).
For example, for $N=2$ both tests show vanishing nonclassicality at $s=14.2~\textrm{cm}$ ($\tau=14.2~\textrm{ms}$).

With increasing the number $N$ of on-off detectors, the method of inequalities  (\ref{eq:ineq}) witness nonclassicality for larger values of the time $\tau$, compared to the Binomial $Q$ parameter.
At the same time, the value of $\tau$ for which $Q_N=0$ decreases and tends to the threshold for $Q=0$ as $N\rightarrow+\infty$.
For example, in the case of $N=3$ such nonclassicality vanishes for $s=19.6~\textrm{cm}$ ($\tau=19.6~\textrm{ms}$) and $s=11.4~\textrm{cm}$ ($\tau=11.4~\textrm{ms}$), while witnessing with Bell-like inequalities and the Binomial $Q$ parameter, respectively.
For $N=5$ nonclassicality of click statistics is always preserved in the range of our simulation parameters, i.e., up to $s=28~\textrm{cm}$ ($\tau=28~\textrm{ms}$), while witnessing with inequalities  (\ref{eq:ineq}).
Meanwhile, the threshold for $Q_5=0$ is $s=9.7~\textrm{cm}$ ($\tau=9.7~\textrm{ms}$).
Therefore, the adaptive real-time selection technique gives a possibility to preserve nonclassical properties of the electromagnetic radiation for long times between classical and quantum pulses.
Moreover, this property can be verified also with realistic click detectors.

It is also worth noting that the photon-number resolution may become more significant for other values of parameters (e.g., $\alpha_0=1.4$ and $\xi=0.16$).
In such cases, the values of $s$ for which $Q_N=0$ are smaller than the values of $s$ for $Q=0$.
However, the method of inequalities  (\ref{eq:ineq}) still gives better results compared to the method based on the parameter $Q_N$ for $N=3$ and $N=5$.

\section{Summary and conclusions}
\label{Sec:Conclusions}

We have demonstrated that correlating atmospheric-channel transmittances at distinct time intervals represent a significant resource for quantum-information transfer. 
Under typical atmospheric conditions, the correlation times can be on the order of a few milliseconds and are significantly influenced by the receiver-aperture radius. 
This paves the ways for the development of efficient transmission protocols. 
A key characteristic of such channels is the multi-time probability distribution of transmittance (PDT), which we have simulated numerically for the two-time case and applied to typical transmission scenarios.

The first scenario is related to transferring continuous- and discrete-variable entanglement of two time-separated pulses.
The use of this resource requires quantum memory in order to store the quantum state of one of them.
On the other hand, it results in additional losses.
Nevertheless, we show that continuous- and discrete-variable entanglement is stable for long-time intervals between light pulses.
Counterintuitively, Gaussian entanglement becomes less stable as the squeezing parameter increases.
Discrete-variable entanglement is preserved by the atmospheric turbulence itself for long times between pulses.
However, in this case, the reading mapping losses of the quantum memory play a key destructive role, pointing to the need for its further improvement for use in such protocols.

In the second scenario, the adaptive real-time selection protocol, the channel transmittance is tested with a bright classical pulse, and if it exceeds a predetermined value, the pulses of nonclassical light are sent.
We have checked whether such a technique can efficiently preserve nonclassicality verified with realistic click detectors.
The use of inequalities testing nonclassicality of click statistics enables us to verify it for significantly long-time intervals between the classical and quantum pulses.
Moreover, there are scenarios where this nonclassicality is preserved for times well beyond those for which the proper photon-number statistics are sub-Poissonian.

We hope that our results will find their application in both theoretical and experimental research for improving the capacity of free-space quantum channels.
In particular, our results show that long correlation times in the atmosphere can also be used to transmit multipartite entanglement between light pulses.
This resource could be used to increase the effective dimension of the Hilbert space for transmitted quantum states.

The authors thank A. Lvovsky, V. Kovtoniuk, and E.~Stolyarov for enlightening discussions.
M.K. and A.A.S. appreciate support from the National Research Foundation of Ukraine through the Project No.~2020.02/0111, Nonclassical and hybrid correlations of quantum systems under realistic conditions.

\bibliography{biblio}

\begin{thebibliography}{135}%
\makeatletter
\providecommand \@ifxundefined [1]{%
 \@ifx{#1\undefined}
}%
\providecommand \@ifnum [1]{%
 \ifnum #1\expandafter \@firstoftwo
 \else \expandafter \@secondoftwo
 \fi
}%
\providecommand \@ifx [1]{%
 \ifx #1\expandafter \@firstoftwo
 \else \expandafter \@secondoftwo
 \fi
}%
\providecommand \natexlab [1]{#1}%
\providecommand \enquote  [1]{``#1''}%
\providecommand \bibnamefont  [1]{#1}%
\providecommand \bibfnamefont [1]{#1}%
\providecommand \citenamefont [1]{#1}%
\providecommand \href@noop [0]{\@secondoftwo}%
\providecommand \href [0]{\begingroup \@sanitize@url \@href}%
\providecommand \@href[1]{\@@startlink{#1}\@@href}%
\providecommand \@@href[1]{\endgroup#1\@@endlink}%
\providecommand \@sanitize@url [0]{\catcode `\\12\catcode `\$12\catcode
  `\&12\catcode `\#12\catcode `\^12\catcode `\_12\catcode `\%12\relax}%
\providecommand \@@startlink[1]{}%
\providecommand \@@endlink[0]{}%
\providecommand \url  [0]{\begingroup\@sanitize@url \@url }%
\providecommand \@url [1]{\endgroup\@href {#1}{\urlprefix }}%
\providecommand \urlprefix  [0]{URL }%
\providecommand \Eprint [0]{\href }%
\providecommand \doibase [0]{https://doi.org/}%
\providecommand \selectlanguage [0]{\@gobble}%
\providecommand \bibinfo  [0]{\@secondoftwo}%
\providecommand \bibfield  [0]{\@secondoftwo}%
\providecommand \translation [1]{[#1]}%
\providecommand \BibitemOpen [0]{}%
\providecommand \bibitemStop [0]{}%
\providecommand \bibitemNoStop [0]{.\EOS\space}%
\providecommand \EOS [0]{\spacefactor3000\relax}%
\providecommand \BibitemShut  [1]{\csname bibitem#1\endcsname}%
\let\auto@bib@innerbib\@empty
\bibitem [{\citenamefont {Gisin}\ \emph {et~al.}(2002)\citenamefont {Gisin},
  \citenamefont {Ribordy}, \citenamefont {Tittel},\ and\ \citenamefont
  {Zbinden}}]{gisin02}%
  \BibitemOpen
  \bibfield  {author} {\bibinfo {author} {\bibfnamefont {N.}~\bibnamefont
  {Gisin}}, \bibinfo {author} {\bibfnamefont {G.}~\bibnamefont {Ribordy}},
  \bibinfo {author} {\bibfnamefont {W.}~\bibnamefont {Tittel}},\ and\ \bibinfo
  {author} {\bibfnamefont {H.}~\bibnamefont {Zbinden}},\ }\bibfield  {title}
  {\bibinfo {title} {Quantum cryptography},\ }\href
  {https://doi.org/10.1103/RevModPhys.74.145} {\bibfield  {journal} {\bibinfo
  {journal} {Rev. Mod. Phys.}\ }\textbf {\bibinfo {volume} {74}},\ \bibinfo
  {pages} {145} (\bibinfo {year} {2002})}\BibitemShut {NoStop}%
\bibitem [{\citenamefont {Xu}\ \emph {et~al.}(2020)\citenamefont {Xu},
  \citenamefont {Ma}, \citenamefont {Zhang}, \citenamefont {Lo},\ and\
  \citenamefont {Pan}}]{Xu2020}%
  \BibitemOpen
  \bibfield  {author} {\bibinfo {author} {\bibfnamefont {F.}~\bibnamefont
  {Xu}}, \bibinfo {author} {\bibfnamefont {X.}~\bibnamefont {Ma}}, \bibinfo
  {author} {\bibfnamefont {Q.}~\bibnamefont {Zhang}}, \bibinfo {author}
  {\bibfnamefont {H.-K.}\ \bibnamefont {Lo}},\ and\ \bibinfo {author}
  {\bibfnamefont {J.-W.}\ \bibnamefont {Pan}},\ }\bibfield  {title} {\bibinfo
  {title} {Secure quantum key distribution with realistic devices},\ }\href
  {https://doi.org/10.1103/RevModPhys.92.025002} {\bibfield  {journal}
  {\bibinfo  {journal} {Rev. Mod. Phys.}\ }\textbf {\bibinfo {volume} {92}},\
  \bibinfo {pages} {025002} (\bibinfo {year} {2020})}\BibitemShut {NoStop}%
\bibitem [{\citenamefont {Pirandola}\ \emph {et~al.}(2020)\citenamefont
  {Pirandola}, \citenamefont {Andersen}, \citenamefont {Banchi}, \citenamefont
  {Berta}, \citenamefont {D.}, \citenamefont {Colbeck}, \citenamefont
  {Englund}, \citenamefont {Gehring}, \citenamefont {Lupo}, \citenamefont
  {Ottaviani}, \citenamefont {Pereira}, \citenamefont {Razavi}, \citenamefont
  {Shamsul~Shaari}, \citenamefont {Tomamichel}, \citenamefont {Usenko},
  \citenamefont {Vallone}, \citenamefont {Villoresi},\ and\ \citenamefont
  {Wallden}}]{Pirandola2020}%
  \BibitemOpen
  \bibfield  {author} {\bibinfo {author} {\bibfnamefont {S.}~\bibnamefont
  {Pirandola}}, \bibinfo {author} {\bibfnamefont {U.~L.}\ \bibnamefont
  {Andersen}}, \bibinfo {author} {\bibfnamefont {L.}~\bibnamefont {Banchi}},
  \bibinfo {author} {\bibfnamefont {M.}~\bibnamefont {Berta}}, \bibinfo
  {author} {\bibfnamefont {B.}~\bibnamefont {D.}}, \bibinfo {author}
  {\bibfnamefont {R.}~\bibnamefont {Colbeck}}, \bibinfo {author} {\bibfnamefont
  {D.}~\bibnamefont {Englund}}, \bibinfo {author} {\bibfnamefont
  {T.}~\bibnamefont {Gehring}}, \bibinfo {author} {\bibfnamefont
  {C.}~\bibnamefont {Lupo}}, \bibinfo {author} {\bibfnamefont {C.}~\bibnamefont
  {Ottaviani}}, \bibinfo {author} {\bibfnamefont {J.~L.}\ \bibnamefont
  {Pereira}}, \bibinfo {author} {\bibfnamefont {M.}~\bibnamefont {Razavi}},
  \bibinfo {author} {\bibfnamefont {J.}~\bibnamefont {Shamsul~Shaari}},
  \bibinfo {author} {\bibfnamefont {M.}~\bibnamefont {Tomamichel}}, \bibinfo
  {author} {\bibfnamefont {V.~C.}\ \bibnamefont {Usenko}}, \bibinfo {author}
  {\bibfnamefont {G.}~\bibnamefont {Vallone}}, \bibinfo {author} {\bibfnamefont
  {P.}~\bibnamefont {Villoresi}},\ and\ \bibinfo {author} {\bibfnamefont
  {P.}~\bibnamefont {Wallden}},\ }\bibfield  {title} {\bibinfo {title}
  {Advances in quantum cryptography},\ }\href
  {https://doi.org/10.1364/AOP.361502} {\bibfield  {journal} {\bibinfo
  {journal} {Adv. Opt. Photon.}\ }\textbf {\bibinfo {volume} {12}},\ \bibinfo
  {pages} {1012} (\bibinfo {year} {2020})}\BibitemShut {NoStop}%
\bibitem [{\citenamefont {Renner}\ and\ \citenamefont
  {Wolf}(2023)}]{Renner2023}%
  \BibitemOpen
  \bibfield  {author} {\bibinfo {author} {\bibfnamefont {R.}~\bibnamefont
  {Renner}}\ and\ \bibinfo {author} {\bibfnamefont {R.}~\bibnamefont {Wolf}},\
  }\bibfield  {title} {\bibinfo {title} {Quantum advantage in cryptography},\
  }\href {https://doi.org/10.2514/1.J062267} {\bibfield  {journal} {\bibinfo
  {journal} {AIAA Journal}\ }\textbf {\bibinfo {volume} {61}},\ \bibinfo
  {pages} {1895} (\bibinfo {year} {2023})}\BibitemShut {NoStop}%
\bibitem [{\citenamefont {Gottesman}\ and\ \citenamefont
  {Chuang}(2001)}]{Gottesman01}%
  \BibitemOpen
  \bibfield  {author} {\bibinfo {author} {\bibfnamefont {D.}~\bibnamefont
  {Gottesman}}\ and\ \bibinfo {author} {\bibfnamefont {I.}~\bibnamefont
  {Chuang}},\ }\href@noop {} {\bibinfo {title} {Quantum digital signatures}}
  (\bibinfo {year} {2001}),\ \Eprint {https://arxiv.org/abs/quant-ph/0105032}
  {arXiv:quant-ph/0105032 [quant-ph]} \BibitemShut {NoStop}%
\bibitem [{\citenamefont {Bennett}\ \emph {et~al.}(1993)\citenamefont
  {Bennett}, \citenamefont {Brassard}, \citenamefont {Cr\'epeau}, \citenamefont
  {Jozsa}, \citenamefont {Peres},\ and\ \citenamefont {Wootters}}]{bennett93}%
  \BibitemOpen
  \bibfield  {author} {\bibinfo {author} {\bibfnamefont {C.~H.}\ \bibnamefont
  {Bennett}}, \bibinfo {author} {\bibfnamefont {G.}~\bibnamefont {Brassard}},
  \bibinfo {author} {\bibfnamefont {C.}~\bibnamefont {Cr\'epeau}}, \bibinfo
  {author} {\bibfnamefont {R.}~\bibnamefont {Jozsa}}, \bibinfo {author}
  {\bibfnamefont {A.}~\bibnamefont {Peres}},\ and\ \bibinfo {author}
  {\bibfnamefont {W.~K.}\ \bibnamefont {Wootters}},\ }\bibfield  {title}
  {\bibinfo {title} {Teleporting an unknown quantum state via dual classical
  and einstein-podolsky-rosen channels},\ }\href
  {https://doi.org/10.1103/PhysRevLett.70.1895} {\bibfield  {journal} {\bibinfo
   {journal} {Phys. Rev. Lett.}\ }\textbf {\bibinfo {volume} {70}},\ \bibinfo
  {pages} {1895} (\bibinfo {year} {1993})}\BibitemShut {NoStop}%
\bibitem [{\citenamefont {Braunstein}\ and\ \citenamefont
  {Kimble}(1998)}]{Braunstein1998}%
  \BibitemOpen
  \bibfield  {author} {\bibinfo {author} {\bibfnamefont {S.~L.}\ \bibnamefont
  {Braunstein}}\ and\ \bibinfo {author} {\bibfnamefont {H.~J.}\ \bibnamefont
  {Kimble}},\ }\bibfield  {title} {\bibinfo {title} {Teleportation of
  continuous quantum variables},\ }\href
  {https://doi.org/10.1103/PhysRevLett.80.869} {\bibfield  {journal} {\bibinfo
  {journal} {Phys. Rev. Lett.}\ }\textbf {\bibinfo {volume} {80}},\ \bibinfo
  {pages} {869} (\bibinfo {year} {1998})}\BibitemShut {NoStop}%
\bibitem [{\citenamefont {\ifmmode~\dot{Z}\else \.{Z}\fi{}ukowski}\ \emph
  {et~al.}(1993)\citenamefont {\ifmmode~\dot{Z}\else \.{Z}\fi{}ukowski},
  \citenamefont {Zeilinger}, \citenamefont {Horne},\ and\ \citenamefont
  {Ekert}}]{Zukowski1993}%
  \BibitemOpen
  \bibfield  {author} {\bibinfo {author} {\bibfnamefont {M.}~\bibnamefont
  {\ifmmode~\dot{Z}\else \.{Z}\fi{}ukowski}}, \bibinfo {author} {\bibfnamefont
  {A.}~\bibnamefont {Zeilinger}}, \bibinfo {author} {\bibfnamefont {M.~A.}\
  \bibnamefont {Horne}},\ and\ \bibinfo {author} {\bibfnamefont {A.~K.}\
  \bibnamefont {Ekert}},\ }\bibfield  {title} {\bibinfo {title}
  {``{E}vent-ready-detectors'' {B}ell experiment via entanglement swapping},\
  }\href {https://doi.org/10.1103/PhysRevLett.71.4287} {\bibfield  {journal}
  {\bibinfo  {journal} {Phys. Rev. Lett.}\ }\textbf {\bibinfo {volume} {71}},\
  \bibinfo {pages} {4287} (\bibinfo {year} {1993})}\BibitemShut {NoStop}%
\bibitem [{\citenamefont {Ursin}\ \emph {et~al.}(2007)\citenamefont {Ursin}
  \emph {et~al.}}]{ursin07}%
  \BibitemOpen
  \bibfield  {author} {\bibinfo {author} {\bibfnamefont {R.}~\bibnamefont
  {Ursin}} \emph {et~al.},\ }\bibfield  {title} {\bibinfo {title}
  {Entanglement-based quantum communication over 144 km},\ }\href
  {https://doi.org/10.1038/nphys629} {\bibfield  {journal} {\bibinfo  {journal}
  {Nat. Phys.}\ }\textbf {\bibinfo {volume} {3}},\ \bibinfo {pages} {481}
  (\bibinfo {year} {2007})}\BibitemShut {NoStop}%
\bibitem [{\citenamefont {Fedrizzi}\ \emph {et~al.}(2009)\citenamefont
  {Fedrizzi}, \citenamefont {Ursin}, \citenamefont {Herbst}, \citenamefont
  {Nespoli}, \citenamefont {Prevedel}, \citenamefont {Scheidl}, \citenamefont
  {Tiefenbacher}, \citenamefont {Jennewein},\ and\ \citenamefont
  {Zeilinger}}]{fedrizzi09}%
  \BibitemOpen
  \bibfield  {author} {\bibinfo {author} {\bibfnamefont {A.}~\bibnamefont
  {Fedrizzi}}, \bibinfo {author} {\bibfnamefont {R.}~\bibnamefont {Ursin}},
  \bibinfo {author} {\bibfnamefont {T.}~\bibnamefont {Herbst}}, \bibinfo
  {author} {\bibfnamefont {M.}~\bibnamefont {Nespoli}}, \bibinfo {author}
  {\bibfnamefont {R.}~\bibnamefont {Prevedel}}, \bibinfo {author}
  {\bibfnamefont {T.}~\bibnamefont {Scheidl}}, \bibinfo {author} {\bibfnamefont
  {F.}~\bibnamefont {Tiefenbacher}}, \bibinfo {author} {\bibfnamefont
  {T.}~\bibnamefont {Jennewein}},\ and\ \bibinfo {author} {\bibfnamefont
  {A.}~\bibnamefont {Zeilinger}},\ }\bibfield  {title} {\bibinfo {title}
  {High-fidelity transmission of entanglement over a high-loss free-space
  channel},\ }\href {https://doi.org/10.1038/nphys1255} {\bibfield  {journal}
  {\bibinfo  {journal} {Nat. Phys.}\ }\textbf {\bibinfo {volume} {5}},\
  \bibinfo {pages} {389} (\bibinfo {year} {2009})}\BibitemShut {NoStop}%
\bibitem [{\citenamefont {Elser}\ \emph {et~al.}(2009)\citenamefont {Elser},
  \citenamefont {Bartley}, \citenamefont {Heim}, \citenamefont {Wittmann},
  \citenamefont {Sych},\ and\ \citenamefont {Leuchs}}]{elser09}%
  \BibitemOpen
  \bibfield  {author} {\bibinfo {author} {\bibfnamefont {D.}~\bibnamefont
  {Elser}}, \bibinfo {author} {\bibfnamefont {T.}~\bibnamefont {Bartley}},
  \bibinfo {author} {\bibfnamefont {B.}~\bibnamefont {Heim}}, \bibinfo {author}
  {\bibfnamefont {C.}~\bibnamefont {Wittmann}}, \bibinfo {author}
  {\bibfnamefont {D.}~\bibnamefont {Sych}},\ and\ \bibinfo {author}
  {\bibfnamefont {G.}~\bibnamefont {Leuchs}},\ }\bibfield  {title} {\bibinfo
  {title} {Feasibility of free space quantum key distribution with coherent
  polarization states},\ }\href
  {http://stacks.iop.org/1367-2630/11/i=4/a=045014} {\bibfield  {journal}
  {\bibinfo  {journal} {New J. Phys.}\ }\textbf {\bibinfo {volume} {11}},\
  \bibinfo {pages} {045014} (\bibinfo {year} {2009})}\BibitemShut {NoStop}%
\bibitem [{\citenamefont {Heim}\ \emph {et~al.}(2010)\citenamefont {Heim},
  \citenamefont {Elser}, \citenamefont {Bartley}, \citenamefont {Sabuncu},
  \citenamefont {Wittmann}, \citenamefont {Sych}, \citenamefont {Marquardt},\
  and\ \citenamefont {Leuchs}}]{heim10}%
  \BibitemOpen
  \bibfield  {author} {\bibinfo {author} {\bibfnamefont {B.}~\bibnamefont
  {Heim}}, \bibinfo {author} {\bibfnamefont {D.}~\bibnamefont {Elser}},
  \bibinfo {author} {\bibfnamefont {T.}~\bibnamefont {Bartley}}, \bibinfo
  {author} {\bibfnamefont {M.}~\bibnamefont {Sabuncu}}, \bibinfo {author}
  {\bibfnamefont {C.}~\bibnamefont {Wittmann}}, \bibinfo {author}
  {\bibfnamefont {D.}~\bibnamefont {Sych}}, \bibinfo {author} {\bibfnamefont
  {C.}~\bibnamefont {Marquardt}},\ and\ \bibinfo {author} {\bibfnamefont
  {G.}~\bibnamefont {Leuchs}},\ }\bibfield  {title} {\bibinfo {title}
  {Atmospheric channel characteristics for quantum communication with
  continuous polarization variables},\ }\href
  {https://doi.org/10.1007/s00340-009-3838-8} {\bibfield  {journal} {\bibinfo
  {journal} {Appl. Phys. B}\ }\textbf {\bibinfo {volume} {98}},\ \bibinfo
  {pages} {635} (\bibinfo {year} {2010})}\BibitemShut {NoStop}%
\bibitem [{\citenamefont {Capraro}\ \emph {et~al.}(2012)\citenamefont
  {Capraro}, \citenamefont {Tomaello}, \citenamefont {Dall'Arche},
  \citenamefont {Gerlin}, \citenamefont {Ursin}, \citenamefont {Vallone},\ and\
  \citenamefont {Villoresi}}]{capraro12}%
  \BibitemOpen
  \bibfield  {author} {\bibinfo {author} {\bibfnamefont {I.}~\bibnamefont
  {Capraro}}, \bibinfo {author} {\bibfnamefont {A.}~\bibnamefont {Tomaello}},
  \bibinfo {author} {\bibfnamefont {A.}~\bibnamefont {Dall'Arche}}, \bibinfo
  {author} {\bibfnamefont {F.}~\bibnamefont {Gerlin}}, \bibinfo {author}
  {\bibfnamefont {R.}~\bibnamefont {Ursin}}, \bibinfo {author} {\bibfnamefont
  {G.}~\bibnamefont {Vallone}},\ and\ \bibinfo {author} {\bibfnamefont
  {P.}~\bibnamefont {Villoresi}},\ }\bibfield  {title} {\bibinfo {title}
  {Impact of turbulence in long range quantum and classical communications},\
  }\href {https://doi.org/10.1103/PhysRevLett.109.200502} {\bibfield  {journal}
  {\bibinfo  {journal} {Phys. Rev. Lett.}\ }\textbf {\bibinfo {volume} {109}},\
  \bibinfo {pages} {200502} (\bibinfo {year} {2012})}\BibitemShut {NoStop}%
\bibitem [{\citenamefont {Yin}\ \emph {et~al.}(2012)\citenamefont {Yin} \emph
  {et~al.}}]{yin12}%
  \BibitemOpen
  \bibfield  {author} {\bibinfo {author} {\bibfnamefont {J.}~\bibnamefont
  {Yin}} \emph {et~al.},\ }\bibfield  {title} {\bibinfo {title} {Quantum
  teleportation and entanglement distribution over 100-kilometre free-space
  channels},\ }\href {https://doi.org/10.1038/nature11332} {\bibfield
  {journal} {\bibinfo  {journal} {Nature (London)}\ }\textbf {\bibinfo {volume}
  {488}},\ \bibinfo {pages} {185} (\bibinfo {year} {2012})}\BibitemShut
  {NoStop}%
\bibitem [{\citenamefont {Ma}\ \emph {et~al.}(2012)\citenamefont {Ma} \emph
  {et~al.}}]{ma12}%
  \BibitemOpen
  \bibfield  {author} {\bibinfo {author} {\bibfnamefont {X.-S.}\ \bibnamefont
  {Ma}} \emph {et~al.},\ }\bibfield  {title} {\bibinfo {title} {Quantum
  teleportation over 143 kilometres using active feed-forward},\ }\href
  {https://doi.org/10.1038/nature11472} {\bibfield  {journal} {\bibinfo
  {journal} {Nature (London)}\ }\textbf {\bibinfo {volume} {489}},\ \bibinfo
  {pages} {269} (\bibinfo {year} {2012})}\BibitemShut {NoStop}%
\bibitem [{\citenamefont {Peuntinger}\ \emph {et~al.}(2014)\citenamefont
  {Peuntinger}, \citenamefont {Heim}, \citenamefont {M\"uller}, \citenamefont
  {Gabriel}, \citenamefont {Marquardt},\ and\ \citenamefont
  {Leuchs}}]{peuntinger14}%
  \BibitemOpen
  \bibfield  {author} {\bibinfo {author} {\bibfnamefont {C.}~\bibnamefont
  {Peuntinger}}, \bibinfo {author} {\bibfnamefont {B.}~\bibnamefont {Heim}},
  \bibinfo {author} {\bibfnamefont {C.~R.}\ \bibnamefont {M\"uller}}, \bibinfo
  {author} {\bibfnamefont {C.}~\bibnamefont {Gabriel}}, \bibinfo {author}
  {\bibfnamefont {C.}~\bibnamefont {Marquardt}},\ and\ \bibinfo {author}
  {\bibfnamefont {G.}~\bibnamefont {Leuchs}},\ }\bibfield  {title} {\bibinfo
  {title} {Distribution of squeezed states through an atmospheric channel},\
  }\href {https://doi.org/10.1103/PhysRevLett.113.060502} {\bibfield  {journal}
  {\bibinfo  {journal} {Phys. Rev. Lett.}\ }\textbf {\bibinfo {volume} {113}},\
  \bibinfo {pages} {060502} (\bibinfo {year} {2014})}\BibitemShut {NoStop}%
\bibitem [{\citenamefont {Vasylyev}\ \emph {et~al.}(2017)\citenamefont
  {Vasylyev}, \citenamefont {Semenov}, \citenamefont {Vogel}, \citenamefont
  {G\"unthner}, \citenamefont {Thurn}, \citenamefont {Bayraktar},\ and\
  \citenamefont {Marquardt}}]{vasylyev17}%
  \BibitemOpen
  \bibfield  {author} {\bibinfo {author} {\bibfnamefont {D.}~\bibnamefont
  {Vasylyev}}, \bibinfo {author} {\bibfnamefont {A.~A.}\ \bibnamefont
  {Semenov}}, \bibinfo {author} {\bibfnamefont {W.}~\bibnamefont {Vogel}},
  \bibinfo {author} {\bibfnamefont {K.}~\bibnamefont {G\"unthner}}, \bibinfo
  {author} {\bibfnamefont {A.}~\bibnamefont {Thurn}}, \bibinfo {author}
  {\bibfnamefont {O.}~\bibnamefont {Bayraktar}},\ and\ \bibinfo {author}
  {\bibfnamefont {C.}~\bibnamefont {Marquardt}},\ }\bibfield  {title} {\bibinfo
  {title} {Free-space quantum links under diverse weather conditions},\ }\href
  {https://doi.org/10.1103/PhysRevA.96.043856} {\bibfield  {journal} {\bibinfo
  {journal} {Phys. Rev. A}\ }\textbf {\bibinfo {volume} {96}},\ \bibinfo
  {pages} {043856} (\bibinfo {year} {2017})}\BibitemShut {NoStop}%
\bibitem [{\citenamefont {Jin}\ \emph {et~al.}(2019)\citenamefont {Jin},
  \citenamefont {Bourgoin}, \citenamefont {Tannous}, \citenamefont {Agne},
  \citenamefont {Pugh}, \citenamefont {Kuntz}, \citenamefont {Higgins},\ and\
  \citenamefont {Jennewein}}]{Jin2019}%
  \BibitemOpen
  \bibfield  {author} {\bibinfo {author} {\bibfnamefont {J.}~\bibnamefont
  {Jin}}, \bibinfo {author} {\bibfnamefont {J.-P.}\ \bibnamefont {Bourgoin}},
  \bibinfo {author} {\bibfnamefont {R.}~\bibnamefont {Tannous}}, \bibinfo
  {author} {\bibfnamefont {S.}~\bibnamefont {Agne}}, \bibinfo {author}
  {\bibfnamefont {C.~J.}\ \bibnamefont {Pugh}}, \bibinfo {author}
  {\bibfnamefont {K.~B.}\ \bibnamefont {Kuntz}}, \bibinfo {author}
  {\bibfnamefont {B.~L.}\ \bibnamefont {Higgins}},\ and\ \bibinfo {author}
  {\bibfnamefont {T.}~\bibnamefont {Jennewein}},\ }\bibfield  {title} {\bibinfo
  {title} {Genuine time-bin-encoded quantum key distribution over a turbulent
  depolarizing free-space channel},\ }\href
  {https://doi.org/10.1364/OE.27.037214} {\bibfield  {journal} {\bibinfo
  {journal} {Opt. Express}\ }\textbf {\bibinfo {volume} {27}},\ \bibinfo
  {pages} {37214} (\bibinfo {year} {2019})}\BibitemShut {NoStop}%
\bibitem [{\citenamefont {Nauerth}\ \emph {et~al.}(2013)\citenamefont
  {Nauerth}, \citenamefont {Moll}, \citenamefont {Rau}, \citenamefont {Fuchs},
  \citenamefont {Horwath}, \citenamefont {Frick},\ and\ \citenamefont
  {Weinfurter}}]{nauerth13}%
  \BibitemOpen
  \bibfield  {author} {\bibinfo {author} {\bibfnamefont {S.}~\bibnamefont
  {Nauerth}}, \bibinfo {author} {\bibfnamefont {F.}~\bibnamefont {Moll}},
  \bibinfo {author} {\bibfnamefont {M.}~\bibnamefont {Rau}}, \bibinfo {author}
  {\bibfnamefont {C.}~\bibnamefont {Fuchs}}, \bibinfo {author} {\bibfnamefont
  {J.}~\bibnamefont {Horwath}}, \bibinfo {author} {\bibfnamefont
  {S.}~\bibnamefont {Frick}},\ and\ \bibinfo {author} {\bibfnamefont
  {H.}~\bibnamefont {Weinfurter}},\ }\bibfield  {title} {\bibinfo {title}
  {Air-to-ground quantum communication},\ }\href
  {http://dx.doi.org/10.1038/nphoton.2013.46} {\bibfield  {journal} {\bibinfo
  {journal} {Nat. Photonics}\ }\textbf {\bibinfo {volume} {7}},\ \bibinfo
  {pages} {382} (\bibinfo {year} {2013})},\ \bibinfo {note}
  {letter}\BibitemShut {NoStop}%
\bibitem [{\citenamefont {Wang}\ \emph {et~al.}(2013)\citenamefont {Wang} \emph
  {et~al.}}]{wang13}%
  \BibitemOpen
  \bibfield  {author} {\bibinfo {author} {\bibfnamefont {J.-Y.}\ \bibnamefont
  {Wang}} \emph {et~al.},\ }\bibfield  {title} {\bibinfo {title} {Direct and
  full-scale experimental verifications towards ground-satellite quantum key
  distribution},\ }\href {http://dx.doi.org/10.1038/nphoton.2013.89} {\bibfield
   {journal} {\bibinfo  {journal} {Nat. Photonics}\ }\textbf {\bibinfo {volume}
  {7}},\ \bibinfo {pages} {387} (\bibinfo {year} {2013})}\BibitemShut {NoStop}%
\bibitem [{\citenamefont {Bourgoin}\ \emph {et~al.}(2015)\citenamefont
  {Bourgoin}, \citenamefont {Higgins}, \citenamefont {Gigov}, \citenamefont
  {Holloway}, \citenamefont {Pugh}, \citenamefont {Kaiser}, \citenamefont
  {Cranmer},\ and\ \citenamefont {Jennewein}}]{bourgoin15}%
  \BibitemOpen
  \bibfield  {author} {\bibinfo {author} {\bibfnamefont {J.-P.}\ \bibnamefont
  {Bourgoin}}, \bibinfo {author} {\bibfnamefont {B.~L.}\ \bibnamefont
  {Higgins}}, \bibinfo {author} {\bibfnamefont {N.}~\bibnamefont {Gigov}},
  \bibinfo {author} {\bibfnamefont {C.}~\bibnamefont {Holloway}}, \bibinfo
  {author} {\bibfnamefont {C.~J.}\ \bibnamefont {Pugh}}, \bibinfo {author}
  {\bibfnamefont {S.}~\bibnamefont {Kaiser}}, \bibinfo {author} {\bibfnamefont
  {M.}~\bibnamefont {Cranmer}},\ and\ \bibinfo {author} {\bibfnamefont
  {T.}~\bibnamefont {Jennewein}},\ }\bibfield  {title} {\bibinfo {title}
  {Free-space quantum key distribution to a moving receiver},\ }\href
  {https://doi.org/10.1364/OE.23.033437} {\bibfield  {journal} {\bibinfo
  {journal} {Opt. Express}\ }\textbf {\bibinfo {volume} {23}},\ \bibinfo
  {pages} {33437} (\bibinfo {year} {2015})}\BibitemShut {NoStop}%
\bibitem [{\citenamefont {Vallone}\ \emph
  {et~al.}(2015{\natexlab{a}})\citenamefont {Vallone}, \citenamefont {Bacco},
  \citenamefont {Dequal}, \citenamefont {Gaiarin}, \citenamefont {Luceri},
  \citenamefont {Bianco},\ and\ \citenamefont {Villoresi}}]{vallone15}%
  \BibitemOpen
  \bibfield  {author} {\bibinfo {author} {\bibfnamefont {G.}~\bibnamefont
  {Vallone}}, \bibinfo {author} {\bibfnamefont {D.}~\bibnamefont {Bacco}},
  \bibinfo {author} {\bibfnamefont {D.}~\bibnamefont {Dequal}}, \bibinfo
  {author} {\bibfnamefont {S.}~\bibnamefont {Gaiarin}}, \bibinfo {author}
  {\bibfnamefont {V.}~\bibnamefont {Luceri}}, \bibinfo {author} {\bibfnamefont
  {G.}~\bibnamefont {Bianco}},\ and\ \bibinfo {author} {\bibfnamefont
  {P.}~\bibnamefont {Villoresi}},\ }\bibfield  {title} {\bibinfo {title}
  {Experimental satellite quantum communications},\ }\href
  {https://doi.org/10.1103/PhysRevLett.115.040502} {\bibfield  {journal}
  {\bibinfo  {journal} {Phys. Rev. Lett.}\ }\textbf {\bibinfo {volume} {115}},\
  \bibinfo {pages} {040502} (\bibinfo {year} {2015}{\natexlab{a}})}\BibitemShut
  {NoStop}%
\bibitem [{\citenamefont {Dequal}\ \emph {et~al.}(2016)\citenamefont {Dequal},
  \citenamefont {Vallone}, \citenamefont {Bacco}, \citenamefont {Gaiarin},
  \citenamefont {Luceri}, \citenamefont {Bianco},\ and\ \citenamefont
  {Villoresi}}]{dequal16}%
  \BibitemOpen
  \bibfield  {author} {\bibinfo {author} {\bibfnamefont {D.}~\bibnamefont
  {Dequal}}, \bibinfo {author} {\bibfnamefont {G.}~\bibnamefont {Vallone}},
  \bibinfo {author} {\bibfnamefont {D.}~\bibnamefont {Bacco}}, \bibinfo
  {author} {\bibfnamefont {S.}~\bibnamefont {Gaiarin}}, \bibinfo {author}
  {\bibfnamefont {V.}~\bibnamefont {Luceri}}, \bibinfo {author} {\bibfnamefont
  {G.}~\bibnamefont {Bianco}},\ and\ \bibinfo {author} {\bibfnamefont
  {P.}~\bibnamefont {Villoresi}},\ }\bibfield  {title} {\bibinfo {title}
  {Experimental single-photon exchange along a space link of 7000 km},\ }\href
  {https://doi.org/10.1103/PhysRevA.93.010301} {\bibfield  {journal} {\bibinfo
  {journal} {Phys. Rev. A}\ }\textbf {\bibinfo {volume} {93}},\ \bibinfo
  {pages} {010301(R)} (\bibinfo {year} {2016})}\BibitemShut {NoStop}%
\bibitem [{\citenamefont {Vallone}\ \emph {et~al.}(2016)\citenamefont
  {Vallone}, \citenamefont {Dequal}, \citenamefont {Tomasin}, \citenamefont
  {Vedovato}, \citenamefont {Schiavon}, \citenamefont {Luceri}, \citenamefont
  {Bianco},\ and\ \citenamefont {Villoresi}}]{vallone16}%
  \BibitemOpen
  \bibfield  {author} {\bibinfo {author} {\bibfnamefont {G.}~\bibnamefont
  {Vallone}}, \bibinfo {author} {\bibfnamefont {D.}~\bibnamefont {Dequal}},
  \bibinfo {author} {\bibfnamefont {M.}~\bibnamefont {Tomasin}}, \bibinfo
  {author} {\bibfnamefont {F.}~\bibnamefont {Vedovato}}, \bibinfo {author}
  {\bibfnamefont {M.}~\bibnamefont {Schiavon}}, \bibinfo {author}
  {\bibfnamefont {V.}~\bibnamefont {Luceri}}, \bibinfo {author} {\bibfnamefont
  {G.}~\bibnamefont {Bianco}},\ and\ \bibinfo {author} {\bibfnamefont
  {P.}~\bibnamefont {Villoresi}},\ }\bibfield  {title} {\bibinfo {title}
  {Interference at the single photon level along satellite-ground channels},\
  }\href {https://doi.org/10.1103/PhysRevLett.116.253601} {\bibfield  {journal}
  {\bibinfo  {journal} {Phys. Rev. Lett.}\ }\textbf {\bibinfo {volume} {116}},\
  \bibinfo {pages} {253601} (\bibinfo {year} {2016})}\BibitemShut {NoStop}%
\bibitem [{\citenamefont {Carrasco-Casado}\ \emph {et~al.}(2016)\citenamefont
  {Carrasco-Casado}, \citenamefont {Kunimori}, \citenamefont {Takenaka},
  \citenamefont {Kubo-Oka}, \citenamefont {Akioka}, \citenamefont {Fuse},
  \citenamefont {Koyama}, \citenamefont {Kolev}, \citenamefont {Munemasa},\
  and\ \citenamefont {Toyoshima}}]{carrasco-casado16}%
  \BibitemOpen
  \bibfield  {author} {\bibinfo {author} {\bibfnamefont {A.}~\bibnamefont
  {Carrasco-Casado}}, \bibinfo {author} {\bibfnamefont {H.}~\bibnamefont
  {Kunimori}}, \bibinfo {author} {\bibfnamefont {H.}~\bibnamefont {Takenaka}},
  \bibinfo {author} {\bibfnamefont {T.}~\bibnamefont {Kubo-Oka}}, \bibinfo
  {author} {\bibfnamefont {M.}~\bibnamefont {Akioka}}, \bibinfo {author}
  {\bibfnamefont {T.}~\bibnamefont {Fuse}}, \bibinfo {author} {\bibfnamefont
  {Y.}~\bibnamefont {Koyama}}, \bibinfo {author} {\bibfnamefont
  {D.}~\bibnamefont {Kolev}}, \bibinfo {author} {\bibfnamefont
  {Y.}~\bibnamefont {Munemasa}},\ and\ \bibinfo {author} {\bibfnamefont
  {M.}~\bibnamefont {Toyoshima}},\ }\bibfield  {title} {\bibinfo {title}
  {{LEO}-to-ground polarization measurements aiming for space qkd using small
  optical transponder ({SOTA})},\ }\href {https://doi.org/10.1364/OE.24.012254}
  {\bibfield  {journal} {\bibinfo  {journal} {Opt. Express}\ }\textbf {\bibinfo
  {volume} {24}},\ \bibinfo {pages} {12254} (\bibinfo {year}
  {2016})}\BibitemShut {NoStop}%
\bibitem [{\citenamefont {Takenaka}\ \emph {et~al.}(2017)\citenamefont
  {Takenaka}, \citenamefont {Carrasco-Casado}, \citenamefont {Fujiwara},
  \citenamefont {Kitamura}, \citenamefont {Sasaki},\ and\ \citenamefont
  {Toyoshima}}]{takenaka17}%
  \BibitemOpen
  \bibfield  {author} {\bibinfo {author} {\bibfnamefont {H.}~\bibnamefont
  {Takenaka}}, \bibinfo {author} {\bibfnamefont {A.}~\bibnamefont
  {Carrasco-Casado}}, \bibinfo {author} {\bibfnamefont {M.}~\bibnamefont
  {Fujiwara}}, \bibinfo {author} {\bibfnamefont {M.}~\bibnamefont {Kitamura}},
  \bibinfo {author} {\bibfnamefont {M.}~\bibnamefont {Sasaki}},\ and\ \bibinfo
  {author} {\bibfnamefont {M.}~\bibnamefont {Toyoshima}},\ }\bibfield  {title}
  {\bibinfo {title} {Satellite-to-ground quantum-limited communication using a
  50-kg-class microsatellite},\ }\href
  {http://dx.doi.org/10.1038/nphoton.2017.107} {\bibfield  {journal} {\bibinfo
  {journal} {Nat. Photonics}\ }\textbf {\bibinfo {volume} {11}},\ \bibinfo
  {pages} {502} (\bibinfo {year} {2017})}\BibitemShut {NoStop}%
\bibitem [{\citenamefont {Liao}\ \emph {et~al.}(2017)\citenamefont {Liao} \emph
  {et~al.}}]{liao17}%
  \BibitemOpen
  \bibfield  {author} {\bibinfo {author} {\bibfnamefont {S.-K.}\ \bibnamefont
  {Liao}} \emph {et~al.},\ }\bibfield  {title} {\bibinfo {title}
  {Satellite-to-ground quantum key distribution},\ }\href
  {http://dx.doi.org/10.1038/nature23655} {\bibfield  {journal} {\bibinfo
  {journal} {Nature (London)}\ }\textbf {\bibinfo {volume} {549}},\ \bibinfo
  {pages} {43} (\bibinfo {year} {2017})}\BibitemShut {NoStop}%
\bibitem [{\citenamefont {Yin}\ \emph {et~al.}(2017{\natexlab{a}})\citenamefont
  {Yin} \emph {et~al.}}]{yin17}%
  \BibitemOpen
  \bibfield  {author} {\bibinfo {author} {\bibfnamefont {J.}~\bibnamefont
  {Yin}} \emph {et~al.},\ }\bibfield  {title} {\bibinfo {title}
  {Satellite-based entanglement distribution over 1200 kilometers},\ }\href
  {https://doi.org/10.1126/science.aan3211} {\bibfield  {journal} {\bibinfo
  {journal} {Science}\ }\textbf {\bibinfo {volume} {356}},\ \bibinfo {pages}
  {1140} (\bibinfo {year} {2017}{\natexlab{a}})}\BibitemShut {NoStop}%
\bibitem [{\citenamefont {G\"{u}nthner}\ \emph {et~al.}(2017)\citenamefont
  {G\"{u}nthner} \emph {et~al.}}]{gunthner17}%
  \BibitemOpen
  \bibfield  {author} {\bibinfo {author} {\bibfnamefont {K.}~\bibnamefont
  {G\"{u}nthner}} \emph {et~al.},\ }\bibfield  {title} {\bibinfo {title}
  {Quantum-limited measurements of optical signals from a geostationary
  satellite},\ }\href {https://doi.org/10.1364/OPTICA.4.000611} {\bibfield
  {journal} {\bibinfo  {journal} {Optica}\ }\textbf {\bibinfo {volume} {4}},\
  \bibinfo {pages} {611} (\bibinfo {year} {2017})}\BibitemShut {NoStop}%
\bibitem [{\citenamefont {Ren}\ \emph {et~al.}(2017)\citenamefont {Ren} \emph
  {et~al.}}]{ren17}%
  \BibitemOpen
  \bibfield  {author} {\bibinfo {author} {\bibfnamefont {J.-G.}\ \bibnamefont
  {Ren}} \emph {et~al.},\ }\bibfield  {title} {\bibinfo {title}
  {Ground-to-satellite quantum teleportation},\ }\href
  {http://dx.doi.org/10.1038/nature23675} {\bibfield  {journal} {\bibinfo
  {journal} {Nature (London)}\ }\textbf {\bibinfo {volume} {549}},\ \bibinfo
  {pages} {70} (\bibinfo {year} {2017})}\BibitemShut {NoStop}%
\bibitem [{\citenamefont {Yin}\ \emph {et~al.}(2017{\natexlab{b}})\citenamefont
  {Yin}, \citenamefont {Cao}, \citenamefont {Li}, \citenamefont {Ren},
  \citenamefont {Liao}, \citenamefont {Zhang} \emph {et~al.}}]{yin17b}%
  \BibitemOpen
  \bibfield  {author} {\bibinfo {author} {\bibfnamefont {J.}~\bibnamefont
  {Yin}}, \bibinfo {author} {\bibfnamefont {Y.}~\bibnamefont {Cao}}, \bibinfo
  {author} {\bibfnamefont {Y.-H.}\ \bibnamefont {Li}}, \bibinfo {author}
  {\bibfnamefont {J.-G.}\ \bibnamefont {Ren}}, \bibinfo {author} {\bibfnamefont
  {S.-K.}\ \bibnamefont {Liao}}, \bibinfo {author} {\bibfnamefont
  {L.}~\bibnamefont {Zhang}}, \emph {et~al.},\ }\bibfield  {title} {\bibinfo
  {title} {Satellite-to-ground entanglement-based quantum key distribution},\
  }\href {https://doi.org/10.1103/PhysRevLett.119.200501} {\bibfield  {journal}
  {\bibinfo  {journal} {Phys. Rev. Lett.}\ }\textbf {\bibinfo {volume} {119}},\
  \bibinfo {pages} {200501} (\bibinfo {year} {2017}{\natexlab{b}})}\BibitemShut
  {NoStop}%
\bibitem [{\citenamefont {Liao}\ \emph {et~al.}(2018)\citenamefont {Liao},
  \citenamefont {Cai}, \citenamefont {Handsteiner}, \citenamefont {Liu},
  \citenamefont {Yin}, \citenamefont {Zhang} \emph {et~al.}}]{Liao2018}%
  \BibitemOpen
  \bibfield  {author} {\bibinfo {author} {\bibfnamefont {S.-K.}\ \bibnamefont
  {Liao}}, \bibinfo {author} {\bibfnamefont {W.-Q.}\ \bibnamefont {Cai}},
  \bibinfo {author} {\bibfnamefont {J.}~\bibnamefont {Handsteiner}}, \bibinfo
  {author} {\bibfnamefont {B.}~\bibnamefont {Liu}}, \bibinfo {author}
  {\bibfnamefont {J.}~\bibnamefont {Yin}}, \bibinfo {author} {\bibfnamefont
  {L.}~\bibnamefont {Zhang}}, \emph {et~al.},\ }\bibfield  {title} {\bibinfo
  {title} {Satellite-relayed intercontinental quantum network},\ }\href
  {https://doi.org/10.1103/PhysRevLett.120.030501} {\bibfield  {journal}
  {\bibinfo  {journal} {Phys. Rev. Lett.}\ }\textbf {\bibinfo {volume} {120}},\
  \bibinfo {pages} {030501} (\bibinfo {year} {2018})}\BibitemShut {NoStop}%
\bibitem [{\citenamefont {Vasylyev}\ \emph {et~al.}(2019)\citenamefont
  {Vasylyev}, \citenamefont {Vogel},\ and\ \citenamefont
  {Moll}}]{Vasylyev2019}%
  \BibitemOpen
  \bibfield  {author} {\bibinfo {author} {\bibfnamefont {D.}~\bibnamefont
  {Vasylyev}}, \bibinfo {author} {\bibfnamefont {W.}~\bibnamefont {Vogel}},\
  and\ \bibinfo {author} {\bibfnamefont {F.}~\bibnamefont {Moll}},\ }\bibfield
  {title} {\bibinfo {title} {Satellite-mediated quantum atmospheric links},\
  }\href {https://doi.org/10.1103/PhysRevA.99.053830} {\bibfield  {journal}
  {\bibinfo  {journal} {Phys. Rev. A}\ }\textbf {\bibinfo {volume} {99}},\
  \bibinfo {pages} {053830} (\bibinfo {year} {2019})}\BibitemShut {NoStop}%
\bibitem [{\citenamefont {Ecker}\ \emph {et~al.}(2021)\citenamefont {Ecker},
  \citenamefont {Liu}, \citenamefont {Handsteiner}, \citenamefont {Fink},
  \citenamefont {Rauch}, \citenamefont {Steinlechner}, \citenamefont {Scheidl},
  \citenamefont {Zeilinger},\ and\ \citenamefont {Ursin}}]{Ecker2021}%
  \BibitemOpen
  \bibfield  {author} {\bibinfo {author} {\bibfnamefont {S.}~\bibnamefont
  {Ecker}}, \bibinfo {author} {\bibfnamefont {B.}~\bibnamefont {Liu}}, \bibinfo
  {author} {\bibfnamefont {J.}~\bibnamefont {Handsteiner}}, \bibinfo {author}
  {\bibfnamefont {M.}~\bibnamefont {Fink}}, \bibinfo {author} {\bibfnamefont
  {D.}~\bibnamefont {Rauch}}, \bibinfo {author} {\bibfnamefont
  {F.}~\bibnamefont {Steinlechner}}, \bibinfo {author} {\bibfnamefont
  {T.}~\bibnamefont {Scheidl}}, \bibinfo {author} {\bibfnamefont
  {A.}~\bibnamefont {Zeilinger}},\ and\ \bibinfo {author} {\bibfnamefont
  {R.}~\bibnamefont {Ursin}},\ }\bibfield  {title} {\bibinfo {title}
  {Strategies for achieving high key rates in satellite-based {QKD}},\ }\href
  {https://doi.org/10.1038/s41534-020-00335-5} {\bibfield  {journal} {\bibinfo
  {journal} {npj Quantum Information}\ }\textbf {\bibinfo {volume} {7}},\
  \bibinfo {pages} {5} (\bibinfo {year} {2021})}\BibitemShut {NoStop}%
\bibitem [{\citenamefont {Ecker}\ \emph {et~al.}(2022)\citenamefont {Ecker},
  \citenamefont {Pseiner}, \citenamefont {Piris},\ and\ \citenamefont
  {Bohmann}}]{Ecker2022}%
  \BibitemOpen
  \bibfield  {author} {\bibinfo {author} {\bibfnamefont {S.}~\bibnamefont
  {Ecker}}, \bibinfo {author} {\bibfnamefont {J.}~\bibnamefont {Pseiner}},
  \bibinfo {author} {\bibfnamefont {J.}~\bibnamefont {Piris}},\ and\ \bibinfo
  {author} {\bibfnamefont {M.}~\bibnamefont {Bohmann}},\ }\href@noop {}
  {\bibinfo {title} {Advances in entanglement-based {QKD} for space
  applications}} (\bibinfo {year} {2022}),\ \Eprint
  {https://arxiv.org/abs/2210.02229} {arXiv:2210.02229 [quant-ph]} \BibitemShut
  {NoStop}%
\bibitem [{\citenamefont {Tatarskii}(1971)}]{Tatarskii}%
  \BibitemOpen
  \bibfield  {author} {\bibinfo {author} {\bibfnamefont {V.}~\bibnamefont
  {Tatarskii}},\ }\href@noop {} {\emph {\bibinfo {title} {The Effect of the
  Turbulent Atmosphere on Wave Propagation}}}\ (\bibinfo  {publisher} {Israel
  Program for Scientific Translations},\ \bibinfo {address} {Jerusalem},\
  \bibinfo {year} {1971})\BibitemShut {NoStop}%
\bibitem [{\citenamefont {Tatarskii}(2016)}]{Tatarskii2016}%
  \BibitemOpen
  \bibfield  {author} {\bibinfo {author} {\bibfnamefont {V.}~\bibnamefont
  {Tatarskii}},\ }\href@noop {} {\emph {\bibinfo {title} {Wave propagation in a
  turbulent medium}}}\ (\bibinfo  {publisher} {Dover Publications Inc.},\
  \bibinfo {address} {Mineola, New York},\ \bibinfo {year} {2016})\BibitemShut
  {NoStop}%
\bibitem [{\citenamefont {Fante}(1975)}]{Fante1975}%
  \BibitemOpen
  \bibfield  {author} {\bibinfo {author} {\bibfnamefont {R.}~\bibnamefont
  {Fante}},\ }\bibfield  {title} {\bibinfo {title} {Electromagnetic beam
  propagation in turbulent media},\ }\href
  {https://doi.org/10.1109/PROC.1975.10035} {\bibfield  {journal} {\bibinfo
  {journal} {Proc. IEEE}\ }\textbf {\bibinfo {volume} {63}},\ \bibinfo {pages}
  {1669} (\bibinfo {year} {1975})}\BibitemShut {NoStop}%
\bibitem [{\citenamefont {Fante}(1980)}]{Fante1980}%
  \BibitemOpen
  \bibfield  {author} {\bibinfo {author} {\bibfnamefont {R.}~\bibnamefont
  {Fante}},\ }\bibfield  {title} {\bibinfo {title} {Electromagnetic beam
  propagation in turbulent media: An update},\ }\href
  {https://doi.org/10.1109/PROC.1980.11882} {\bibfield  {journal} {\bibinfo
  {journal} {Proc. IEEE}\ }\textbf {\bibinfo {volume} {68}},\ \bibinfo {pages}
  {1424} (\bibinfo {year} {1980})}\BibitemShut {NoStop}%
\bibitem [{\citenamefont {Andrews}\ and\ \citenamefont
  {Phillips}(2005)}]{Andrews_book}%
  \BibitemOpen
  \bibfield  {author} {\bibinfo {author} {\bibfnamefont {L.}~\bibnamefont
  {Andrews}}\ and\ \bibinfo {author} {\bibfnamefont {R.}~\bibnamefont
  {Phillips}},\ }\href@noop {} {\emph {\bibinfo {title} {Laser Beam Propagation
  Through Random Media}}},\ Online access with subscription: SPIE Digital
  Library\ (\bibinfo  {publisher} {Society of Photo Optical},\ \bibinfo {year}
  {2005})\BibitemShut {NoStop}%
\bibitem [{\citenamefont {Usenko}\ \emph {et~al.}(2012)\citenamefont {Usenko},
  \citenamefont {Heim}, \citenamefont {Peuntinger}, \citenamefont {Wittmann},
  \citenamefont {Marquardt}, \citenamefont {Leuchs},\ and\ \citenamefont
  {Filip}}]{usenko12}%
  \BibitemOpen
  \bibfield  {author} {\bibinfo {author} {\bibfnamefont {V.~C.}\ \bibnamefont
  {Usenko}}, \bibinfo {author} {\bibfnamefont {B.}~\bibnamefont {Heim}},
  \bibinfo {author} {\bibfnamefont {C.}~\bibnamefont {Peuntinger}}, \bibinfo
  {author} {\bibfnamefont {C.}~\bibnamefont {Wittmann}}, \bibinfo {author}
  {\bibfnamefont {C.}~\bibnamefont {Marquardt}}, \bibinfo {author}
  {\bibfnamefont {G.}~\bibnamefont {Leuchs}},\ and\ \bibinfo {author}
  {\bibfnamefont {R.}~\bibnamefont {Filip}},\ }\bibfield  {title} {\bibinfo
  {title} {Entanglement of {G}aussian states and the applicability to quantum
  key distribution over fading channels},\ }\href
  {http://stacks.iop.org/1367-2630/14/i=9/a=093048} {\bibfield  {journal}
  {\bibinfo  {journal} {New J. Phys.}\ }\textbf {\bibinfo {volume} {14}},\
  \bibinfo {pages} {093048} (\bibinfo {year} {2012})}\BibitemShut {NoStop}%
\bibitem [{\citenamefont {Guo}\ \emph {et~al.}(2017)\citenamefont {Guo},
  \citenamefont {Xie}, \citenamefont {Liao}, \citenamefont {Zhao},
  \citenamefont {Zeng},\ and\ \citenamefont {Huang}}]{guo2017}%
  \BibitemOpen
  \bibfield  {author} {\bibinfo {author} {\bibfnamefont {Y.}~\bibnamefont
  {Guo}}, \bibinfo {author} {\bibfnamefont {C.}~\bibnamefont {Xie}}, \bibinfo
  {author} {\bibfnamefont {Q.}~\bibnamefont {Liao}}, \bibinfo {author}
  {\bibfnamefont {W.}~\bibnamefont {Zhao}}, \bibinfo {author} {\bibfnamefont
  {G.}~\bibnamefont {Zeng}},\ and\ \bibinfo {author} {\bibfnamefont
  {D.}~\bibnamefont {Huang}},\ }\bibfield  {title} {\bibinfo {title}
  {Entanglement-distillation attack on continuous-variable quantum key
  distribution in a turbulent atmospheric channel},\ }\href
  {https://doi.org/10.1103/PhysRevA.96.022320} {\bibfield  {journal} {\bibinfo
  {journal} {Phys. Rev. A}\ }\textbf {\bibinfo {volume} {96}},\ \bibinfo
  {pages} {022320} (\bibinfo {year} {2017})}\BibitemShut {NoStop}%
\bibitem [{\citenamefont {Papanastasiou}\ \emph {et~al.}(2018)\citenamefont
  {Papanastasiou}, \citenamefont {Weedbrook},\ and\ \citenamefont
  {Pirandola}}]{papanastasiou2018}%
  \BibitemOpen
  \bibfield  {author} {\bibinfo {author} {\bibfnamefont {P.}~\bibnamefont
  {Papanastasiou}}, \bibinfo {author} {\bibfnamefont {C.}~\bibnamefont
  {Weedbrook}},\ and\ \bibinfo {author} {\bibfnamefont {S.}~\bibnamefont
  {Pirandola}},\ }\bibfield  {title} {\bibinfo {title} {Continuous-variable
  quantum key distribution in uniform fast-fading channels},\ }\href
  {https://doi.org/10.1103/PhysRevA.97.032311} {\bibfield  {journal} {\bibinfo
  {journal} {Phys. Rev. A}\ }\textbf {\bibinfo {volume} {97}},\ \bibinfo
  {pages} {032311} (\bibinfo {year} {2018})}\BibitemShut {NoStop}%
\bibitem [{\citenamefont {Derkach}\ \emph {et~al.}(2020)\citenamefont
  {Derkach}, \citenamefont {Usenko},\ and\ \citenamefont
  {Filip}}]{derkach2020b}%
  \BibitemOpen
  \bibfield  {author} {\bibinfo {author} {\bibfnamefont {I.}~\bibnamefont
  {Derkach}}, \bibinfo {author} {\bibfnamefont {V.~C.}\ \bibnamefont
  {Usenko}},\ and\ \bibinfo {author} {\bibfnamefont {R.}~\bibnamefont
  {Filip}},\ }\bibfield  {title} {\bibinfo {title} {Squeezing-enhanced quantum
  key distribution over atmospheric channels},\ }\href
  {https://doi.org/10.1088/1367-2630/ab7f8f} {\bibfield  {journal} {\bibinfo
  {journal} {New J. Phys.}\ }\textbf {\bibinfo {volume} {22}},\ \bibinfo
  {pages} {053006} (\bibinfo {year} {2020})}\BibitemShut {NoStop}%
\bibitem [{\citenamefont {Hosseinidehaj}\ \emph {et~al.}(2019)\citenamefont
  {Hosseinidehaj}, \citenamefont {Babar}, \citenamefont {Malaney},
  \citenamefont {Ng},\ and\ \citenamefont {Hanzo}}]{hosseinidehaj2019}%
  \BibitemOpen
  \bibfield  {author} {\bibinfo {author} {\bibfnamefont {N.}~\bibnamefont
  {Hosseinidehaj}}, \bibinfo {author} {\bibfnamefont {Z.}~\bibnamefont
  {Babar}}, \bibinfo {author} {\bibfnamefont {R.}~\bibnamefont {Malaney}},
  \bibinfo {author} {\bibfnamefont {S.~X.}\ \bibnamefont {Ng}},\ and\ \bibinfo
  {author} {\bibfnamefont {L.}~\bibnamefont {Hanzo}},\ }\bibfield  {title}
  {\bibinfo {title} {Satellite-based continuous-variable quantum
  communications: State-of-the-art and a predictive outlook},\ }\href
  {https://doi.org/10.1109/COMST.2018.2864557} {\bibfield  {journal} {\bibinfo
  {journal} {IEEE Commun. Surv. Tutor.}\ }\textbf {\bibinfo {volume} {21}},\
  \bibinfo {pages} {881} (\bibinfo {year} {2019})}\BibitemShut {NoStop}%
\bibitem [{\citenamefont {Wang}\ \emph
  {et~al.}(2018{\natexlab{a}})\citenamefont {Wang}, \citenamefont {Huang},
  \citenamefont {Wang},\ and\ \citenamefont {Zeng}}]{ShiyuWang2018}%
  \BibitemOpen
  \bibfield  {author} {\bibinfo {author} {\bibfnamefont {S.}~\bibnamefont
  {Wang}}, \bibinfo {author} {\bibfnamefont {P.}~\bibnamefont {Huang}},
  \bibinfo {author} {\bibfnamefont {T.}~\bibnamefont {Wang}},\ and\ \bibinfo
  {author} {\bibfnamefont {G.}~\bibnamefont {Zeng}},\ }\bibfield  {title}
  {\bibinfo {title} {Atmospheric effects on continuous-variable quantum key
  distribution},\ }\href {https://doi.org/10.1088/1367-2630/aad9c4} {\bibfield
  {journal} {\bibinfo  {journal} {New J. Phys.}\ }\textbf {\bibinfo {volume}
  {20}},\ \bibinfo {pages} {083037} (\bibinfo {year}
  {2018}{\natexlab{a}})}\BibitemShut {NoStop}%
\bibitem [{\citenamefont {Chai}\ \emph {et~al.}(2019)\citenamefont {Chai},
  \citenamefont {Cao}, \citenamefont {Liu}, \citenamefont {Wang}, \citenamefont
  {Huang},\ and\ \citenamefont {Zeng}}]{chai2019}%
  \BibitemOpen
  \bibfield  {author} {\bibinfo {author} {\bibfnamefont {G.}~\bibnamefont
  {Chai}}, \bibinfo {author} {\bibfnamefont {Z.}~\bibnamefont {Cao}}, \bibinfo
  {author} {\bibfnamefont {W.}~\bibnamefont {Liu}}, \bibinfo {author}
  {\bibfnamefont {S.}~\bibnamefont {Wang}}, \bibinfo {author} {\bibfnamefont
  {P.}~\bibnamefont {Huang}},\ and\ \bibinfo {author} {\bibfnamefont
  {G.}~\bibnamefont {Zeng}},\ }\bibfield  {title} {\bibinfo {title} {Parameter
  estimation of atmospheric continuous-variable quantum key distribution},\
  }\href {https://doi.org/10.1103/PhysRevA.99.032326} {\bibfield  {journal}
  {\bibinfo  {journal} {Phys. Rev. A}\ }\textbf {\bibinfo {volume} {99}},\
  \bibinfo {pages} {032326} (\bibinfo {year} {2019})}\BibitemShut {NoStop}%
\bibitem [{\citenamefont {Derkach}\ and\ \citenamefont
  {Usenko}(2021)}]{Derkach2021}%
  \BibitemOpen
  \bibfield  {author} {\bibinfo {author} {\bibfnamefont {I.}~\bibnamefont
  {Derkach}}\ and\ \bibinfo {author} {\bibfnamefont {V.~C.}\ \bibnamefont
  {Usenko}},\ }\bibfield  {title} {\bibinfo {title} {Applicability of squeezed-
  and coherent-state continuous-variable quantum key distribution over
  satellite links},\ }\bibfield  {journal} {\bibinfo  {journal} {Entropy}\
  }\textbf {\bibinfo {volume} {23}},\ \href {https://doi.org/10.3390/e23010055}
  {10.3390/e23010055} (\bibinfo {year} {2021})\BibitemShut {NoStop}%
\bibitem [{\citenamefont {Pirandola}(2021{\natexlab{a}})}]{pirandola2021}%
  \BibitemOpen
  \bibfield  {author} {\bibinfo {author} {\bibfnamefont {S.}~\bibnamefont
  {Pirandola}},\ }\bibfield  {title} {\bibinfo {title} {Composable security for
  continuous variable quantum key distribution: Trust levels and practical key
  rates in wired and wireless networks},\ }\href
  {https://doi.org/10.1103/PhysRevResearch.3.043014} {\bibfield  {journal}
  {\bibinfo  {journal} {Phys. Rev. Research}\ }\textbf {\bibinfo {volume}
  {3}},\ \bibinfo {pages} {043014} (\bibinfo {year}
  {2021}{\natexlab{a}})}\BibitemShut {NoStop}%
\bibitem [{\citenamefont {Hosseinidehaj}\ \emph {et~al.}(2021)\citenamefont
  {Hosseinidehaj}, \citenamefont {Walk},\ and\ \citenamefont
  {Ralph}}]{hosseinidehaj2021}%
  \BibitemOpen
  \bibfield  {author} {\bibinfo {author} {\bibfnamefont {N.}~\bibnamefont
  {Hosseinidehaj}}, \bibinfo {author} {\bibfnamefont {N.}~\bibnamefont
  {Walk}},\ and\ \bibinfo {author} {\bibfnamefont {T.~C.}\ \bibnamefont
  {Ralph}},\ }\bibfield  {title} {\bibinfo {title} {Composable finite-size
  effects in free-space continuous-variable quantum-key-distribution systems},\
  }\href {https://doi.org/10.1103/PhysRevA.103.012605} {\bibfield  {journal}
  {\bibinfo  {journal} {Phys. Rev. A}\ }\textbf {\bibinfo {volume} {103}},\
  \bibinfo {pages} {012605} (\bibinfo {year} {2021})}\BibitemShut {NoStop}%
\bibitem [{\citenamefont {Pirandola}(2021{\natexlab{b}})}]{pirandola2021b}%
  \BibitemOpen
  \bibfield  {author} {\bibinfo {author} {\bibfnamefont {S.}~\bibnamefont
  {Pirandola}},\ }\bibfield  {title} {\bibinfo {title} {Satellite quantum
  communications: Fundamental bounds and practical security},\ }\href
  {https://doi.org/10.1103/PhysRevResearch.3.023130} {\bibfield  {journal}
  {\bibinfo  {journal} {Phys. Rev. Research}\ }\textbf {\bibinfo {volume}
  {3}},\ \bibinfo {pages} {023130} (\bibinfo {year}
  {2021}{\natexlab{b}})}\BibitemShut {NoStop}%
\bibitem [{\citenamefont {Pirandola}(2021{\natexlab{c}})}]{pirandola2021c}%
  \BibitemOpen
  \bibfield  {author} {\bibinfo {author} {\bibfnamefont {S.}~\bibnamefont
  {Pirandola}},\ }\bibfield  {title} {\bibinfo {title} {Limits and security of
  free-space quantum communications},\ }\href
  {https://doi.org/10.1103/PhysRevResearch.3.013279} {\bibfield  {journal}
  {\bibinfo  {journal} {Phys. Rev. Research}\ }\textbf {\bibinfo {volume}
  {3}},\ \bibinfo {pages} {013279} (\bibinfo {year}
  {2021}{\natexlab{c}})}\BibitemShut {NoStop}%
\bibitem [{\citenamefont {Wang}\ \emph
  {et~al.}(2018{\natexlab{b}})\citenamefont {Wang}, \citenamefont {Xu},\ and\
  \citenamefont {Lo}}]{Wang2018}%
  \BibitemOpen
  \bibfield  {author} {\bibinfo {author} {\bibfnamefont {W.}~\bibnamefont
  {Wang}}, \bibinfo {author} {\bibfnamefont {F.}~\bibnamefont {Xu}},\ and\
  \bibinfo {author} {\bibfnamefont {H.-K.}\ \bibnamefont {Lo}},\ }\bibfield
  {title} {\bibinfo {title} {Prefixed-threshold real-time selection method in
  free-space quantum key distribution},\ }\href
  {https://doi.org/10.1103/PhysRevA.97.032337} {\bibfield  {journal} {\bibinfo
  {journal} {Phys. Rev. A}\ }\textbf {\bibinfo {volume} {97}},\ \bibinfo
  {pages} {032337} (\bibinfo {year} {2018}{\natexlab{b}})}\BibitemShut
  {NoStop}%
\bibitem [{\citenamefont {Hofmann}\ \emph {et~al.}(2019)\citenamefont
  {Hofmann}, \citenamefont {Semenov}, \citenamefont {Vogel},\ and\
  \citenamefont {Bohmann}}]{hofmann2019}%
  \BibitemOpen
  \bibfield  {author} {\bibinfo {author} {\bibfnamefont {K.}~\bibnamefont
  {Hofmann}}, \bibinfo {author} {\bibfnamefont {A.~A.}\ \bibnamefont
  {Semenov}}, \bibinfo {author} {\bibfnamefont {W.}~\bibnamefont {Vogel}},\
  and\ \bibinfo {author} {\bibfnamefont {M.}~\bibnamefont {Bohmann}},\
  }\bibfield  {title} {\bibinfo {title} {Quantum teleportation through
  atmospheric channels},\ }\href {https://doi.org/10.1088/1402-4896/ab36e0}
  {\bibfield  {journal} {\bibinfo  {journal} {Phys. Scr.}\ }\textbf {\bibinfo
  {volume} {94}},\ \bibinfo {pages} {125104} (\bibinfo {year}
  {2019})}\BibitemShut {NoStop}%
\bibitem [{\citenamefont {Sheng-Li}\ \emph {et~al.}(2017)\citenamefont
  {Sheng-Li}, \citenamefont {Jin}, \citenamefont {Shi}, \citenamefont {Guo},
  \citenamefont {Zou},\ and\ \citenamefont {Guo}}]{zhang2017}%
  \BibitemOpen
  \bibfield  {author} {\bibinfo {author} {\bibnamefont {Sheng-Li}}, \bibinfo
  {author} {\bibfnamefont {C.-H.}\ \bibnamefont {Jin}}, \bibinfo {author}
  {\bibfnamefont {J.-H.}\ \bibnamefont {Shi}}, \bibinfo {author} {\bibfnamefont
  {J.-S.}\ \bibnamefont {Guo}}, \bibinfo {author} {\bibfnamefont {X.-B.}\
  \bibnamefont {Zou}},\ and\ \bibinfo {author} {\bibfnamefont {G.-C.}\
  \bibnamefont {Guo}},\ }\bibfield  {title} {\bibinfo {title} {Continuous
  variable quantum teleportation in beam-wandering modeled atmosphere
  channel},\ }\href {https://doi.org/10.1088/0256-307x/34/4/040302} {\bibfield
  {journal} {\bibinfo  {journal} {Chinese Phys. Lett.}\ }\textbf {\bibinfo
  {volume} {34}},\ \bibinfo {pages} {040302} (\bibinfo {year}
  {2017})}\BibitemShut {NoStop}%
\bibitem [{\citenamefont {Villase\~nor}\ \emph {et~al.}(2021)\citenamefont
  {Villase\~nor}, \citenamefont {He}, \citenamefont {Wang}, \citenamefont
  {Malaney},\ and\ \citenamefont {Win}}]{villasenor2021}%
  \BibitemOpen
  \bibfield  {author} {\bibinfo {author} {\bibfnamefont {E.}~\bibnamefont
  {Villase\~nor}}, \bibinfo {author} {\bibfnamefont {M.}~\bibnamefont {He}},
  \bibinfo {author} {\bibfnamefont {Z.}~\bibnamefont {Wang}}, \bibinfo {author}
  {\bibfnamefont {R.}~\bibnamefont {Malaney}},\ and\ \bibinfo {author}
  {\bibfnamefont {M.~Z.}\ \bibnamefont {Win}},\ }\bibfield  {title} {\bibinfo
  {title} {Enhanced uplink quantum communication with satellites via downlink
  channels},\ }\href {https://doi.org/10.1109/TQE.2021.3091709} {\bibfield
  {journal} {\bibinfo  {journal} {IEEE Trans. Quant. Eng.}\ }\textbf {\bibinfo
  {volume} {2}},\ \bibinfo {pages} {1} (\bibinfo {year} {2021})}\BibitemShut
  {NoStop}%
\bibitem [{\citenamefont {Bohmann}\ \emph {et~al.}(2016)\citenamefont
  {Bohmann}, \citenamefont {Semenov}, \citenamefont {Sperling},\ and\
  \citenamefont {Vogel}}]{bohmann16}%
  \BibitemOpen
  \bibfield  {author} {\bibinfo {author} {\bibfnamefont {M.}~\bibnamefont
  {Bohmann}}, \bibinfo {author} {\bibfnamefont {A.~A.}\ \bibnamefont
  {Semenov}}, \bibinfo {author} {\bibfnamefont {J.}~\bibnamefont {Sperling}},\
  and\ \bibinfo {author} {\bibfnamefont {W.}~\bibnamefont {Vogel}},\ }\bibfield
   {title} {\bibinfo {title} {Gaussian entanglement in the turbulent
  atmosphere},\ }\href {https://doi.org/10.1103/PhysRevA.94.010302} {\bibfield
  {journal} {\bibinfo  {journal} {Phys. Rev. A}\ }\textbf {\bibinfo {volume}
  {94}},\ \bibinfo {pages} {010302(R)} (\bibinfo {year} {2016})}\BibitemShut
  {NoStop}%
\bibitem [{\citenamefont {Hosseinidehaj}\ and\ \citenamefont
  {Malaney}(2015)}]{hosseinidehaj15a}%
  \BibitemOpen
  \bibfield  {author} {\bibinfo {author} {\bibfnamefont {N.}~\bibnamefont
  {Hosseinidehaj}}\ and\ \bibinfo {author} {\bibfnamefont {R.}~\bibnamefont
  {Malaney}},\ }\bibfield  {title} {\bibinfo {title} {Gaussian entanglement
  distribution via satellite},\ }\href
  {https://doi.org/10.1103/PhysRevA.91.022304} {\bibfield  {journal} {\bibinfo
  {journal} {Phys. Rev. A}\ }\textbf {\bibinfo {volume} {91}},\ \bibinfo
  {pages} {022304} (\bibinfo {year} {2015})}\BibitemShut {NoStop}%
\bibitem [{\citenamefont {Heim}\ \emph {et~al.}(2014)\citenamefont {Heim},
  \citenamefont {Peuntinger}, \citenamefont {Killoran}, \citenamefont {Khan},
  \citenamefont {Wittmann}, \citenamefont {Marquardt},\ and\ \citenamefont
  {Leuchs}}]{heim14}%
  \BibitemOpen
  \bibfield  {author} {\bibinfo {author} {\bibfnamefont {B.}~\bibnamefont
  {Heim}}, \bibinfo {author} {\bibfnamefont {C.}~\bibnamefont {Peuntinger}},
  \bibinfo {author} {\bibfnamefont {N.}~\bibnamefont {Killoran}}, \bibinfo
  {author} {\bibfnamefont {I.}~\bibnamefont {Khan}}, \bibinfo {author}
  {\bibfnamefont {C.}~\bibnamefont {Wittmann}}, \bibinfo {author}
  {\bibfnamefont {C.}~\bibnamefont {Marquardt}},\ and\ \bibinfo {author}
  {\bibfnamefont {G.}~\bibnamefont {Leuchs}},\ }\bibfield  {title} {\bibinfo
  {title} {Atmospheric continuous-variable quantum communication},\ }\href
  {http://stacks.iop.org/1367-2630/16/i=11/a=113018} {\bibfield  {journal}
  {\bibinfo  {journal} {New J. Phys.}\ }\textbf {\bibinfo {volume} {16}},\
  \bibinfo {pages} {113018} (\bibinfo {year} {2014})}\BibitemShut {NoStop}%
\bibitem [{\citenamefont {Pe\ifmmode~\check{r}\else \v{r}\fi{}ina}\ \emph
  {et~al.}(1973)\citenamefont {Pe\ifmmode~\check{r}\else \v{r}\fi{}ina},
  \citenamefont {Pe\ifmmode~\check{r}\else \v{r}\fi{}inov\'a}, \citenamefont
  {Teich},\ and\ \citenamefont {Diament}}]{perina73}%
  \BibitemOpen
  \bibfield  {author} {\bibinfo {author} {\bibfnamefont {J.}~\bibnamefont
  {Pe\ifmmode~\check{r}\else \v{r}\fi{}ina}}, \bibinfo {author} {\bibfnamefont
  {V.}~\bibnamefont {Pe\ifmmode~\check{r}\else \v{r}\fi{}inov\'a}}, \bibinfo
  {author} {\bibfnamefont {M.~C.}\ \bibnamefont {Teich}},\ and\ \bibinfo
  {author} {\bibfnamefont {P.}~\bibnamefont {Diament}},\ }\bibfield  {title}
  {\bibinfo {title} {Two descriptions for the photocounting detection of
  radiation passed through a random medium: A comparison for the turbulent
  atmosphere},\ }\href {https://doi.org/10.1103/PhysRevA.7.1732} {\bibfield
  {journal} {\bibinfo  {journal} {Phys. Rev. A}\ }\textbf {\bibinfo {volume}
  {7}},\ \bibinfo {pages} {1732} (\bibinfo {year} {1973})}\BibitemShut
  {NoStop}%
\bibitem [{\citenamefont {Milonni}\ \emph {et~al.}(2004)\citenamefont
  {Milonni}, \citenamefont {Carter}, \citenamefont {Peterson},\ and\
  \citenamefont {Hughes}}]{milonni04}%
  \BibitemOpen
  \bibfield  {author} {\bibinfo {author} {\bibfnamefont {P.~W.}\ \bibnamefont
  {Milonni}}, \bibinfo {author} {\bibfnamefont {J.~H.}\ \bibnamefont {Carter}},
  \bibinfo {author} {\bibfnamefont {C.~G.}\ \bibnamefont {Peterson}},\ and\
  \bibinfo {author} {\bibfnamefont {R.~J.}\ \bibnamefont {Hughes}},\ }\bibfield
   {title} {\bibinfo {title} {Effects of propagation through atmospheric
  turbulence on photon statistics},\ }\href
  {http://stacks.iop.org/1464-4266/6/i=8/a=018} {\bibfield  {journal} {\bibinfo
   {journal} {J. Opt. B: Quantum Semiclass. Opt.}\ }\textbf {\bibinfo {volume}
  {6}},\ \bibinfo {pages} {S742} (\bibinfo {year} {2004})}\BibitemShut
  {NoStop}%
\bibitem [{\citenamefont {Semenov}\ and\ \citenamefont
  {Vogel}(2009)}]{semenov09}%
  \BibitemOpen
  \bibfield  {author} {\bibinfo {author} {\bibfnamefont {A.~A.}\ \bibnamefont
  {Semenov}}\ and\ \bibinfo {author} {\bibfnamefont {W.}~\bibnamefont
  {Vogel}},\ }\bibfield  {title} {\bibinfo {title} {Quantum light in the
  turbulent atmosphere},\ }\href {https://doi.org/10.1103/PhysRevA.80.021802}
  {\bibfield  {journal} {\bibinfo  {journal} {Phys. Rev. A}\ }\textbf {\bibinfo
  {volume} {80}},\ \bibinfo {pages} {021802(R)} (\bibinfo {year}
  {2009})}\BibitemShut {NoStop}%
\bibitem [{\citenamefont {Semenov}\ \emph {et~al.}(2012)\citenamefont
  {Semenov}, \citenamefont {T\"oppel}, \citenamefont {Vasylyev}, \citenamefont
  {Gomonay},\ and\ \citenamefont {Vogel}}]{semenov12}%
  \BibitemOpen
  \bibfield  {author} {\bibinfo {author} {\bibfnamefont {A.~A.}\ \bibnamefont
  {Semenov}}, \bibinfo {author} {\bibfnamefont {F.}~\bibnamefont {T\"oppel}},
  \bibinfo {author} {\bibfnamefont {D.~Y.}\ \bibnamefont {Vasylyev}}, \bibinfo
  {author} {\bibfnamefont {H.~V.}\ \bibnamefont {Gomonay}},\ and\ \bibinfo
  {author} {\bibfnamefont {W.}~\bibnamefont {Vogel}},\ }\bibfield  {title}
  {\bibinfo {title} {Homodyne detection for atmosphere channels},\ }\href
  {https://doi.org/10.1103/PhysRevA.85.013826} {\bibfield  {journal} {\bibinfo
  {journal} {Phys. Rev. A}\ }\textbf {\bibinfo {volume} {85}},\ \bibinfo
  {pages} {013826} (\bibinfo {year} {2012})}\BibitemShut {NoStop}%
\bibitem [{\citenamefont {Vasylyev}\ \emph {et~al.}(2012)\citenamefont
  {Vasylyev}, \citenamefont {Semenov},\ and\ \citenamefont
  {Vogel}}]{vasylyev12}%
  \BibitemOpen
  \bibfield  {author} {\bibinfo {author} {\bibfnamefont {D.~Y.}\ \bibnamefont
  {Vasylyev}}, \bibinfo {author} {\bibfnamefont {A.~A.}\ \bibnamefont
  {Semenov}},\ and\ \bibinfo {author} {\bibfnamefont {W.}~\bibnamefont
  {Vogel}},\ }\bibfield  {title} {\bibinfo {title} {Toward global quantum
  communication: Beam wandering preserves nonclassicality},\ }\href
  {https://doi.org/10.1103/PhysRevLett.108.220501} {\bibfield  {journal}
  {\bibinfo  {journal} {Phys. Rev. Lett.}\ }\textbf {\bibinfo {volume} {108}},\
  \bibinfo {pages} {220501} (\bibinfo {year} {2012})}\BibitemShut {NoStop}%
\bibitem [{\citenamefont {Vasylyev}\ \emph {et~al.}(2016)\citenamefont
  {Vasylyev}, \citenamefont {Semenov},\ and\ \citenamefont
  {Vogel}}]{vasylyev16}%
  \BibitemOpen
  \bibfield  {author} {\bibinfo {author} {\bibfnamefont {D.}~\bibnamefont
  {Vasylyev}}, \bibinfo {author} {\bibfnamefont {A.~A.}\ \bibnamefont
  {Semenov}},\ and\ \bibinfo {author} {\bibfnamefont {W.}~\bibnamefont
  {Vogel}},\ }\bibfield  {title} {\bibinfo {title} {Atmospheric quantum
  channels with weak and strong turbulence},\ }\href
  {https://doi.org/10.1103/PhysRevLett.117.090501} {\bibfield  {journal}
  {\bibinfo  {journal} {Phys. Rev. Lett.}\ }\textbf {\bibinfo {volume} {117}},\
  \bibinfo {pages} {090501} (\bibinfo {year} {2016})}\BibitemShut {NoStop}%
\bibitem [{\citenamefont {Vasylyev}\ \emph {et~al.}(2018)\citenamefont
  {Vasylyev}, \citenamefont {Vogel},\ and\ \citenamefont
  {Semenov}}]{vasylyev18}%
  \BibitemOpen
  \bibfield  {author} {\bibinfo {author} {\bibfnamefont {D.}~\bibnamefont
  {Vasylyev}}, \bibinfo {author} {\bibfnamefont {W.}~\bibnamefont {Vogel}},\
  and\ \bibinfo {author} {\bibfnamefont {A.~A.}\ \bibnamefont {Semenov}},\
  }\bibfield  {title} {\bibinfo {title} {Theory of atmospheric quantum channels
  based on the law of total probability},\ }\href
  {https://doi.org/10.1103/PhysRevA.97.063852} {\bibfield  {journal} {\bibinfo
  {journal} {Phys. Rev. A}\ }\textbf {\bibinfo {volume} {97}},\ \bibinfo
  {pages} {063852} (\bibinfo {year} {2018})}\BibitemShut {NoStop}%
\bibitem [{\citenamefont {Semenov}\ \emph {et~al.}(2018)\citenamefont
  {Semenov}, \citenamefont {Bohmann}, \citenamefont {Vasylyev},\ and\
  \citenamefont {Vogel}}]{semenov2018}%
  \BibitemOpen
  \bibfield  {author} {\bibinfo {author} {\bibfnamefont {A.~A.}\ \bibnamefont
  {Semenov}}, \bibinfo {author} {\bibfnamefont {M.}~\bibnamefont {Bohmann}},
  \bibinfo {author} {\bibfnamefont {D.}~\bibnamefont {Vasylyev}},\ and\
  \bibinfo {author} {\bibfnamefont {W.}~\bibnamefont {Vogel}},\ }\bibfield
  {title} {\bibinfo {title} {{Nonclassicality and Bell nonlocality in
  atmospheric links}},\ }in\ \href {https://doi.org/10.1117/12.2319475} {\emph
  {\bibinfo {booktitle} {Quantum Communications and Quantum Imaging XVI}}},\
  Vol.\ \bibinfo {volume} {10771},\ \bibinfo {editor} {edited by\ \bibinfo
  {editor} {\bibfnamefont {R.~E.}\ \bibnamefont {Meyers}}, \bibinfo {editor}
  {\bibfnamefont {Y.}~\bibnamefont {Shih}},\ and\ \bibinfo {editor}
  {\bibfnamefont {K.~S.}\ \bibnamefont {Deacon}}},\ \bibinfo {organization}
  {International Society for Optics and Photonics}\ (\bibinfo  {publisher}
  {SPIE},\ \bibinfo {year} {2018})\ pp.\ \bibinfo {pages} {155 --
  170}\BibitemShut {NoStop}%
\bibitem [{\citenamefont {Klen}\ and\ \citenamefont
  {Semenov}(2023)}]{klen2023}%
  \BibitemOpen
  \bibfield  {author} {\bibinfo {author} {\bibfnamefont {M.}~\bibnamefont
  {Klen}}\ and\ \bibinfo {author} {\bibfnamefont {A.~A.}\ \bibnamefont
  {Semenov}},\ }\bibfield  {title} {\bibinfo {title} {Numerical simulations of
  atmospheric quantum channels},\ }\href
  {https://doi.org/10.1103/PhysRevA.108.033718} {\bibfield  {journal} {\bibinfo
   {journal} {Phys. Rev. A}\ }\textbf {\bibinfo {volume} {108}},\ \bibinfo
  {pages} {033718} (\bibinfo {year} {2023})}\BibitemShut {NoStop}%
\bibitem [{\citenamefont {Semenov}\ and\ \citenamefont
  {Vogel}(2010)}]{semenov10}%
  \BibitemOpen
  \bibfield  {author} {\bibinfo {author} {\bibfnamefont {A.~A.}\ \bibnamefont
  {Semenov}}\ and\ \bibinfo {author} {\bibfnamefont {W.}~\bibnamefont
  {Vogel}},\ }\bibfield  {title} {\bibinfo {title} {Entanglement transfer
  through the turbulent atmosphere},\ }\href
  {https://doi.org/10.1103/PhysRevA.81.023835} {\bibfield  {journal} {\bibinfo
  {journal} {Phys. Rev. A}\ }\textbf {\bibinfo {volume} {81}},\ \bibinfo
  {pages} {023835} (\bibinfo {year} {2010})}\BibitemShut {NoStop}%
\bibitem [{\citenamefont {Gumberidze}\ \emph {et~al.}(2016)\citenamefont
  {Gumberidze}, \citenamefont {Semenov}, \citenamefont {Vasylyev},\ and\
  \citenamefont {Vogel}}]{gumberidze16}%
  \BibitemOpen
  \bibfield  {author} {\bibinfo {author} {\bibfnamefont {M.~O.}\ \bibnamefont
  {Gumberidze}}, \bibinfo {author} {\bibfnamefont {A.~A.}\ \bibnamefont
  {Semenov}}, \bibinfo {author} {\bibfnamefont {D.}~\bibnamefont {Vasylyev}},\
  and\ \bibinfo {author} {\bibfnamefont {W.}~\bibnamefont {Vogel}},\ }\bibfield
   {title} {\bibinfo {title} {Bell nonlocality in the turbulent atmosphere},\
  }\href {https://doi.org/10.1103/PhysRevA.94.053801} {\bibfield  {journal}
  {\bibinfo  {journal} {Phys. Rev. A}\ }\textbf {\bibinfo {volume} {94}},\
  \bibinfo {pages} {053801} (\bibinfo {year} {2016})}\BibitemShut {NoStop}%
\bibitem [{\citenamefont {Klug}\ \emph {et~al.}(2023)\citenamefont {Klug},
  \citenamefont {Peters},\ and\ \citenamefont {Forbes}}]{Klug2023}%
  \BibitemOpen
  \bibfield  {author} {\bibinfo {author} {\bibfnamefont {A.}~\bibnamefont
  {Klug}}, \bibinfo {author} {\bibfnamefont {C.}~\bibnamefont {Peters}},\ and\
  \bibinfo {author} {\bibfnamefont {A.}~\bibnamefont {Forbes}},\ }\bibfield
  {title} {\bibinfo {title} {{Robust structured light in atmospheric
  turbulence}},\ }\href {https://doi.org/10.1117/1.AP.5.1.016006} {\bibfield
  {journal} {\bibinfo  {journal} {Adv. Phot.}\ }\textbf {\bibinfo {volume}
  {5}},\ \bibinfo {pages} {016006} (\bibinfo {year} {2023})}\BibitemShut
  {NoStop}%
\bibitem [{\citenamefont {Bachmann}\ \emph {et~al.}(2023)\citenamefont
  {Bachmann}, \citenamefont {Isoard}, \citenamefont {Shatokhin}, \citenamefont
  {Sorelli}, \citenamefont {Treps},\ and\ \citenamefont
  {Buchleitner}}]{Bachmann2023}%
  \BibitemOpen
  \bibfield  {author} {\bibinfo {author} {\bibfnamefont {D.}~\bibnamefont
  {Bachmann}}, \bibinfo {author} {\bibfnamefont {M.}~\bibnamefont {Isoard}},
  \bibinfo {author} {\bibfnamefont {V.}~\bibnamefont {Shatokhin}}, \bibinfo
  {author} {\bibfnamefont {G.}~\bibnamefont {Sorelli}}, \bibinfo {author}
  {\bibfnamefont {N.}~\bibnamefont {Treps}},\ and\ \bibinfo {author}
  {\bibfnamefont {A.}~\bibnamefont {Buchleitner}},\ }\bibfield  {title}
  {\bibinfo {title} {Highly transmitting modes of light in dynamic atmospheric
  turbulence},\ }\href {https://doi.org/10.1103/PhysRevLett.130.073801}
  {\bibfield  {journal} {\bibinfo  {journal} {Phys. Rev. Lett.}\ }\textbf
  {\bibinfo {volume} {130}},\ \bibinfo {pages} {073801} (\bibinfo {year}
  {2023})}\BibitemShut {NoStop}%
\bibitem [{\citenamefont {Paterson}(2005)}]{paterson05}%
  \BibitemOpen
  \bibfield  {author} {\bibinfo {author} {\bibfnamefont {C.}~\bibnamefont
  {Paterson}},\ }\bibfield  {title} {\bibinfo {title} {Atmospheric turbulence
  and orbital angular momentum of single photons for optical communication},\
  }\href {https://doi.org/10.1103/PhysRevLett.94.153901} {\bibfield  {journal}
  {\bibinfo  {journal} {Phys. Rev. Lett.}\ }\textbf {\bibinfo {volume} {94}},\
  \bibinfo {pages} {153901} (\bibinfo {year} {2005})}\BibitemShut {NoStop}%
\bibitem [{\citenamefont {Pors}\ \emph {et~al.}(2011)\citenamefont {Pors},
  \citenamefont {Monken}, \citenamefont {Eliel},\ and\ \citenamefont
  {Woerdman}}]{pors11}%
  \BibitemOpen
  \bibfield  {author} {\bibinfo {author} {\bibfnamefont {B.-J.}\ \bibnamefont
  {Pors}}, \bibinfo {author} {\bibfnamefont {C.~H.}\ \bibnamefont {Monken}},
  \bibinfo {author} {\bibfnamefont {E.~R.}\ \bibnamefont {Eliel}},\ and\
  \bibinfo {author} {\bibfnamefont {J.~P.}\ \bibnamefont {Woerdman}},\
  }\bibfield  {title} {\bibinfo {title} {Transport of orbital-angular-momentum
  entanglement through a turbulent atmosphere},\ }\href
  {https://doi.org/10.1364/OE.19.006671} {\bibfield  {journal} {\bibinfo
  {journal} {Opt. Express}\ }\textbf {\bibinfo {volume} {19}},\ \bibinfo
  {pages} {6671} (\bibinfo {year} {2011})}\BibitemShut {NoStop}%
\bibitem [{\citenamefont {Roux}(2011)}]{Roux2013}%
  \BibitemOpen
  \bibfield  {author} {\bibinfo {author} {\bibfnamefont {F.~S.}\ \bibnamefont
  {Roux}},\ }\bibfield  {title} {\bibinfo {title} {Infinitesimal-propagation
  equation for decoherence of an orbital-angular-momentum-entangled biphoton
  state in atmospheric turbulence},\ }\href
  {https://doi.org/10.1103/PhysRevA.83.053822} {\bibfield  {journal} {\bibinfo
  {journal} {Phys. Rev. A}\ }\textbf {\bibinfo {volume} {83}},\ \bibinfo
  {pages} {053822} (\bibinfo {year} {2011})}\BibitemShut {NoStop}%
\bibitem [{\citenamefont {Roux}\ \emph {et~al.}(2015)\citenamefont {Roux},
  \citenamefont {Wellens},\ and\ \citenamefont {Shatokhin}}]{Roux2015}%
  \BibitemOpen
  \bibfield  {author} {\bibinfo {author} {\bibfnamefont {F.~S.}\ \bibnamefont
  {Roux}}, \bibinfo {author} {\bibfnamefont {T.}~\bibnamefont {Wellens}},\ and\
  \bibinfo {author} {\bibfnamefont {V.~N.}\ \bibnamefont {Shatokhin}},\
  }\bibfield  {title} {\bibinfo {title} {Entanglement evolution of twisted
  photons in strong atmospheric turbulence},\ }\href
  {https://doi.org/10.1103/PhysRevA.92.012326} {\bibfield  {journal} {\bibinfo
  {journal} {Phys. Rev. A}\ }\textbf {\bibinfo {volume} {92}},\ \bibinfo
  {pages} {012326} (\bibinfo {year} {2015})}\BibitemShut {NoStop}%
\bibitem [{\citenamefont {Leonhard}\ \emph {et~al.}(2015)\citenamefont
  {Leonhard}, \citenamefont {Shatokhin},\ and\ \citenamefont
  {Buchleitner}}]{Leonhard2015}%
  \BibitemOpen
  \bibfield  {author} {\bibinfo {author} {\bibfnamefont {N.~D.}\ \bibnamefont
  {Leonhard}}, \bibinfo {author} {\bibfnamefont {V.~N.}\ \bibnamefont
  {Shatokhin}},\ and\ \bibinfo {author} {\bibfnamefont {A.}~\bibnamefont
  {Buchleitner}},\ }\bibfield  {title} {\bibinfo {title} {Universal
  entanglement decay of photonic-orbital-angular-momentum qubit states in
  atmospheric turbulence},\ }\href {https://doi.org/10.1103/PhysRevA.91.012345}
  {\bibfield  {journal} {\bibinfo  {journal} {Phys. Rev. A}\ }\textbf {\bibinfo
  {volume} {91}},\ \bibinfo {pages} {012345} (\bibinfo {year}
  {2015})}\BibitemShut {NoStop}%
\bibitem [{\citenamefont {Lavery}(2018)}]{Lavery2018}%
  \BibitemOpen
  \bibfield  {author} {\bibinfo {author} {\bibfnamefont {M.~P.~J.}\
  \bibnamefont {Lavery}},\ }\bibfield  {title} {\bibinfo {title} {Vortex
  instability in turbulent free-space propagation},\ }\href
  {https://doi.org/10.1088/1367-2630/aaae9e} {\bibfield  {journal} {\bibinfo
  {journal} {New J. Phys.}\ }\textbf {\bibinfo {volume} {20}},\ \bibinfo
  {pages} {043023} (\bibinfo {year} {2018})}\BibitemShut {NoStop}%
\bibitem [{\citenamefont {Sorelli}\ \emph {et~al.}(2019)\citenamefont
  {Sorelli}, \citenamefont {Leonhard}, \citenamefont {Shatokhin}, \citenamefont
  {Reinlein},\ and\ \citenamefont {Buchleitner}}]{Sorelli2019}%
  \BibitemOpen
  \bibfield  {author} {\bibinfo {author} {\bibfnamefont {G.}~\bibnamefont
  {Sorelli}}, \bibinfo {author} {\bibfnamefont {N.}~\bibnamefont {Leonhard}},
  \bibinfo {author} {\bibfnamefont {V.~N.}\ \bibnamefont {Shatokhin}}, \bibinfo
  {author} {\bibfnamefont {C.}~\bibnamefont {Reinlein}},\ and\ \bibinfo
  {author} {\bibfnamefont {A.}~\bibnamefont {Buchleitner}},\ }\bibfield
  {title} {\bibinfo {title} {Entanglement protection of high-dimensional states
  by adaptive optics},\ }\href {https://doi.org/10.1088/1367-2630/ab006e}
  {\bibfield  {journal} {\bibinfo  {journal} {New J. Phys.}\ }\textbf {\bibinfo
  {volume} {21}},\ \bibinfo {pages} {023003} (\bibinfo {year}
  {2019})}\BibitemShut {NoStop}%
\bibitem [{\citenamefont {Bulla}\ \emph
  {et~al.}(2023{\natexlab{a}})\citenamefont {Bulla}, \citenamefont {Pivoluska},
  \citenamefont {Hjorth}, \citenamefont {Kohout}, \citenamefont {Lang},
  \citenamefont {Ecker}, \citenamefont {Neumann}, \citenamefont {Bittermann},
  \citenamefont {Kindler}, \citenamefont {Huber}, \citenamefont {Bohmann},\
  and\ \citenamefont {Ursin}}]{Bulla2023a}%
  \BibitemOpen
  \bibfield  {author} {\bibinfo {author} {\bibfnamefont {L.}~\bibnamefont
  {Bulla}}, \bibinfo {author} {\bibfnamefont {M.}~\bibnamefont {Pivoluska}},
  \bibinfo {author} {\bibfnamefont {K.}~\bibnamefont {Hjorth}}, \bibinfo
  {author} {\bibfnamefont {O.}~\bibnamefont {Kohout}}, \bibinfo {author}
  {\bibfnamefont {J.}~\bibnamefont {Lang}}, \bibinfo {author} {\bibfnamefont
  {S.}~\bibnamefont {Ecker}}, \bibinfo {author} {\bibfnamefont {S.~P.}\
  \bibnamefont {Neumann}}, \bibinfo {author} {\bibfnamefont {J.}~\bibnamefont
  {Bittermann}}, \bibinfo {author} {\bibfnamefont {R.}~\bibnamefont {Kindler}},
  \bibinfo {author} {\bibfnamefont {M.}~\bibnamefont {Huber}}, \bibinfo
  {author} {\bibfnamefont {M.}~\bibnamefont {Bohmann}},\ and\ \bibinfo {author}
  {\bibfnamefont {R.}~\bibnamefont {Ursin}},\ }\bibfield  {title} {\bibinfo
  {title} {Nonlocal temporal interferometry for highly resilient free-space
  quantum communication},\ }\href {https://doi.org/10.1103/PhysRevX.13.021001}
  {\bibfield  {journal} {\bibinfo  {journal} {Phys. Rev. X}\ }\textbf {\bibinfo
  {volume} {13}},\ \bibinfo {pages} {021001} (\bibinfo {year}
  {2023}{\natexlab{a}})}\BibitemShut {NoStop}%
\bibitem [{\citenamefont {Bulla}\ \emph
  {et~al.}(2023{\natexlab{b}})\citenamefont {Bulla}, \citenamefont {Hjorth},
  \citenamefont {Kohout}, \citenamefont {Lang}, \citenamefont {Ecker},
  \citenamefont {Neumann}, \citenamefont {Bittermann}, \citenamefont {Kindler},
  \citenamefont {Huber}, \citenamefont {Bohmann}, \citenamefont {Ursin},\ and\
  \citenamefont {Pivoluska}}]{Bulla2023b}%
  \BibitemOpen
  \bibfield  {author} {\bibinfo {author} {\bibfnamefont {L.}~\bibnamefont
  {Bulla}}, \bibinfo {author} {\bibfnamefont {K.}~\bibnamefont {Hjorth}},
  \bibinfo {author} {\bibfnamefont {O.}~\bibnamefont {Kohout}}, \bibinfo
  {author} {\bibfnamefont {J.}~\bibnamefont {Lang}}, \bibinfo {author}
  {\bibfnamefont {S.}~\bibnamefont {Ecker}}, \bibinfo {author} {\bibfnamefont
  {S.~P.}\ \bibnamefont {Neumann}}, \bibinfo {author} {\bibfnamefont
  {J.}~\bibnamefont {Bittermann}}, \bibinfo {author} {\bibfnamefont
  {R.}~\bibnamefont {Kindler}}, \bibinfo {author} {\bibfnamefont
  {M.}~\bibnamefont {Huber}}, \bibinfo {author} {\bibfnamefont
  {M.}~\bibnamefont {Bohmann}}, \bibinfo {author} {\bibfnamefont
  {R.}~\bibnamefont {Ursin}},\ and\ \bibinfo {author} {\bibfnamefont
  {M.}~\bibnamefont {Pivoluska}},\ }\bibfield  {title} {\bibinfo {title}
  {Distribution of genuine high-dimensional entanglement over 10.2 km of noisy
  metropolitan atmosphere},\ }\href
  {https://doi.org/10.1103/PhysRevA.107.L050402} {\bibfield  {journal}
  {\bibinfo  {journal} {Phys. Rev. A}\ }\textbf {\bibinfo {volume} {107}},\
  \bibinfo {pages} {L050402} (\bibinfo {year}
  {2023}{\natexlab{b}})}\BibitemShut {NoStop}%
\bibitem [{\citenamefont {Cozzolino}\ \emph {et~al.}(2019)\citenamefont
  {Cozzolino}, \citenamefont {Da~Lio}, \citenamefont {Bacco},\ and\
  \citenamefont {Oxenløwe}}]{Cozzolino2019}%
  \BibitemOpen
  \bibfield  {author} {\bibinfo {author} {\bibfnamefont {D.}~\bibnamefont
  {Cozzolino}}, \bibinfo {author} {\bibfnamefont {B.}~\bibnamefont {Da~Lio}},
  \bibinfo {author} {\bibfnamefont {D.}~\bibnamefont {Bacco}},\ and\ \bibinfo
  {author} {\bibfnamefont {L.~K.}\ \bibnamefont {Oxenløwe}},\ }\bibfield
  {title} {\bibinfo {title} {High-dimensional quantum communication: Benefits,
  progress, and future challenges},\ }\href
  {https://doi.org/https://doi.org/10.1002/qute.201900038} {\bibfield
  {journal} {\bibinfo  {journal} {Adv. Quantum Technol.}\ }\textbf {\bibinfo
  {volume} {2}},\ \bibinfo {pages} {1900038} (\bibinfo {year}
  {2019})}\BibitemShut {NoStop}%
\bibitem [{\citenamefont {Tang}\ and\ \citenamefont {Zhu}(2013)}]{Tang2013}%
  \BibitemOpen
  \bibfield  {author} {\bibinfo {author} {\bibfnamefont {F.}~\bibnamefont
  {Tang}}\ and\ \bibinfo {author} {\bibfnamefont {B.}~\bibnamefont {Zhu}},\
  }\bibfield  {title} {\bibinfo {title} {Scintillation discriminator improves
  free-space quantum key distribution},\ }\href
  {https://doi.org/10.3788/col201311.090101} {\bibfield  {journal} {\bibinfo
  {journal} {Chin. Opt. Lett.}\ }\textbf {\bibinfo {volume} {11}},\ \bibinfo
  {pages} {090101} (\bibinfo {year} {2013})}\BibitemShut {NoStop}%
\bibitem [{\citenamefont {Vallone}\ \emph
  {et~al.}(2015{\natexlab{b}})\citenamefont {Vallone}, \citenamefont
  {Marangon}, \citenamefont {Canale}, \citenamefont {Savorgnan}, \citenamefont
  {Bacco}, \citenamefont {Barbieri}, \citenamefont {Calimani}, \citenamefont
  {Barbieri}, \citenamefont {Laurenti},\ and\ \citenamefont
  {Villoresi}}]{vallone2015}%
  \BibitemOpen
  \bibfield  {author} {\bibinfo {author} {\bibfnamefont {G.}~\bibnamefont
  {Vallone}}, \bibinfo {author} {\bibfnamefont {D.~G.}\ \bibnamefont
  {Marangon}}, \bibinfo {author} {\bibfnamefont {M.}~\bibnamefont {Canale}},
  \bibinfo {author} {\bibfnamefont {I.}~\bibnamefont {Savorgnan}}, \bibinfo
  {author} {\bibfnamefont {D.}~\bibnamefont {Bacco}}, \bibinfo {author}
  {\bibfnamefont {M.}~\bibnamefont {Barbieri}}, \bibinfo {author}
  {\bibfnamefont {S.}~\bibnamefont {Calimani}}, \bibinfo {author}
  {\bibfnamefont {C.}~\bibnamefont {Barbieri}}, \bibinfo {author}
  {\bibfnamefont {N.}~\bibnamefont {Laurenti}},\ and\ \bibinfo {author}
  {\bibfnamefont {P.}~\bibnamefont {Villoresi}},\ }\bibfield  {title} {\bibinfo
  {title} {Adaptive real time selection for quantum key distribution in lossy
  and turbulent free-space channels},\ }\href
  {https://doi.org/10.1103/PhysRevA.91.042320} {\bibfield  {journal} {\bibinfo
  {journal} {Phys. Rev. A}\ }\textbf {\bibinfo {volume} {91}},\ \bibinfo
  {pages} {042320} (\bibinfo {year} {2015}{\natexlab{b}})}\BibitemShut
  {NoStop}%
\bibitem [{\citenamefont {Semenov}\ and\ \citenamefont
  {Klimov}(2021)}]{semenov2021}%
  \BibitemOpen
  \bibfield  {author} {\bibinfo {author} {\bibfnamefont {A.~A.}\ \bibnamefont
  {Semenov}}\ and\ \bibinfo {author} {\bibfnamefont {A.~B.}\ \bibnamefont
  {Klimov}},\ }\bibfield  {title} {\bibinfo {title} {Dual form of the
  phase-space classical simulation problem in quantum optics},\ }\href
  {https://doi.org/10.1088/1367-2630/ac40cc} {\bibfield  {journal} {\bibinfo
  {journal} {New J. Phys.}\ }\textbf {\bibinfo {volume} {23}},\ \bibinfo
  {pages} {123046} (\bibinfo {year} {2021})}\BibitemShut {NoStop}%
\bibitem [{\citenamefont {Kovtoniuk}\ \emph {et~al.}(2023)\citenamefont
  {Kovtoniuk}, \citenamefont {Stolyarov}, \citenamefont {Kliushnichenko},\ and\
  \citenamefont {Semenov}}]{kovtoniuk2023}%
  \BibitemOpen
  \bibfield  {author} {\bibinfo {author} {\bibfnamefont {V.~S.}\ \bibnamefont
  {Kovtoniuk}}, \bibinfo {author} {\bibfnamefont {E.~V.}\ \bibnamefont
  {Stolyarov}}, \bibinfo {author} {\bibfnamefont {O.~V.}\ \bibnamefont
  {Kliushnichenko}},\ and\ \bibinfo {author} {\bibfnamefont {A.~A.}\
  \bibnamefont {Semenov}},\ }\href@noop {} {\bibinfo {title} {Tight
  inequalities for nonclassicality of measurement statistics}} (\bibinfo {year}
  {2023}),\ \Eprint {https://arxiv.org/abs/2310.14263} {arXiv:2310.14263
  [quant-ph]} \BibitemShut {NoStop}%
\bibitem [{sup()}]{supplement}%
  \BibitemOpen
  \href@noop {} {}\bibinfo {howpublished} {{See Ancillary files for the
  numerically simulated data and the corresponding Python 3
  codes.}}\BibitemShut {Stop}%
\bibitem [{\citenamefont {Glauber}(1963)}]{glauber63c}%
  \BibitemOpen
  \bibfield  {author} {\bibinfo {author} {\bibfnamefont {R.~J.}\ \bibnamefont
  {Glauber}},\ }\bibfield  {title} {\bibinfo {title} {Coherent and incoherent
  states of the radiation field},\ }\href
  {https://doi.org/10.1103/PhysRev.131.2766} {\bibfield  {journal} {\bibinfo
  {journal} {Phys. Rev.}\ }\textbf {\bibinfo {volume} {131}},\ \bibinfo {pages}
  {2766} (\bibinfo {year} {1963})}\BibitemShut {NoStop}%
\bibitem [{\citenamefont {Sudarshan}(1963)}]{sudarshan63}%
  \BibitemOpen
  \bibfield  {author} {\bibinfo {author} {\bibfnamefont {E.~C.~G.}\
  \bibnamefont {Sudarshan}},\ }\bibfield  {title} {\bibinfo {title}
  {Equivalence of semiclassical and quantum mechanical descriptions of
  statistical light beams},\ }\href
  {https://doi.org/10.1103/PhysRevLett.10.277} {\bibfield  {journal} {\bibinfo
  {journal} {Phys. Rev. Lett.}\ }\textbf {\bibinfo {volume} {10}},\ \bibinfo
  {pages} {277} (\bibinfo {year} {1963})}\BibitemShut {NoStop}%
\bibitem [{\citenamefont {Taylor}(1938)}]{taylor1938}%
  \BibitemOpen
  \bibfield  {author} {\bibinfo {author} {\bibfnamefont {G.~I.}\ \bibnamefont
  {Taylor}},\ }\bibfield  {title} {\bibinfo {title} {The spectrum of
  turbulence},\ }\href {https://doi.org/10.1098/rspa.1938.0032} {\bibfield
  {journal} {\bibinfo  {journal} {Proc. R. Soc. London A}\ }\textbf {\bibinfo
  {volume} {164}},\ \bibinfo {pages} {476} (\bibinfo {year}
  {1938})}\BibitemShut {NoStop}%
\bibitem [{\citenamefont {Fleck}\ \emph {et~al.}(1976)\citenamefont {Fleck},
  \citenamefont {Morris},\ and\ \citenamefont {Feit}}]{Fleck1976}%
  \BibitemOpen
  \bibfield  {author} {\bibinfo {author} {\bibfnamefont {J.~A.}\ \bibnamefont
  {Fleck}}, \bibinfo {author} {\bibfnamefont {J.~R.}\ \bibnamefont {Morris}},\
  and\ \bibinfo {author} {\bibfnamefont {M.~D.}\ \bibnamefont {Feit}},\
  }\bibfield  {title} {\bibinfo {title} {Time-dependent propagation of high
  energy laser beams through the atmosphere},\ }\href
  {https://doi.org/10.1007/BF00896333} {\bibfield  {journal} {\bibinfo
  {journal} {Appl. Phys.}\ }\textbf {\bibinfo {volume} {10}},\ \bibinfo {pages}
  {129} (\bibinfo {year} {1976})}\BibitemShut {NoStop}%
\bibitem [{\citenamefont {Frehlich}(2000)}]{Frehlich2000}%
  \BibitemOpen
  \bibfield  {author} {\bibinfo {author} {\bibfnamefont {R.}~\bibnamefont
  {Frehlich}},\ }\bibfield  {title} {\bibinfo {title} {Simulation of laser
  propagation in a turbulent atmosphere},\ }\href
  {https://doi.org/10.1364/AO.39.000393} {\bibfield  {journal} {\bibinfo
  {journal} {Appl. Opt.}\ }\textbf {\bibinfo {volume} {39}},\ \bibinfo {pages}
  {393} (\bibinfo {year} {2000})}\BibitemShut {NoStop}%
\bibitem [{\citenamefont {Lukin}\ and\ \citenamefont
  {Fortes}(2002)}]{Lukin_book}%
  \BibitemOpen
  \bibfield  {author} {\bibinfo {author} {\bibfnamefont {V.~P.}\ \bibnamefont
  {Lukin}}\ and\ \bibinfo {author} {\bibfnamefont {B.~V.}\ \bibnamefont
  {Fortes}},\ }\href {https://doi.org/10.1117/3.452443} {\emph {\bibinfo
  {title} {Adaptive Beaming and Imaging in the Turbulent Atmosphere}}}\
  (\bibinfo  {publisher} {{SPIE}},\ \bibinfo {year} {2002})\BibitemShut
  {NoStop}%
\bibitem [{\citenamefont {Schmidt}(2010)}]{Schmidt_book}%
  \BibitemOpen
  \bibfield  {author} {\bibinfo {author} {\bibfnamefont {J.}~\bibnamefont
  {Schmidt}},\ }\href@noop {} {\emph {\bibinfo {title} {Numerical simulations
  of optical wave propagation with examples in {MATLAB}}}}\ (\bibinfo
  {publisher} {SPIE},\ \bibinfo {address} {Bellingham},\ \bibinfo {year}
  {2010})\BibitemShut {NoStop}%
\bibitem [{\citenamefont {Charnotskii}(2013{\natexlab{a}})}]{Charnotskii2013a}%
  \BibitemOpen
  \bibfield  {author} {\bibinfo {author} {\bibfnamefont {M.}~\bibnamefont
  {Charnotskii}},\ }\bibfield  {title} {\bibinfo {title} {Sparse spectrum model
  for a turbulent phase},\ }\href {https://doi.org/10.1364/JOSAA.30.000479}
  {\bibfield  {journal} {\bibinfo  {journal} {J. Opt. Soc. Am. A}\ }\textbf
  {\bibinfo {volume} {30}},\ \bibinfo {pages} {479} (\bibinfo {year}
  {2013}{\natexlab{a}})}\BibitemShut {NoStop}%
\bibitem [{\citenamefont {Charnotskii}(2013{\natexlab{b}})}]{Charnotskii2013b}%
  \BibitemOpen
  \bibfield  {author} {\bibinfo {author} {\bibfnamefont {M.}~\bibnamefont
  {Charnotskii}},\ }\bibfield  {title} {\bibinfo {title} {Statistics of the
  sparse spectrum turbulent phase},\ }\href
  {https://doi.org/10.1364/JOSAA.30.002455} {\bibfield  {journal} {\bibinfo
  {journal} {J. Opt. Soc. Am. A}\ }\textbf {\bibinfo {volume} {30}},\ \bibinfo
  {pages} {2455} (\bibinfo {year} {2013}{\natexlab{b}})}\BibitemShut {NoStop}%
\bibitem [{\citenamefont {Charnotskii}(2020)}]{Charnotskii2020}%
  \BibitemOpen
  \bibfield  {author} {\bibinfo {author} {\bibfnamefont {M.}~\bibnamefont
  {Charnotskii}},\ }\bibfield  {title} {\bibinfo {title} {Comparison of four
  techniques for turbulent phase screens simulation},\ }\href
  {https://doi.org/10.1364/JOSAA.385754} {\bibfield  {journal} {\bibinfo
  {journal} {J. Opt. Soc. Am. A}\ }\textbf {\bibinfo {volume} {37}},\ \bibinfo
  {pages} {738} (\bibinfo {year} {2020})}\BibitemShut {NoStop}%
\bibitem [{\citenamefont {Martin}\ and\ \citenamefont
  {Flatt\'{e}}(1988)}]{Martin1988}%
  \BibitemOpen
  \bibfield  {author} {\bibinfo {author} {\bibfnamefont {J.~M.}\ \bibnamefont
  {Martin}}\ and\ \bibinfo {author} {\bibfnamefont {S.~M.}\ \bibnamefont
  {Flatt\'{e}}},\ }\bibfield  {title} {\bibinfo {title} {Intensity images and
  statistics from numerical simulation of wave propagation in 3-d random
  media},\ }\href {https://doi.org/10.1364/AO.27.002111} {\bibfield  {journal}
  {\bibinfo  {journal} {Appl. Opt.}\ }\textbf {\bibinfo {volume} {27}},\
  \bibinfo {pages} {2111} (\bibinfo {year} {1988})}\BibitemShut {NoStop}%
\bibitem [{\citenamefont {Julsgaard}\ \emph {et~al.}(2004)\citenamefont
  {Julsgaard}, \citenamefont {Sherson}, \citenamefont {Cirac}, \citenamefont
  {Fiur{\'a}{\v{s}}ek},\ and\ \citenamefont {Polzik}}]{Julsgaard2004}%
  \BibitemOpen
  \bibfield  {author} {\bibinfo {author} {\bibfnamefont {B.}~\bibnamefont
  {Julsgaard}}, \bibinfo {author} {\bibfnamefont {J.}~\bibnamefont {Sherson}},
  \bibinfo {author} {\bibfnamefont {J.~I.}\ \bibnamefont {Cirac}}, \bibinfo
  {author} {\bibfnamefont {J.}~\bibnamefont {Fiur{\'a}{\v{s}}ek}},\ and\
  \bibinfo {author} {\bibfnamefont {E.~S.}\ \bibnamefont {Polzik}},\ }\bibfield
   {title} {\bibinfo {title} {Experimental demonstration of quantum memory for
  light},\ }\href {https://doi.org/10.1038/nature03064} {\bibfield  {journal}
  {\bibinfo  {journal} {Nature}\ }\textbf {\bibinfo {volume} {432}},\ \bibinfo
  {pages} {482} (\bibinfo {year} {2004})}\BibitemShut {NoStop}%
\bibitem [{\citenamefont {Lvovsky}\ \emph {et~al.}(2009)\citenamefont
  {Lvovsky}, \citenamefont {Sanders},\ and\ \citenamefont
  {Tittel}}]{Lvovsky2009}%
  \BibitemOpen
  \bibfield  {author} {\bibinfo {author} {\bibfnamefont {A.~I.}\ \bibnamefont
  {Lvovsky}}, \bibinfo {author} {\bibfnamefont {B.~C.}\ \bibnamefont
  {Sanders}},\ and\ \bibinfo {author} {\bibfnamefont {W.}~\bibnamefont
  {Tittel}},\ }\bibfield  {title} {\bibinfo {title} {Optical quantum memory},\
  }\href {https://doi.org/10.1038/nphoton.2009.231} {\bibfield  {journal}
  {\bibinfo  {journal} {Nat. Photonics}\ }\textbf {\bibinfo {volume} {3}},\
  \bibinfo {pages} {706} (\bibinfo {year} {2009})}\BibitemShut {NoStop}%
\bibitem [{\citenamefont {Hedges}\ \emph {et~al.}(2010)\citenamefont {Hedges},
  \citenamefont {Longdell}, \citenamefont {Li},\ and\ \citenamefont
  {Sellars}}]{Hedges2010}%
  \BibitemOpen
  \bibfield  {author} {\bibinfo {author} {\bibfnamefont {M.~P.}\ \bibnamefont
  {Hedges}}, \bibinfo {author} {\bibfnamefont {J.~J.}\ \bibnamefont
  {Longdell}}, \bibinfo {author} {\bibfnamefont {Y.}~\bibnamefont {Li}},\ and\
  \bibinfo {author} {\bibfnamefont {M.~J.}\ \bibnamefont {Sellars}},\
  }\bibfield  {title} {\bibinfo {title} {Efficient quantum memory for light},\
  }\href {https://doi.org/10.1038/nature09081} {\bibfield  {journal} {\bibinfo
  {journal} {Nature}\ }\textbf {\bibinfo {volume} {465}},\ \bibinfo {pages}
  {1052} (\bibinfo {year} {2010})}\BibitemShut {NoStop}%
\bibitem [{\citenamefont {Jensen}\ \emph {et~al.}(2011)\citenamefont {Jensen},
  \citenamefont {Wasilewski}, \citenamefont {Krauter}, \citenamefont
  {Fernholz}, \citenamefont {Nielsen}, \citenamefont {Owari}, \citenamefont
  {Plenio}, \citenamefont {Serafini}, \citenamefont {Wolf},\ and\ \citenamefont
  {Polzik}}]{Jensen2011}%
  \BibitemOpen
  \bibfield  {author} {\bibinfo {author} {\bibfnamefont {K.}~\bibnamefont
  {Jensen}}, \bibinfo {author} {\bibfnamefont {W.}~\bibnamefont {Wasilewski}},
  \bibinfo {author} {\bibfnamefont {H.}~\bibnamefont {Krauter}}, \bibinfo
  {author} {\bibfnamefont {T.}~\bibnamefont {Fernholz}}, \bibinfo {author}
  {\bibfnamefont {B.~M.}\ \bibnamefont {Nielsen}}, \bibinfo {author}
  {\bibfnamefont {M.}~\bibnamefont {Owari}}, \bibinfo {author} {\bibfnamefont
  {M.~B.}\ \bibnamefont {Plenio}}, \bibinfo {author} {\bibfnamefont
  {A.}~\bibnamefont {Serafini}}, \bibinfo {author} {\bibfnamefont {M.~M.}\
  \bibnamefont {Wolf}},\ and\ \bibinfo {author} {\bibfnamefont {E.~S.}\
  \bibnamefont {Polzik}},\ }\bibfield  {title} {\bibinfo {title} {Quantum
  memory for entangled continuous-variable states},\ }\href
  {https://doi.org/10.1038/nphys1819} {\bibfield  {journal} {\bibinfo
  {journal} {Nat. Physics}\ }\textbf {\bibinfo {volume} {7}},\ \bibinfo {pages}
  {13} (\bibinfo {year} {2011})}\BibitemShut {NoStop}%
\bibitem [{\citenamefont {Yan}\ \emph {et~al.}(2018)\citenamefont {Yan},
  \citenamefont {Liu}, \citenamefont {Yan},\ and\ \citenamefont
  {Jia}}]{Yan2018}%
  \BibitemOpen
  \bibfield  {author} {\bibinfo {author} {\bibfnamefont {Z.}~\bibnamefont
  {Yan}}, \bibinfo {author} {\bibfnamefont {Y.}~\bibnamefont {Liu}}, \bibinfo
  {author} {\bibfnamefont {J.}~\bibnamefont {Yan}},\ and\ \bibinfo {author}
  {\bibfnamefont {X.}~\bibnamefont {Jia}},\ }\bibfield  {title} {\bibinfo
  {title} {Deterministically entangling multiple remote quantum memories inside
  an optical cavity},\ }\href {https://doi.org/10.1103/PhysRevA.97.013856}
  {\bibfield  {journal} {\bibinfo  {journal} {Phys. Rev. A}\ }\textbf {\bibinfo
  {volume} {97}},\ \bibinfo {pages} {013856} (\bibinfo {year}
  {2018})}\BibitemShut {NoStop}%
\bibitem [{\citenamefont {Ma}\ \emph {et~al.}(2022)\citenamefont {Ma},
  \citenamefont {Lei}, \citenamefont {Yan}, \citenamefont {Li}, \citenamefont
  {Chai}, \citenamefont {Yan}, \citenamefont {Jia}, \citenamefont {Xie},\ and\
  \citenamefont {Peng}}]{Ma2022}%
  \BibitemOpen
  \bibfield  {author} {\bibinfo {author} {\bibfnamefont {L.}~\bibnamefont
  {Ma}}, \bibinfo {author} {\bibfnamefont {X.}~\bibnamefont {Lei}}, \bibinfo
  {author} {\bibfnamefont {J.}~\bibnamefont {Yan}}, \bibinfo {author}
  {\bibfnamefont {R.}~\bibnamefont {Li}}, \bibinfo {author} {\bibfnamefont
  {T.}~\bibnamefont {Chai}}, \bibinfo {author} {\bibfnamefont {Z.}~\bibnamefont
  {Yan}}, \bibinfo {author} {\bibfnamefont {X.}~\bibnamefont {Jia}}, \bibinfo
  {author} {\bibfnamefont {C.}~\bibnamefont {Xie}},\ and\ \bibinfo {author}
  {\bibfnamefont {K.}~\bibnamefont {Peng}},\ }\bibfield  {title} {\bibinfo
  {title} {High-performance cavity-enhanced quantum memory with warm atomic
  cell},\ }\href {https://doi.org/10.1038/s41467-022-30077-1} {\bibfield
  {journal} {\bibinfo  {journal} {Nat. Commun.}\ }\textbf {\bibinfo {volume}
  {13}},\ \bibinfo {pages} {2368} (\bibinfo {year} {2022})}\BibitemShut
  {NoStop}%
\bibitem [{\citenamefont {Simon}(2000)}]{simon00}%
  \BibitemOpen
  \bibfield  {author} {\bibinfo {author} {\bibfnamefont {R.}~\bibnamefont
  {Simon}},\ }\bibfield  {title} {\bibinfo {title} {Peres-horodecki
  separability criterion for continuous variable systems},\ }\href
  {https://doi.org/10.1103/PhysRevLett.84.2726} {\bibfield  {journal} {\bibinfo
   {journal} {Phys. Rev. Lett.}\ }\textbf {\bibinfo {volume} {84}},\ \bibinfo
  {pages} {2726} (\bibinfo {year} {2000})}\BibitemShut {NoStop}%
\bibitem [{\citenamefont {Scarani}\ \emph {et~al.}(2009)\citenamefont
  {Scarani}, \citenamefont {Bechmann-Pasquinucci}, \citenamefont {Cerf},
  \citenamefont {Du\ifmmode~\check{s}\else \v{s}\fi{}ek}, \citenamefont
  {L\"utkenhaus},\ and\ \citenamefont {Peev}}]{scarani09}%
  \BibitemOpen
  \bibfield  {author} {\bibinfo {author} {\bibfnamefont {V.}~\bibnamefont
  {Scarani}}, \bibinfo {author} {\bibfnamefont {H.}~\bibnamefont
  {Bechmann-Pasquinucci}}, \bibinfo {author} {\bibfnamefont {N.~J.}\
  \bibnamefont {Cerf}}, \bibinfo {author} {\bibfnamefont {M.}~\bibnamefont
  {Du\ifmmode~\check{s}\else \v{s}\fi{}ek}}, \bibinfo {author} {\bibfnamefont
  {N.}~\bibnamefont {L\"utkenhaus}},\ and\ \bibinfo {author} {\bibfnamefont
  {M.}~\bibnamefont {Peev}},\ }\bibfield  {title} {\bibinfo {title} {The
  security of practical quantum key distribution},\ }\href
  {https://doi.org/10.1103/RevModPhys.81.1301} {\bibfield  {journal} {\bibinfo
  {journal} {Rev. Mod. Phys.}\ }\textbf {\bibinfo {volume} {81}},\ \bibinfo
  {pages} {1301} (\bibinfo {year} {2009})}\BibitemShut {NoStop}%
\bibitem [{\citenamefont {Kok}\ and\ \citenamefont
  {Braunstein}(2000)}]{Kok2000}%
  \BibitemOpen
  \bibfield  {author} {\bibinfo {author} {\bibfnamefont {P.}~\bibnamefont
  {Kok}}\ and\ \bibinfo {author} {\bibfnamefont {S.~L.}\ \bibnamefont
  {Braunstein}},\ }\bibfield  {title} {\bibinfo {title} {Postselected versus
  nonpostselected quantum teleportation using parametric down-conversion},\
  }\href {https://doi.org/10.1103/PhysRevA.61.042304} {\bibfield  {journal}
  {\bibinfo  {journal} {Phys. Rev. A}\ }\textbf {\bibinfo {volume} {61}},\
  \bibinfo {pages} {042304} (\bibinfo {year} {2000})}\BibitemShut {NoStop}%
\bibitem [{\citenamefont {Ma}\ \emph {et~al.}(2007)\citenamefont {Ma},
  \citenamefont {Fung},\ and\ \citenamefont {Lo}}]{Ma2007}%
  \BibitemOpen
  \bibfield  {author} {\bibinfo {author} {\bibfnamefont {X.}~\bibnamefont
  {Ma}}, \bibinfo {author} {\bibfnamefont {C.-H.~F.}\ \bibnamefont {Fung}},\
  and\ \bibinfo {author} {\bibfnamefont {H.-K.}\ \bibnamefont {Lo}},\
  }\bibfield  {title} {\bibinfo {title} {Quantum key distribution with
  entangled photon sources},\ }\href
  {https://doi.org/10.1103/PhysRevA.76.012307} {\bibfield  {journal} {\bibinfo
  {journal} {Phys. Rev. A}\ }\textbf {\bibinfo {volume} {76}},\ \bibinfo
  {pages} {012307} (\bibinfo {year} {2007})}\BibitemShut {NoStop}%
\bibitem [{\citenamefont {Semenov}\ and\ \citenamefont
  {Vogel}(2011)}]{Semenov2011}%
  \BibitemOpen
  \bibfield  {author} {\bibinfo {author} {\bibfnamefont {A.~A.}\ \bibnamefont
  {Semenov}}\ and\ \bibinfo {author} {\bibfnamefont {W.}~\bibnamefont
  {Vogel}},\ }\bibfield  {title} {\bibinfo {title} {Fake violations of the
  quantum bell-parameter bound},\ }\href
  {https://doi.org/10.1103/PhysRevA.83.032119} {\bibfield  {journal} {\bibinfo
  {journal} {Phys. Rev. A}\ }\textbf {\bibinfo {volume} {83}},\ \bibinfo
  {pages} {032119} (\bibinfo {year} {2011})}\BibitemShut {NoStop}%
\bibitem [{\citenamefont {Beaudry}\ \emph {et~al.}(2008)\citenamefont
  {Beaudry}, \citenamefont {Moroder},\ and\ \citenamefont
  {L\"utkenhaus}}]{Beaudry2008}%
  \BibitemOpen
  \bibfield  {author} {\bibinfo {author} {\bibfnamefont {N.~J.}\ \bibnamefont
  {Beaudry}}, \bibinfo {author} {\bibfnamefont {T.}~\bibnamefont {Moroder}},\
  and\ \bibinfo {author} {\bibfnamefont {N.}~\bibnamefont {L\"utkenhaus}},\
  }\bibfield  {title} {\bibinfo {title} {Squashing models for optical
  measurements in quantum communication},\ }\href
  {https://doi.org/10.1103/PhysRevLett.101.093601} {\bibfield  {journal}
  {\bibinfo  {journal} {Phys. Rev. Lett.}\ }\textbf {\bibinfo {volume} {101}},\
  \bibinfo {pages} {093601} (\bibinfo {year} {2008})}\BibitemShut {NoStop}%
\bibitem [{\citenamefont {Moroder}\ \emph {et~al.}(2010)\citenamefont
  {Moroder}, \citenamefont {G\"uhne}, \citenamefont {Beaudry}, \citenamefont
  {Piani},\ and\ \citenamefont {L\"utkenhaus}}]{Moroder2010}%
  \BibitemOpen
  \bibfield  {author} {\bibinfo {author} {\bibfnamefont {T.}~\bibnamefont
  {Moroder}}, \bibinfo {author} {\bibfnamefont {O.}~\bibnamefont {G\"uhne}},
  \bibinfo {author} {\bibfnamefont {N.}~\bibnamefont {Beaudry}}, \bibinfo
  {author} {\bibfnamefont {M.}~\bibnamefont {Piani}},\ and\ \bibinfo {author}
  {\bibfnamefont {N.}~\bibnamefont {L\"utkenhaus}},\ }\bibfield  {title}
  {\bibinfo {title} {Entanglement verification with realistic measurement
  devices via squashing operations},\ }\href
  {https://doi.org/10.1103/PhysRevA.81.052342} {\bibfield  {journal} {\bibinfo
  {journal} {Phys. Rev. A}\ }\textbf {\bibinfo {volume} {81}},\ \bibinfo
  {pages} {052342} (\bibinfo {year} {2010})}\BibitemShut {NoStop}%
\bibitem [{\citenamefont {Fung}\ \emph {et~al.}(2011)\citenamefont {Fung},
  \citenamefont {Chau},\ and\ \citenamefont {Lo}}]{Fung2011}%
  \BibitemOpen
  \bibfield  {author} {\bibinfo {author} {\bibfnamefont {C.-H.~F.}\
  \bibnamefont {Fung}}, \bibinfo {author} {\bibfnamefont {H.~F.}\ \bibnamefont
  {Chau}},\ and\ \bibinfo {author} {\bibfnamefont {H.-K.}\ \bibnamefont {Lo}},\
  }\bibfield  {title} {\bibinfo {title} {Universal squash model for optical
  communications using linear optics and threshold detectors},\ }\href
  {https://doi.org/10.1103/PhysRevA.84.020303} {\bibfield  {journal} {\bibinfo
  {journal} {Phys. Rev. A}\ }\textbf {\bibinfo {volume} {84}},\ \bibinfo
  {pages} {020303(R)} (\bibinfo {year} {2011})}\BibitemShut {NoStop}%
\bibitem [{\citenamefont {Clauser}\ \emph {et~al.}(1969)\citenamefont
  {Clauser}, \citenamefont {Horne}, \citenamefont {Shimony},\ and\
  \citenamefont {Holt}}]{CHSH}%
  \BibitemOpen
  \bibfield  {author} {\bibinfo {author} {\bibfnamefont {J.~F.}\ \bibnamefont
  {Clauser}}, \bibinfo {author} {\bibfnamefont {M.~A.}\ \bibnamefont {Horne}},
  \bibinfo {author} {\bibfnamefont {A.}~\bibnamefont {Shimony}},\ and\ \bibinfo
  {author} {\bibfnamefont {R.~A.}\ \bibnamefont {Holt}},\ }\bibfield  {title}
  {\bibinfo {title} {Proposed experiment to test local hidden-variable
  theories},\ }\href {https://doi.org/10.1103/PhysRevLett.23.880} {\bibfield
  {journal} {\bibinfo  {journal} {Phys. Rev. Lett.}\ }\textbf {\bibinfo
  {volume} {23}},\ \bibinfo {pages} {880} (\bibinfo {year} {1969})}\BibitemShut
  {NoStop}%
\bibitem [{\citenamefont {Pratt}(1969)}]{Pratt1969}%
  \BibitemOpen
  \bibfield  {author} {\bibinfo {author} {\bibfnamefont {W.~K.}\ \bibnamefont
  {Pratt}},\ }\href@noop {} {\emph {\bibinfo {title} {Laser Communication
  Systems}}}\ (\bibinfo  {publisher} {Wiley},\ \bibinfo {address} {New York},\
  \bibinfo {year} {1969})\BibitemShut {NoStop}%
\bibitem [{\citenamefont {Karp}\ \emph {et~al.}(1970)\citenamefont {Karp},
  \citenamefont {O'Neill},\ and\ \citenamefont {Gagliardi}}]{Karp1970}%
  \BibitemOpen
  \bibfield  {author} {\bibinfo {author} {\bibfnamefont {S.}~\bibnamefont
  {Karp}}, \bibinfo {author} {\bibfnamefont {E.~L.}\ \bibnamefont {O'Neill}},\
  and\ \bibinfo {author} {\bibfnamefont {R.~M.}\ \bibnamefont {Gagliardi}},\
  }\bibfield  {title} {\bibinfo {title} {Communication theory for the
  free-space optical channel},\ }\href {https://doi.org/10.1109/PROC.1970.7985}
  {\bibfield  {journal} {\bibinfo  {journal} {Proceedings of the IEEE}\
  }\textbf {\bibinfo {volume} {58}},\ \bibinfo {pages} {1611} (\bibinfo {year}
  {1970})}\BibitemShut {NoStop}%
\bibitem [{\citenamefont {Lee}\ \emph {et~al.}(2004)\citenamefont {Lee},
  \citenamefont {Yurtsever}, \citenamefont {Kok}, \citenamefont {Hockney},
  \citenamefont {Adami}, \citenamefont {Braunstein},\ and\ \citenamefont
  {Dowling}}]{Lee2005}%
  \BibitemOpen
  \bibfield  {author} {\bibinfo {author} {\bibfnamefont {H.}~\bibnamefont
  {Lee}}, \bibinfo {author} {\bibfnamefont {U.}~\bibnamefont {Yurtsever}},
  \bibinfo {author} {\bibfnamefont {P.}~\bibnamefont {Kok}}, \bibinfo {author}
  {\bibfnamefont {G.~M.}\ \bibnamefont {Hockney}}, \bibinfo {author}
  {\bibfnamefont {C.}~\bibnamefont {Adami}}, \bibinfo {author} {\bibfnamefont
  {S.~L.}\ \bibnamefont {Braunstein}},\ and\ \bibinfo {author} {\bibfnamefont
  {J.~P.}\ \bibnamefont {Dowling}},\ }\bibfield  {title} {\bibinfo {title}
  {Towards photostatistics from photon-number discriminating detectors},\
  }\href {https://doi.org/10.1080/09500340408235289} {\bibfield  {journal}
  {\bibinfo  {journal} {J. Mod. Opt.}\ }\textbf {\bibinfo {volume} {51}},\
  \bibinfo {pages} {1517} (\bibinfo {year} {2004})}\BibitemShut {NoStop}%
\bibitem [{\citenamefont {Semenov}\ \emph {et~al.}(2008)\citenamefont
  {Semenov}, \citenamefont {Turchin},\ and\ \citenamefont
  {Gomonay}}]{Semenov2008}%
  \BibitemOpen
  \bibfield  {author} {\bibinfo {author} {\bibfnamefont {A.~A.}\ \bibnamefont
  {Semenov}}, \bibinfo {author} {\bibfnamefont {A.~V.}\ \bibnamefont
  {Turchin}},\ and\ \bibinfo {author} {\bibfnamefont {H.~V.}\ \bibnamefont
  {Gomonay}},\ }\bibfield  {title} {\bibinfo {title} {Detection of quantum
  light in the presence of noise},\ }\href
  {https://doi.org/10.1103/PhysRevA.78.055803} {\bibfield  {journal} {\bibinfo
  {journal} {Phys. Rev. A}\ }\textbf {\bibinfo {volume} {78}},\ \bibinfo
  {pages} {055803} (\bibinfo {year} {2008})}\BibitemShut {NoStop}%
\bibitem [{\citenamefont {Mandel}\ and\ \citenamefont
  {Wolf}(1995)}]{mandel_book}%
  \BibitemOpen
  \bibfield  {author} {\bibinfo {author} {\bibfnamefont {L.}~\bibnamefont
  {Mandel}}\ and\ \bibinfo {author} {\bibfnamefont {E.}~\bibnamefont {Wolf}},\
  }\href@noop {} {\emph {\bibinfo {title} {Optical Coherence and Quantum
  Optics}}}\ (\bibinfo  {publisher} {Cambridge University Press, Cambridge},\
  \bibinfo {year} {1995})\BibitemShut {NoStop}%
\bibitem [{\citenamefont {Schnabel}(2017)}]{Schnabel2017}%
  \BibitemOpen
  \bibfield  {author} {\bibinfo {author} {\bibfnamefont {R.}~\bibnamefont
  {Schnabel}},\ }\bibfield  {title} {\bibinfo {title} {Squeezed states of light
  and their applications in laser interferometers},\ }\href
  {https://doi.org/https://doi.org/10.1016/j.physrep.2017.04.001} {\bibfield
  {journal} {\bibinfo  {journal} {Phys. Rep.}\ }\textbf {\bibinfo {volume}
  {684}},\ \bibinfo {pages} {1} (\bibinfo {year} {2017})}\BibitemShut {NoStop}%
\bibitem [{\citenamefont {Paul}\ \emph {et~al.}(1996)\citenamefont {Paul},
  \citenamefont {T\"orm\"a}, \citenamefont {Kiss},\ and\ \citenamefont
  {Jex}}]{paul1996}%
  \BibitemOpen
  \bibfield  {author} {\bibinfo {author} {\bibfnamefont {H.}~\bibnamefont
  {Paul}}, \bibinfo {author} {\bibfnamefont {P.}~\bibnamefont {T\"orm\"a}},
  \bibinfo {author} {\bibfnamefont {T.}~\bibnamefont {Kiss}},\ and\ \bibinfo
  {author} {\bibfnamefont {I.}~\bibnamefont {Jex}},\ }\bibfield  {title}
  {\bibinfo {title} {Photon chopping: New way to measure the quantum state of
  light},\ }\href {https://doi.org/10.1103/PhysRevLett.76.2464} {\bibfield
  {journal} {\bibinfo  {journal} {Phys. Rev. Lett.}\ }\textbf {\bibinfo
  {volume} {76}},\ \bibinfo {pages} {2464} (\bibinfo {year}
  {1996})}\BibitemShut {NoStop}%
\bibitem [{\citenamefont {Castelletto}\ \emph {et~al.}(2007)\citenamefont
  {Castelletto}, \citenamefont {Degiovanni}, \citenamefont {Schettini},\ and\
  \citenamefont {Migdall}}]{castelletto2007}%
  \BibitemOpen
  \bibfield  {author} {\bibinfo {author} {\bibfnamefont {S.~A.}\ \bibnamefont
  {Castelletto}}, \bibinfo {author} {\bibfnamefont {I.~P.}\ \bibnamefont
  {Degiovanni}}, \bibinfo {author} {\bibfnamefont {V.}~\bibnamefont
  {Schettini}},\ and\ \bibinfo {author} {\bibfnamefont {A.~L.}\ \bibnamefont
  {Migdall}},\ }\bibfield  {title} {\bibinfo {title} {Reduced deadtime and
  higher rate photon-counting detection using a multiplexed detector array},\
  }\href {https://doi.org/10.1080/09500340600779579} {\bibfield  {journal}
  {\bibinfo  {journal} {J. Mod. Opt.}\ }\textbf {\bibinfo {volume} {54}},\
  \bibinfo {pages} {337} (\bibinfo {year} {2007})}\BibitemShut {NoStop}%
\bibitem [{\citenamefont {Schettini}\ \emph {et~al.}(2007)\citenamefont
  {Schettini}, \citenamefont {Polyakov}, \citenamefont {Degiovanni},
  \citenamefont {Brida}, \citenamefont {Castelletto},\ and\ \citenamefont
  {Migdall}}]{schettini2007}%
  \BibitemOpen
  \bibfield  {author} {\bibinfo {author} {\bibfnamefont {V.}~\bibnamefont
  {Schettini}}, \bibinfo {author} {\bibfnamefont {S.~V.}\ \bibnamefont
  {Polyakov}}, \bibinfo {author} {\bibfnamefont {I.~P.}\ \bibnamefont
  {Degiovanni}}, \bibinfo {author} {\bibfnamefont {G.}~\bibnamefont {Brida}},
  \bibinfo {author} {\bibfnamefont {S.}~\bibnamefont {Castelletto}},\ and\
  \bibinfo {author} {\bibfnamefont {A.~L.}\ \bibnamefont {Migdall}},\
  }\bibfield  {title} {\bibinfo {title} {Implementing a multiplexed system of
  detectors for higher photon counting rates},\ }\href
  {https://doi.org/10.1109/JSTQE.2007.902846} {\bibfield  {journal} {\bibinfo
  {journal} {IEEE J. Sel. Top. Quantum Electron.}\ }\textbf {\bibinfo {volume}
  {13}},\ \bibinfo {pages} {978} (\bibinfo {year} {2007})}\BibitemShut
  {NoStop}%
\bibitem [{\citenamefont {Blanchet}\ \emph {et~al.}(2008)\citenamefont
  {Blanchet}, \citenamefont {Devaux}, \citenamefont {Furfaro},\ and\
  \citenamefont {Lantz}}]{blanchet08}%
  \BibitemOpen
  \bibfield  {author} {\bibinfo {author} {\bibfnamefont {J.-L.}\ \bibnamefont
  {Blanchet}}, \bibinfo {author} {\bibfnamefont {F.}~\bibnamefont {Devaux}},
  \bibinfo {author} {\bibfnamefont {L.}~\bibnamefont {Furfaro}},\ and\ \bibinfo
  {author} {\bibfnamefont {E.}~\bibnamefont {Lantz}},\ }\bibfield  {title}
  {\bibinfo {title} {Measurement of sub-shot-noise correlations of spatial
  fluctuations in the photon-counting regime},\ }\href
  {https://doi.org/10.1103/PhysRevLett.101.233604} {\bibfield  {journal}
  {\bibinfo  {journal} {Phys. Rev. Lett.}\ }\textbf {\bibinfo {volume} {101}},\
  \bibinfo {pages} {233604} (\bibinfo {year} {2008})}\BibitemShut {NoStop}%
\bibitem [{\citenamefont {Achilles}\ \emph {et~al.}(2003)\citenamefont
  {Achilles}, \citenamefont {Silberhorn}, \citenamefont {\'{S}liwa},
  \citenamefont {Banaszek},\ and\ \citenamefont {Walmsley}}]{achilles03}%
  \BibitemOpen
  \bibfield  {author} {\bibinfo {author} {\bibfnamefont {D.}~\bibnamefont
  {Achilles}}, \bibinfo {author} {\bibfnamefont {C.}~\bibnamefont
  {Silberhorn}}, \bibinfo {author} {\bibfnamefont {C.}~\bibnamefont
  {\'{S}liwa}}, \bibinfo {author} {\bibfnamefont {K.}~\bibnamefont
  {Banaszek}},\ and\ \bibinfo {author} {\bibfnamefont {I.~A.}\ \bibnamefont
  {Walmsley}},\ }\bibfield  {title} {\bibinfo {title} {Fiber-assisted detection
  with photon number resolution},\ }\href
  {https://doi.org/10.1364/OL.28.002387} {\bibfield  {journal} {\bibinfo
  {journal} {Opt. Lett.}\ }\textbf {\bibinfo {volume} {28}},\ \bibinfo {pages}
  {2387} (\bibinfo {year} {2003})}\BibitemShut {NoStop}%
\bibitem [{\citenamefont {Fitch}\ \emph {et~al.}(2003)\citenamefont {Fitch},
  \citenamefont {Jacobs}, \citenamefont {Pittman},\ and\ \citenamefont
  {Franson}}]{fitch03}%
  \BibitemOpen
  \bibfield  {author} {\bibinfo {author} {\bibfnamefont {M.~J.}\ \bibnamefont
  {Fitch}}, \bibinfo {author} {\bibfnamefont {B.~C.}\ \bibnamefont {Jacobs}},
  \bibinfo {author} {\bibfnamefont {T.~B.}\ \bibnamefont {Pittman}},\ and\
  \bibinfo {author} {\bibfnamefont {J.~D.}\ \bibnamefont {Franson}},\
  }\bibfield  {title} {\bibinfo {title} {Photon-number resolution using
  time-multiplexed single-photon detectors},\ }\href
  {https://doi.org/10.1103/PhysRevA.68.043814} {\bibfield  {journal} {\bibinfo
  {journal} {Phys. Rev. A}\ }\textbf {\bibinfo {volume} {68}},\ \bibinfo
  {pages} {043814} (\bibinfo {year} {2003})}\BibitemShut {NoStop}%
\bibitem [{\citenamefont {\ifmmode \check{R}\else
  \v{R}\fi{}eh\'a\ifmmode~\check{c}\else \v{c}\fi{}ek}\ \emph
  {et~al.}(2003)\citenamefont {\ifmmode \check{R}\else
  \v{R}\fi{}eh\'a\ifmmode~\check{c}\else \v{c}\fi{}ek}, \citenamefont {Hradil},
  \citenamefont {Haderka}, \citenamefont {Pe\ifmmode~\check{r}\else
  \v{r}\fi{}ina},\ and\ \citenamefont {Hamar}}]{rehacek03}%
  \BibitemOpen
  \bibfield  {author} {\bibinfo {author} {\bibfnamefont {J.}~\bibnamefont
  {\ifmmode \check{R}\else \v{R}\fi{}eh\'a\ifmmode~\check{c}\else
  \v{c}\fi{}ek}}, \bibinfo {author} {\bibfnamefont {Z.}~\bibnamefont {Hradil}},
  \bibinfo {author} {\bibfnamefont {O.}~\bibnamefont {Haderka}}, \bibinfo
  {author} {\bibfnamefont {J.}~\bibnamefont {Pe\ifmmode~\check{r}\else
  \v{r}\fi{}ina}},\ and\ \bibinfo {author} {\bibfnamefont {M.}~\bibnamefont
  {Hamar}},\ }\bibfield  {title} {\bibinfo {title} {Multiple-photon resolving
  fiber-loop detector},\ }\href {https://doi.org/10.1103/PhysRevA.67.061801}
  {\bibfield  {journal} {\bibinfo  {journal} {Phys. Rev. A}\ }\textbf {\bibinfo
  {volume} {67}},\ \bibinfo {pages} {061801(R)} (\bibinfo {year}
  {2003})}\BibitemShut {NoStop}%
\bibitem [{\citenamefont {Sperling}\ \emph
  {et~al.}(2012{\natexlab{a}})\citenamefont {Sperling}, \citenamefont {Vogel},\
  and\ \citenamefont {Agarwal}}]{sperling12a}%
  \BibitemOpen
  \bibfield  {author} {\bibinfo {author} {\bibfnamefont {J.}~\bibnamefont
  {Sperling}}, \bibinfo {author} {\bibfnamefont {W.}~\bibnamefont {Vogel}},\
  and\ \bibinfo {author} {\bibfnamefont {G.~S.}\ \bibnamefont {Agarwal}},\
  }\bibfield  {title} {\bibinfo {title} {True photocounting statistics of
  multiple on-off detectors},\ }\href
  {https://doi.org/10.1103/PhysRevA.85.023820} {\bibfield  {journal} {\bibinfo
  {journal} {Phys. Rev. A}\ }\textbf {\bibinfo {volume} {85}},\ \bibinfo
  {pages} {023820} (\bibinfo {year} {2012}{\natexlab{a}})}\BibitemShut
  {NoStop}%
\bibitem [{\citenamefont {Sperling}\ \emph
  {et~al.}(2012{\natexlab{b}})\citenamefont {Sperling}, \citenamefont {Vogel},\
  and\ \citenamefont {Agarwal}}]{sperling12c}%
  \BibitemOpen
  \bibfield  {author} {\bibinfo {author} {\bibfnamefont {J.}~\bibnamefont
  {Sperling}}, \bibinfo {author} {\bibfnamefont {W.}~\bibnamefont {Vogel}},\
  and\ \bibinfo {author} {\bibfnamefont {G.~S.}\ \bibnamefont {Agarwal}},\
  }\bibfield  {title} {\bibinfo {title} {Sub-binomial light},\ }\href
  {https://doi.org/10.1103/PhysRevLett.109.093601} {\bibfield  {journal}
  {\bibinfo  {journal} {Phys. Rev. Lett.}\ }\textbf {\bibinfo {volume} {109}},\
  \bibinfo {pages} {093601} (\bibinfo {year} {2012}{\natexlab{b}})}\BibitemShut
  {NoStop}%
\bibitem [{\citenamefont {Titulaer}\ and\ \citenamefont
  {Glauber}(1965)}]{titulaer65}%
  \BibitemOpen
  \bibfield  {author} {\bibinfo {author} {\bibfnamefont {U.~M.}\ \bibnamefont
  {Titulaer}}\ and\ \bibinfo {author} {\bibfnamefont {R.~J.}\ \bibnamefont
  {Glauber}},\ }\bibfield  {title} {\bibinfo {title} {Correlation functions for
  coherent fields},\ }\href {https://doi.org/10.1103/PhysRev.140.B676}
  {\bibfield  {journal} {\bibinfo  {journal} {Phys. Rev.}\ }\textbf {\bibinfo
  {volume} {140}},\ \bibinfo {pages} {B676} (\bibinfo {year}
  {1965})}\BibitemShut {NoStop}%
\bibitem [{\citenamefont {Mandel}(1986)}]{mandel86}%
  \BibitemOpen
  \bibfield  {author} {\bibinfo {author} {\bibfnamefont {L.}~\bibnamefont
  {Mandel}},\ }\bibfield  {title} {\bibinfo {title} {Non-classical states of
  the electromagnetic field},\ }\href
  {http://stacks.iop.org/1402-4896/1986/i=T12/a=005} {\bibfield  {journal}
  {\bibinfo  {journal} {Phys. Scr.}\ }\textbf {\bibinfo {volume} {1986}},\
  \bibinfo {pages} {34} (\bibinfo {year} {1986})}\BibitemShut {NoStop}%
\bibitem [{\citenamefont {Vogel}\ and\ \citenamefont
  {Welsch}(2006)}]{vogel_book}%
  \BibitemOpen
  \bibfield  {author} {\bibinfo {author} {\bibfnamefont {W.}~\bibnamefont
  {Vogel}}\ and\ \bibinfo {author} {\bibfnamefont {D.-G.}\ \bibnamefont
  {Welsch}},\ }\href@noop {} {\emph {\bibinfo {title} {Quantum Optics}}}\
  (\bibinfo  {publisher} {Wiley-VCH Verlag GmbH \& Co. KGaA},\ \bibinfo {year}
  {2006})\BibitemShut {NoStop}%
\bibitem [{\citenamefont {Agarwal}(2013)}]{agarwal_book}%
  \BibitemOpen
  \bibfield  {author} {\bibinfo {author} {\bibfnamefont {G.~S.}\ \bibnamefont
  {Agarwal}},\ }\href@noop {} {\emph {\bibinfo {title} {Quantum Optics}}}\
  (\bibinfo  {publisher} {Cambridge University Press, Cambridge},\ \bibinfo
  {year} {2013})\BibitemShut {NoStop}%
\bibitem [{\citenamefont {Sperling}\ and\ \citenamefont
  {Walmsley}(2018{\natexlab{a}})}]{sperling2018a}%
  \BibitemOpen
  \bibfield  {author} {\bibinfo {author} {\bibfnamefont {J.}~\bibnamefont
  {Sperling}}\ and\ \bibinfo {author} {\bibfnamefont {I.~A.}\ \bibnamefont
  {Walmsley}},\ }\bibfield  {title} {\bibinfo {title} {Quasiprobability
  representation of quantum coherence},\ }\href
  {https://doi.org/10.1103/PhysRevA.97.062327} {\bibfield  {journal} {\bibinfo
  {journal} {Phys. Rev. A}\ }\textbf {\bibinfo {volume} {97}},\ \bibinfo
  {pages} {062327} (\bibinfo {year} {2018}{\natexlab{a}})}\BibitemShut
  {NoStop}%
\bibitem [{\citenamefont {Sperling}\ and\ \citenamefont
  {Walmsley}(2018{\natexlab{b}})}]{sperling2018b}%
  \BibitemOpen
  \bibfield  {author} {\bibinfo {author} {\bibfnamefont {J.}~\bibnamefont
  {Sperling}}\ and\ \bibinfo {author} {\bibfnamefont {I.~A.}\ \bibnamefont
  {Walmsley}},\ }\bibfield  {title} {\bibinfo {title} {Quasistates and
  quasiprobabilities},\ }\href {https://doi.org/10.1103/PhysRevA.98.042122}
  {\bibfield  {journal} {\bibinfo  {journal} {Phys. Rev. A}\ }\textbf {\bibinfo
  {volume} {98}},\ \bibinfo {pages} {042122} (\bibinfo {year}
  {2018}{\natexlab{b}})}\BibitemShut {NoStop}%
\bibitem [{\citenamefont {Sperling}\ and\ \citenamefont
  {Vogel}(2020)}]{sperling2020}%
  \BibitemOpen
  \bibfield  {author} {\bibinfo {author} {\bibfnamefont {J.}~\bibnamefont
  {Sperling}}\ and\ \bibinfo {author} {\bibfnamefont {W.}~\bibnamefont
  {Vogel}},\ }\bibfield  {title} {\bibinfo {title} {Quasiprobability
  distributions for quantum-optical coherence and beyond},\ }\href
  {https://doi.org/10.1088/1402-4896/ab5501} {\bibfield  {journal} {\bibinfo
  {journal} {Phys. Scr.}\ }\textbf {\bibinfo {volume} {95}},\ \bibinfo {pages}
  {034007} (\bibinfo {year} {2020})}\BibitemShut {NoStop}%
\end{thebibliography}%
\end{document}